\documentclass[preprintnumbers,amsmath,11pt,amssymb,floatfix,superscriptaddress,noshowpacs,nofootinbib]{revtex4}
\usepackage{graphicx}
\usepackage{epsfig}
\usepackage{bm}
\usepackage{amsfonts}
\usepackage{xcolor}
\usepackage{hyperref}

\input epsf

\begin{document}
\title{\Large Scalar Field Reconstructions of Holographic Dark Energy Models with Applications to Chaplygin Gas, DBI, Yang-Mills, and NLED Frameworks}

\author{Antonio Pasqua}
\email{toto.pasqua@gmail.com} \affiliation{Department of Physics,
University of Trieste, Via Valerio, 2 34127 Trieste, Italy.}

\date{\today}

\begin{abstract}
\textbf{Abstract}: In this study, we investigate the cosmological implications of two DE models, introduced by Chen \& Jing \cite{modelhigher} and by Granda \& Oliveros \cite{gohnde}. The first model comprises three principal components: one term proportional to the Hubble parameter $ H$ squared, and two additional terms proportional to the first and second time derivatives of $ H $, respectively. The second model, known as New Holographic Dark Energy (NHDE) model, can be considered a generalization of the Ricci DE model and it contains a term proportional to the Hubble parameter $H$ squared and one to the first time derivative of $H$. We derive the analytical expressions for the reduced Hubble parameter squared $h^2$, the Equation of State (EoS) parameter of Dark Energy (DE) $\omega_D $, the pressure of DE $p_D$ and of the deceleration parameter $q $ considering both non-interacting and later on interacting DM and DE. We also consider some limiting cases for the integration constants obtained. Furthermore, we explore the limiting scenario of a flat, dark energy-dominated Universe and establish a correspondence between the proposed DE models and various scalar field frameworks. Specifically, we examine their connection with the Generalized Chaplygin Gas, the Modified Chaplygin Gas, the Modified Variable Chaplygin Gas, the Viscous Generalized Chaplygin Gas, as well as scalar field models based on Dirac-Born-Infeld theory, Yang-Mills theory and Nonlinear Electrodynamics.
\end{abstract}

\maketitle

\tableofcontents

\section{Introduction}
{Recent observational data from sources such as the Wilkinson Microwave Anisotropy Probe (WMAP) \cite{cmb1,cmb2}, the Supernova Cosmology Project \cite{sn2,sn4}, the Sloan Digital Sky Survey (SDSS) \cite{sds1,sds3,sds4}, the Planck satellite mission \cite{planck}, and several X-ray studies \cite{xray} provide compelling evidence that the Universe is experiencing an accelerated phase of expansion in its late-time evolution. To explain this behavior, cosmologists have proposed the existence of an exotic form of energy characterized by negative pressure or equation of state and this is  referred to as dark energy (DE). Although the cosmological constant $\Lambda_{CC}$ is the most conventional and observationally consistent candidate for DE, it is has some unresolved theoretical challenges, including the cosmological constant problem and the coincidence problem, associated with it. These issues have motivated the development of various alternative theoretical models \cite{cosm1,cosm3,cosm4,cosm5}.

{The cosmological constant $\Lambda_{CC}$, introduced as a vacuum energy term in Einstein’s field equations, faces a major challenge: quantum field theory estimates based on Planck or electroweak cut-offs overshoot the observed value by $\sim10^{123}$ or $10^{55}$, respectively. This discrepancy is known as the cosmological constant problem. The coincidence problem further questions why dark energy and matter densities are comparable today despite their distinct evolution. For detailed discussions, see \cite{cosm1,cosm3,cosm4,cosm5}; for a review of early-Universe inflation, refer to \cite{cosm6}.

Within the framework of Einstein's General Relativity, the Universe is believed to be permeated by dark energy (DE). Recent observations indicate that DE accounts for approximately two-thirds of the current total cosmic energy density $\rho_{\text{tot}}$ \cite{twothirds}, with the remaining portion primarily composed of dark matter (DM) and a smaller fraction of baryonic matter. Despite extensive investigation, the fundamental nature of DE remains elusive.
As mentioned earlier, one alternative to the cosmological constant $\Lambda_{CC}$ is a class of dynamical DE models, characterized by a time-dependent equation of state (EoS) parameter $\omega$. Some analyses of Type Ia supernova (SNe Ia) data suggest that these models may offer a better fit to the observational data compared to the constant-$\Lambda$ scenario.
A wide variety of theoretical candidates have been proposed to model DE dynamics. These include scalar field theories such as quintessence \cite{quint1,quint4,quint5,quint6,quint7}, k-essence \cite{kess2,kess3,kess4,kess5}, tachyon fields \cite{tac1,tac2,tac4}, phantom energy \cite{pha2,pha5,pha6}, dilaton fields \cite{dil1,dil2,dil3} and quintom models \cite{qui2,qui3,qui8,qui12}. In addition to non-interacting models, interacting dark energy scenarios have also been explored, including models based on the Chaplygin gas \cite{cgas2,cgas3}, as well as Agegraphic Dark Energy (ADE) and its extension, New ADE (NADE) \cite{ade1,ade2}.
In the context of multifluid cosmologies, a phase-space analysis was performed in \cite{cosm6}, where various complex forms of the EoS parameter were considered to understand the dynamical evolution of the Universe in greater detail.

In addition to scalar field and Chaplygin gas models, another class of dynamical dark energy (DE) theories has emerged based on the holographic principle ~\cite{holo1,holo2,holo5}. Among these, the Holographic Dark Energy (HDE) model, initially proposed by Li~\cite{li}, has gained substantial interest and it has been well studied in different works \cite{nood1,nood2,nood4,nood6,nood7}. The holographic principle, relevant in black hole thermodynamics and string theory, states that the entropy of a physical system scales not with its volume \( V \), but with the surface area \( A \) of the enclosing region, i.e., \( S \sim A \sim L^2 \), where \( L \) is the characteristic length scale.

In Li’s formulation, the HDE density is expressed as:
\begin{eqnarray}
\rho_\Lambda = 3\alpha M_p^2 L^{-2},
\end{eqnarray}
where \( \alpha \) is a dimensionless parameter and \( M_p = \left(8\pi G_N\right)^{-1/2} \) denotes the reduced Planck mass, with \( G_N \) being Newton’s gravitational constant.

The idea of holographic DE was originally introduced by Cohen et al.~\cite{coh1}, who suggested that the energy density of DE should be constrained by the size of the Universe to avoid the formation of black holes. However, their model, in which \( \rho_\Lambda \propto H^2 \), could not drive cosmic acceleration, as it led to an effective equation of state (EoS) parameter \( w = 0 \).

In an attempt to address the shortcomings of earlier HDE models, Li~\cite{li} introduced the idea of identifying the infrared (IR) cutoff with the future event horizon. This modification proved successful in accounting for the Universe's late-time acceleration. Building on this, Gao et al.~\cite{gaoprimo} proposed an alternative approach by relating the IR scale to the Ricci scalar, specifically setting \( L \propto R^{-1/2} \), which led to the formulation of the Holographic Ricci Dark Energy (HRDE) model. Further advancements were made by Granda and collaborators~\cite{go1,go4}, who proposed a generalized HDE density incorporating both the Hubble parameter \( H \) and its time derivative \( \dot{H} \), thus offering a more flexible framework for cosmic evolution. These various formulations have undergone extensive observational scrutiny using datasets such as Type Ia Supernovae, the Cosmic Microwave Background, and Baryon Acoustic Oscillations \cite{cons1,cons2,cons3,cons4,cons6,cons7,cons8,cons9}. A more comprehensive discussion of theoretical developments and observational tests related to HDE models can be found in .~\cite{hde1,hde2,hde7,hde9,hde10,hde13,hde17,hde18,
hde22,hde23,hde24,hde26,hde28,hde30,mioviscuous,hde33,hde34,hde35,saridakis11,saridakis22,33,41,42,30,15,miogup1,saridakis2}.

In this paper, we explore a recently proposed dark energy (DE) density model introduced by Chen \& Jing~\cite{modelhigher}, which incorporates three distinct contributions: one proportional to the second time derivative of the Hubble parameter $H$ (i.e. $\ddot{H}$), one proportional to the first time derivative of the Hubble parameter (i.e. $\dot{H}$) and another proportional to the square of the Hubble parameter (i.e. $H^2$). The model is governed by three free parameters, $\alpha$, $\beta$, and $\gamma$, which serve as the respective coefficients of each term in the energy density expression. In this model, the energy density of DE is given by:
\begin{eqnarray}
    \rho_{D,higher} = 3\left[  \alpha \left(\frac{\ddot{H}}{H}\right)+\beta \dot{H}   + \gamma H^2    \right]
    \end{eqnarray}
Further details regarding this model, including its theoretical motivations and cosmological behavior, will be discussed in the following section.

Moreover, we consider  a DE model known as New Holographic Dark Energy (NHDE) model, which was introduced by Granda \& Oliveros \cite{gohnde} and studied in many other works \cite{hnde1,hnde2,hnde3,mio1,mio2,mio3}.\\
The GO cutoff involves both the square of the Hubble parameter and its time derivative, and is defined as
\begin{equation}
L_{GO} = \big(\mu H^{2} + \lambda \dot{H}\big)^{-1/2}. 
\end{equation}
This choice of cut-off was introduced by Granda \& Oliveros based on dimensional grounds.
The density of the NHDE model can be expressed as
\begin{equation}
\rho_{D,NHDE} = 3c^{2}\big(\mu H^{2} + \lambda \dot{H}\big),\label{coccolo1}
\end{equation}
where $\mu$ and $\lambda$ are positive dimensionless constants. In the limiting case corresponding to $\mu=2$ and $\lambda=1$, the expression of $\rho_{NHDE}$ given in Eq. (\ref{coccolo1}) reduces to an energy density proportional to the Ricci scalar when the spatial curvature parameter is $k=0$, i.e. for a flat Universe.\\ 
Dark energy models based on this cutoff fall within the generalized Nojiri-Odintsov holographic dark energy class \citep{valerossi}. \\
A detailed observational study by Wang \& Xu \citep{wangalfa} constrained the parameters $(\mu,\lambda)$ in non-flat cosmologies, obtaining the best-fit values 
$\mu = 0.8824^{+0.2180}_{-0.1163}$ and $\lambda = 0.5016^{+0.0973}_{-0.0871}$ at the $1\sigma$ confidence level, while in the flat case the constraints are slightly tighter: 
$\mu = 0.8502^{+0.0984}_{-0.0875}$ and $\lambda = 0.4817^{+0.0842}_{-0.0773}$.

There are several motivations for adopting the GO cut-off. If the particle horizon is taken as the infrared scale, the resulting HDE model cannot drive an accelerated expansion \citep{hsunuovo}. 
On the other hand, if the future event horizon is selected, the theory faces a causality problem. 
By contrast, the GO cutoff depends only on local cosmological quantities such as $H$ and $\dot{H}$, which eliminates causality issues and still yields a consistent late-time acceleration of the Universe.

In Sections~2 and 3, we outline the physical setup and derive key cosmological quantities for the proposed models in a non-flat Universe, focusing on the reduced Hubble parameter squared $h^2$, the equation of state parameter $\omega_D$, the pressure of DE $p_D$ and the deceleration parameter $q$. We also obtained the expressions of the quantities we study for some limiting cases of the integration constants involved and we considered the limiting case of a flat Dark Dominated Universe, which is obtained for $\Omega_k=\Omega_m=0$ and $\Omega_D=1$.  In Section 4, we explore the scalar field equivalence of the DE models we are studying by establishing correspondences with a variety of scalar field models, including the Generalized Chaplygin Gas (GCG), the Modified Chaplygin Gas (MCG), the Modified Variable Chaplygin Gas (MVCG), the Viscous Generalized Chaplygin Gas (VGCG), as well as scalar field models based on Dirac-Born-Infeld (DBI) theory, Yang-Mills (YM) theory and Nonlinear Electrodynamics (NLED) theory. Finally, Section 5 presents the main Conclusions of this paper.

\section{Holographic Model in a Non-Flat Universe}
In this Section, we outline the main features of the first Dark Energy (DE) model under investigation and derive several fundamental cosmological quantities. \\
We consider a Universe described by a spatially homogeneous and isotropic Friedmann–Lemaitre–Robertson–Walker (FLRW) spacetime, whose line element can be written as:
\begin{eqnarray}
    ds^2 = -dt^2 + a^2(t)\left[\frac{dr^2}{1 - kr^2} + r^2 (d\theta^2 + \sin^2\theta\, d\varphi^2)\right], \label{6}
\end{eqnarray}
where $t$ denotes the cosmic time, $a(t)$ is the scale factor describing the expansion history of the Universe, $r$ is the comoving radial coordinate, and $\theta$ and $\varphi$ are the usual angular coordinates in spherical symmetry, with $0 \leq \theta \leq \pi$ and $0 \leq \varphi < 2\pi$. The constant $k$ represents the spatial curvature parameter and can take the values $-1$, $0$, or $+1$, corresponding to open, flat, and closed geometries, respectively. 

The evolution of a homogeneous and isotropic Universe within the framework of General Relativity is determined by the Friedmann equations, which, in the presence of both dark energy (DE) and dark matter (DM), can be written as:
\begin{eqnarray}
    H^2 + \frac{k}{a^2} &=& \frac{1}{3M^2_p}\left( \rho_D + \rho_m \right), \label{7} \\
    \dot{H} + 2H^2 + \frac{k}{a^2} &=& \frac{8\pi G}{6}\, p_D, \label{7fri2}
\end{eqnarray}
where $H = \dot{a}/a$ is the Hubble parameter, $\rho_D$ and $p_D$ denote the energy density and pressure of dark energy, respectively, and $\rho_m$ is the energy density of pressureless dark matter (i.e., $p_m = 0$). \\

The fractional energy densities of the matter component, the dark energy component, and the spatial curvature are defined as:
\begin{eqnarray}
    \Omega_m &=& \frac{\rho_m}{\rho_{\text{cr}}} = \frac{\rho_m}{3M^2_p H^2}, \label{8} \\
    \Omega_D &=& \frac{\rho_D}{\rho_{\text{cr}}} = \frac{\rho_D}{3M^2_p H^2}, \label{10} \\
    \Omega_k &=& \frac{k}{a^2 H^2}, \label{10k}
\end{eqnarray}
where $\rho_{\text{cr}}$ is the critical energy density required for a spatially flat Universe, and is given by:
\begin{eqnarray}
    \rho_{\text{cr}} = 3M^2_p H^2.
\end{eqnarray}

Using the results of Eqs.~(\ref{8}) and (\ref{10}), the Friedmann equation given in Eq.~(\ref{7}) can be rewritten in the following dimensionless form:
\begin{eqnarray}
    \Omega_m + \Omega_D + \Omega_k = 1. \label{11}
\end{eqnarray}
In order to ensure the validity of the Bianchi identity, or equivalently the local conservation of energy-momentum, the total energy density $\rho_{\text{tot}} = \rho_D + \rho_m$ must satisfy the continuity equation:
\begin{eqnarray}
    \dot{\rho}_{\text{tot}} + 3H\left( \rho_{\text{tot}} + p_{\text{tot}} \right) = 0, \label{12old}
\end{eqnarray}
where $\rho_{\text{tot}}$ and $p_{\text{tot}}$ denote the total energy density and total pressure of the cosmic fluid, respectively, and are defined by:
\begin{eqnarray}
    \rho_{\text{tot}} &=& \rho_m + \rho_D, \\
    p_{\text{tot}} &=& p_D,
\end{eqnarray}
since DM is assumed to be pressureless.\\
The continuity equation written in Eq.~(\ref{12old}) can alternatively be expressed in terms of the total equation of state (EoS) parameter $\omega_{\text{tot}} = p_{\text{tot}} / \rho_{\text{tot}}$ as:
\begin{eqnarray}
    \dot{\rho}_{\text{tot}} + 3H\left(1 + \omega_{\text{tot}} \right) \rho_{\text{tot}} = 0. \label{12rho}
\end{eqnarray}

Since the energy densities of DM and DE are assumed to be conserved separately, we have that Eq.~(\ref{12rho}) can be decomposed into two independent continuity equations. In the non-interacting case, these read:
\begin{eqnarray}
    \dot{\rho}_D + 3H\left( 1 + \omega_D \right)\rho_D &=& 0, \label{12deold} \\
    \dot{\rho}_m + 3H\rho_m &=& 0. \label{12dm}
\end{eqnarray}
Using the general definition of the EoS parameter $w$ for DE:
\begin{eqnarray}
    \omega_D = \frac{p_D}{\rho_D}, \label{mammud}
\end{eqnarray}
we can equivalently express Eq.~(\ref{12deold}) in the following form:
\begin{eqnarray}
    \dot{\rho}_D + 3H\left( p_D + \rho_D \right) = 0. \label{12de}
\end{eqnarray}

Furthermore, Eqs.~(\ref{12deold}), (\ref{12dm}) and (\ref{12de}) can be rewritten in terms of the new variable $x = \ln a$, using the relation $\frac{d}{dt} = H \frac{d}{dx}$. Defining derivatives with respect to $x$ by a prime, we obtain:
\begin{eqnarray}
    \rho'_D + 3\left( 1 + \omega_D \right)\rho_D &=& 0, \label{12deoldprime} \\
    \rho'_D + 3\left( p_D + \rho_D \right) &=& 0, \label{12deprime} \\
    \rho'_m + 3\rho_m &=& 0. \label{12dmprime}
\end{eqnarray}

We now broaden our framework by allowing for a possible interaction between the dark components. In this picture, dark matter (DM) and dark energy (DE) are not entirely independent but may exchange energy and/or momentum. Such a coupling is frequently invoked as a way to address the so-called coincidence problem, namely the puzzling fact that the energy densities of DM and DE are of the same order today despite their very different redshift evolutions. Introducing an interaction modifies the usual cosmological dynamics and can give rise to distinctive signatures, including variations in the expansion history, altered growth of structure, and potential shifts in the anisotropy pattern of the cosmic microwave background (CMB). For this reason, interacting dark sector models have been extensively discussed as extensions of the concordance $\Lambda$CDM scenario.

When a coupling is present, the conservation laws for the two fluids, with densities $\rho_{D}$ and $\rho_{m}$, become
\begin{eqnarray}
\dot{\rho}_{D} + 3H \rho_{D}(1+\omega_{D}) &=& -Q, \label{eq:DE_cons}\\
\dot{\rho}_{m} + 3H \rho_{m} &=& Q, \label{eq:DM_cons}
\end{eqnarray}
where $Q$ describes the rate of energy exchange between DM and DE. In general, $Q$ may depend on several cosmological quantities, such as the Hubble rate $H$, the deceleration parameter $q$, and the energy densities themselves, $\rho_{m}$ and $\rho_{D}$, i.e. $Q=Q(\rho_{m},\rho_{D},H,q)$. Several functional choices for $Q$ have been put forward in the literature. In this work, we focus on the simple phenomenological prescription
\begin{equation}
Q = 3 d^{2} H \rho_{m}, \label{Q}
\end{equation}
where $d^{2}$ is a dimensionless parameter controlling the interaction strength, often called the transfer rate or coupling constant~\cite{ref144d,ref145d,ref146d}.

From Eqs. (\ref{eq:DE_cons}) and (\ref{eq:DM_cons}), we also obtain the following relations:
\begin{eqnarray}
    \rho'_D + 3\left( 1 + \omega_D \right)\rho_D &=& -Q, \label{12deoldprimeI} \\
    \rho'_D + 3\left( p_D + \rho_D \right) &=& -Q, \label{12deprimeI} \\
    \rho'_m + 3\rho_m &=& Q. \label{12dmprimeI}
\end{eqnarray}

Joint observational analyses combining Type Ia supernovae, WMAP CMB data, and BAO measurements from SDSS generally favor small and positive values of $d^{2}$. Such constraints are consistent with the expectations from the coincidence problem and with thermodynamic considerations~\cite{ref147d}. Further bounds obtained from CMB anisotropies and galaxy cluster observations indicate a viable range $0<d^{2}<0.025$~\cite{ref148d}. In broader terms, the coupling is usually considered within $0 \leq d^{2} \leq 1$, with the special case $d^{2}=0$ recovering the standard non-interacting FRW cosmology. It should also be emphasized that many other functional forms of $Q$ exist in the literature, each leading to distinct phenomenological consequences.

Since dark energy (DE) accounts for roughly two-thirds of the present-day total energy density of the Universe, while its contribution was essentially negligible in the early stages after the Big Bang, it is reasonable to postulate that the DE density evolves with the expansion of the Universe. In this context, it is natural to explore DE models in which the energy density depends on the Hubble parameter $H$ and its derivatives with respect to the cosmic time $t$, given that $H$ encapsulates the expansion rate of the Universe.

Chen \& Jing \cite{modelhigher} recently proposed a dark energy (DE) model in which the energy density $\rho_D$ contains three terms: one proportional to the Hubble parameter $H$, and two others proportional to the first and second time derivatives of $H$, respectively. The expression for the DE energy density is given by:
\begin{eqnarray}
    \rho_{higher} =  3\left[  \alpha \left(\frac{\ddot{H}}{H}\right)+\beta \dot{H}   + \gamma H^2    \right], \label{model}
\end{eqnarray}
where $\alpha$, $\beta$, and $\gamma$ are dimensionless parameters. The numerical factor 3 in Eq.~(\ref{model}) is introduced for convenience, as it simplifies subsequent calculations. For mathematical simplicity, we set the reduced Planck mass to unity, i.e. $M_p = 1$. We note that the inverse of the Hubble parameter, $H^{-1}$, is included in the first term to ensure that all terms have consistent physical dimensions.

The cosmological behavior and main features of this DE model depend crucially on the parameters $\alpha$, $\beta$, and $\gamma$. The energy density given in Eq.~(\ref{model}) can be regarded as a generalization of several previously proposed DE models. For instance, by setting $\alpha = 0$, the model reduces to the form of DE with the Granda-Oliveros infrared (IR) cut-off \cite{gohnde}. Furthermore, for the particular choice $\alpha=0$, $\beta=1$, and $\gamma=2$, the model reproduces the energy density of DE with the IR cut-off given by the average radius of the Ricci scalar, valid for a flat Universe ($k=0$). Since the model introduced here contains an additional free parameter $\alpha$, it is more general than the Ricci Dark Energy (RDE) model. Similar DE models have been studied in detail in \cite{altri1,altri2,altri3}.

In the following Sections, we derive expressions for several important cosmological quantities as functions of the variable $x = \ln a$:  the reduced Hubble parameter squared $h^2$, the DE energy density $\rho_D$, the DE pressure $p_D$, the DE equation of state (EoS) parameter $\omega_D$, and the deceleration parameter $q$. We also consider some limiting cases of the integration constants we obtain during calculations. Moreover, we calculate the same quantities for the limiting case of a flat Dark Dominated Universe, i.e. for $\Omega_m=\Omega_k=0$ and $\Omega_D=1$.\\

\subsection{Non-Interacting Case}
We start considering the non interacting case.\\
Using the variable $x = \ln a$ and substituting the dark energy (DE) density $\rho_D$ expressed in Eq. (\ref{model}) into the Friedmann equation given by Eq. (\ref{7}), we obtain the following second-order differential equation for $h^2$:
\begin{eqnarray}
\left(\frac{\alpha}{2}\right)\frac{d^2h^2}{dx^2} + \left(\frac{\beta}{2}\right)\frac{dh^2}{dx} + \left(\gamma - 1\right) h^2 + \Omega_{k0} e^{-2x} + \Omega_{m0} e^{-3x} = 0 , \label{diff1}
\end{eqnarray}
where $h$ is defined as $h = H / H_0$, with $H_0$ representing the Hubble constant, i.e. the present day value of $H$.
To derive Eq. (\ref{diff1}), we used the expression of $\rho_m$ obtained by solving the continuity equation for dark matter (DM) given in Eq. (\ref{12dm}), which leads to 
\begin{eqnarray}
\rho_m = \rho_{m0} a^{-3} = \rho_{m0} e^{-3x},
\end{eqnarray}
or equivalently to
\begin{eqnarray}
\rho_m = 3H_0^2 \Omega_{m0} a^{-3} = 3H_0^2\Omega_{m0} e^{-3x},
\end{eqnarray}
where $\rho_{m0}$ and $\Omega_{m0}$ are the current values of the DM energy density and DM density parameter, respectively. \\
The general solution of Eq. (\ref{diff1}) can be written as:
\begin{eqnarray}
h^2_{higher} &=& \Omega_{k0} e^{-2x} + \Omega_{m0} e^{-3x} + f_0 e^{-\left(\frac{\beta - \sqrt{\beta^2 - 8\alpha (\gamma - 1)}}{2\alpha}\right) x} + f_1 e^{-\left(\frac{\beta + \sqrt{\beta^2 - 8\alpha (\gamma - 1)}}{2\alpha}\right) x} \nonumber \\
&& + \left(\frac{2\alpha - \beta + \gamma}{1 - 2\alpha + \beta - \gamma}\right) \Omega_{k0} e^{-2x} + \left(\frac{9\alpha - 3\beta + 2\gamma}{2 - 9\alpha + 3\beta - 2\gamma}\right) \Omega_{m0} e^{-3x} , \label{accanon}
\end{eqnarray}
where $f_0$ and $f_1$ are two integration constants. From Eq. (\ref{accanon}), we note that the first two terms on the right-hand side represent the contributions of DM and curvature, while the other terms correspond to the DE contribution. \\
Therefore, we conclude that the dark energy density $\rho_D$ can be expressed as:
\begin{eqnarray}
\rho_{D,higher}&=& 3 H_0^2 \left[ f_0 e^{-\left(\frac{\beta - \sqrt{\beta^2 - 8\alpha (\gamma - 1)}}{2\alpha}\right) x} + f_1 e^{-\left(\frac{\beta + \sqrt{\beta^2 - 8\alpha (\gamma - 1)}}{2\alpha}\right) x} \right. \nonumber \\
&& \left. + \left(\frac{2\alpha - \beta + \gamma}{1 - 2\alpha + \beta - \gamma}\right) \Omega_{k0} e^{-2x} + \left(\frac{9\alpha - 3\beta + 2\gamma}{2 - 9\alpha + 3\beta - 2\gamma}\right) \Omega_{m0} e^{-3x} \right]. \label{PICU1}
\end{eqnarray}
Moreover, from the DE continuity equation given by Eq. (\ref{12deprime}), the general relation for the DE pressure $p_{D,higher}$ as a function of $\rho_{D,higher}$ and its first derivative with respect to $x$ is:
\begin{eqnarray}
p_{D,higher} = -\rho_{D,higher} - \frac{\rho_{D,higher}'}{3} \label{PICU3}.
\end{eqnarray}
Differentiating the expression of $\rho_{D,higher}$ with respect to $x$ yields:
\begin{eqnarray}
\rho_{D,higher}' &=& -3 H_0^2 \left[ \left(\frac{\beta - \sqrt{\beta^2 - 8\alpha (\gamma - 1)}}{2\alpha}\right) f_0 e^{-\left(\frac{\beta - \sqrt{\beta^2 - 8\alpha (\gamma - 1)}}{2\alpha}\right) x} \right. \nonumber \\
&& \left. + \left(\frac{\beta + \sqrt{\beta^2 - 8\alpha (\gamma - 1)}}{2\alpha}\right) f_1 e^{-\left(\frac{\beta + \sqrt{\beta^2 - 8\alpha (\gamma - 1)}}{2\alpha}\right) x} \right. \nonumber \\
&& \left. + 2 \left(\frac{2\alpha - \beta + \gamma}{1 - 2\alpha + \beta - \gamma}\right) \Omega_{k0} e^{-2x} + 3 \left(\frac{9\alpha - 3\beta + 2\gamma}{2 - 9\alpha + 3\beta - 2\gamma}\right) \Omega_{m0} e^{-3x} \right]. \label{PICU2}
\end{eqnarray}
Substituting in Eq. (\ref{PICU3}) the expressions of $\rho_{D,higher}$ and $\rho_{D,higher}'$ given in Eqs. (\ref{PICU1}) and (\ref{PICU2}), we obtain the following expression for $p_{D,higher}$:
\begin{eqnarray}
p_{D,higher} &=& -\frac{H_0^2}{2\alpha} \left[ \left(6\alpha - \beta + \sqrt{\beta^2 - 8\alpha (\gamma - 1)} \right) f_0 e^{-\left(\frac{\beta - \sqrt{\beta^2 - 8\alpha (\gamma - 1)}}{2\alpha}\right) x} \right. \nonumber \\
&& + \left(6\alpha - \beta - \sqrt{\beta^2 - 8\alpha (\gamma - 1)} \right) f_1 e^{-\left(\frac{\beta + \sqrt{\beta^2 - 8\alpha (\gamma - 1)}}{2\alpha}\right) x} \nonumber \\
&& \left. + 2 \alpha \left(\frac{2\alpha - \beta + \gamma}{1 - 2\alpha + \beta - \gamma}\right) \Omega_{k0} e^{-2x} \right]. \label{pdnon}
\end{eqnarray}
From the DE continuity equation given in Eq. (\ref{12deoldprime}), we obtain that the general relation for the DE equation of state (EoS) parameter $\omega_{D,higher} $ is given by:
\begin{eqnarray}
\omega_{D,higher} = -1 - \frac{\rho'_{D,higher}}{3 \rho_{D,higher} }  \label{eosnongen}.
\end{eqnarray}
Using the expressions of $\rho_{D,higher}$ and $\rho_{D,higher}'$ we obtained in Eqs. (\ref{PICU1}) and (\ref{PICU2}), we obtain the following expression for $\omega_{D,higher} $:

\begin{eqnarray}
\omega_{D,higher} &=& -1 + \frac{1}{3}\cdot \left\{   \left(\frac{\beta - \sqrt{\beta^2 - 8\alpha (\gamma - 1)}}{2\alpha}\right) f_0 e^{-\left(\frac{\beta - \sqrt{\beta^2 - 8\alpha (\gamma - 1)}}{2\alpha}\right) x} \right. \nonumber \\
&& \left. + \left(\frac{\beta + \sqrt{\beta^2 - 8\alpha (\gamma - 1)}}{2\alpha}\right) f_1 e^{-\left(\frac{\beta + \sqrt{\beta^2 - 8\alpha (\gamma - 1)}}{2\alpha}\right) x} \right. \nonumber \\
&& \left. + 2 \left(\frac{2\alpha - \beta + \gamma}{1 - 2\alpha + \beta - \gamma}\right) \Omega_{k0} e^{-2x} + 3 \left(\frac{9\alpha - 3\beta + 2\gamma}{2 - 9\alpha + 3\beta - 2\gamma}\right) \Omega_{m0} e^{-3x}    \right\}\times\nonumber \\
&&\left\{   f_0 e^{-\left(\frac{\beta - \sqrt{\beta^2 - 8\alpha (\gamma - 1)}}{2\alpha}\right) x} + f_1 e^{-\left(\frac{\beta + \sqrt{\beta^2 - 8\alpha (\gamma - 1)}}{2\alpha}\right) x} \right. \nonumber \\
&& \left. + \left(\frac{2\alpha - \beta + \gamma}{1 - 2\alpha + \beta - \gamma}\right) \Omega_{k0} e^{-2x} + \left(\frac{9\alpha - 3\beta + 2\gamma}{2 - 9\alpha + 3\beta - 2\gamma}\right) \Omega_{m0} e^{-3x}     \right\}^{-1}   \label{eosnongen}.
\end{eqnarray}

Next, we derive the expression for the deceleration parameter $q_{higher}$, generally defined as:
\begin{eqnarray}
q_{higher} &=& -1 - \frac{1}{2 h_{higher}^2} \frac{d h_{higher}^2}{dx}. \label{}
\end{eqnarray}

Differentiating the expression of $h_{higher}^2$ given by Eq. (\ref{accanon}) with respect to $x$, we obtain:
\begin{eqnarray}
 \frac{d h_{higher}^2}{dx} &=& -2\Omega_{k0} e^{-2x} -3 \Omega_{m0} e^{-3x} 
\nonumber \\
&&-\left(\frac{\beta - \sqrt{\beta^2 - 8\alpha (\gamma - 1)}}{2\alpha}\right) f_0 e^{-\left(\frac{\beta - \sqrt{\beta^2 - 8\alpha (\gamma - 1)}}{2\alpha}\right) x} \nonumber \\
&&-\left(\frac{\beta + \sqrt{\beta^2 - 8\alpha (\gamma - 1)}}{2\alpha}\right)f_1 e^{-\left(\frac{\beta + \sqrt{\beta^2 - 8\alpha (\gamma - 1)}}{2\alpha}\right) x} \nonumber \\
&& -2 \left(\frac{2\alpha - \beta + \gamma}{1 - 2\alpha + \beta - \gamma}\right) \Omega_{k0} e^{-2x} -3 \left(\frac{9\alpha - 3\beta + 2\gamma}{2 - 9\alpha + 3\beta - 2\gamma}\right) \Omega_{m0} e^{-3x} , \label{}
\end{eqnarray}

Therefore, the final expression of $q_{higher}$ is given by:
\begin{eqnarray}
q_{higher} &=& -1 + \frac{1}{2}\cdot \left\{2\Omega_{k0} e^{-2x} +3 \Omega_{m0} e^{-3x} \right.
\nonumber \\
&&\left.+\left(\frac{\beta - \sqrt{\beta^2 - 8\alpha (\gamma - 1)}}{2\alpha}\right) f_0 e^{-\left(\frac{\beta - \sqrt{\beta^2 - 8\alpha (\gamma - 1)}}{2\alpha}\right) x} \right.\nonumber \\
&&\left.+\left(\frac{\beta + \sqrt{\beta^2 - 8\alpha (\gamma - 1)}}{2\alpha}\right)f_1 e^{-\left(\frac{\beta + \sqrt{\beta^2 - 8\alpha (\gamma - 1)}}{2\alpha}\right) x}\right. \nonumber \\
&&\left. +2 \left(\frac{2\alpha - \beta + \gamma}{1 - 2\alpha + \beta - \gamma}\right) \Omega_{k0} e^{-2x} +3 \left(\frac{9\alpha - 3\beta + 2\gamma}{2 - 9\alpha + 3\beta - 2\gamma}\right) \Omega_{m0} e^{-3x}  \right\} \times\nonumber \\
&&\left\{\Omega_{k0} e^{-2x} + \Omega_{m0} e^{-3x} + f_0 e^{-\left(\frac{\beta - \sqrt{\beta^2 - 8\alpha (\gamma - 1)}}{2\alpha}\right) x} + f_1 e^{-\left(\frac{\beta + \sqrt{\beta^2 - 8\alpha (\gamma - 1)}}{2\alpha}\right) x} \nonumber \right.\\
&&\left. + \left(\frac{2\alpha - \beta + \gamma}{1 - 2\alpha + \beta - \gamma}\right) \Omega_{k0} e^{-2x} + \left(\frac{9\alpha - 3\beta + 2\gamma}{2 - 9\alpha + 3\beta - 2\gamma}\right) \Omega_{m0} e^{-3x}  \right\}^{-1}, \label{decelerationnon}
\end{eqnarray}

From the initial constraint, namely that the sum of the current density parameters equals one, the constants $f_0$ and $f_1$ must satisfy:
\begin{eqnarray}
f_0 + f_1 + \frac{\Omega_{k0}}{1 - 2\alpha + \beta - \gamma} + \frac{2 \Omega_{m0}}{2 - 9\alpha + 3\beta - 2\gamma} = 1, \label{condition}
\end{eqnarray}
which implies the existence of one free integration constant between $f_0$ and $f_1$. For this reason, as proposed in the original work by Chen \& Jing \cite{modelhigher}, in this paper we consider three distinct cases: $f_1 = 0$, $f_0 = 0$, and $f_1 = f_0 \equiv f_{01}$.

\subsubsection{Case with $f_1=0$}
We start by considering the limiting case corresponding to $f_1 = 0$.\\
Using this assumption in Eq. (\ref{condition}), we derive the following expression for $f_0$:
\begin{eqnarray}
f_0=1-  \frac{\Omega_{k0}}{1-2\alpha + \beta -\gamma}- \frac{2\Omega_{m0}}{2-9\alpha + 3\beta -2\gamma}. \label{esempio1}
\end{eqnarray}
In this case, the expression of $h^2$ is given by the following relation:
\begin{eqnarray}
h^2_{higher,1} &=& \Omega_{k0}e^{-2x}+\Omega_{m0}e^{-3x}   + f_0e^{-\left(\frac{\beta - \sqrt{\beta^2 -8\alpha\left(\gamma -1\right)}}{2\alpha}\right)x} + \nonumber \\
&& + \left(\frac{2\alpha-\beta+\gamma}{1-2\alpha +\beta -\gamma}\right)\Omega_{k0}e^{-2x}+ \left(\frac{9\alpha - 3\beta +2\gamma}{2-9\alpha + 3\beta -2\gamma}\right)\Omega_{m0}e^{-3x}. \label{accanones1}
\end{eqnarray}
We can easily derive that the density of DE $\rho_{D,higher,1}$ can be expressed in the following form:
\begin{eqnarray}
\rho_{D,higher,1} &=& 3H_0^2\left[ f_0 e^{-\left(\frac{\beta-\sqrt{\beta^2-8\alpha\left(\gamma-1 \right)}}{2\alpha}\right)x} +\left(\frac{2\alpha-\beta+\gamma}{1-2\alpha +\beta -\gamma}\right)\Omega_{k0}e^{-2x}\right.  \nonumber \\
&&\left.\, \, \, \, \, \, \, \, \, \, \, \, \, \,+    \left(\frac{9\alpha - 3\beta +2\gamma}{2-9\alpha + 3\beta -2\gamma}\right) \Omega_{m0}e^{-3x} \right]. \label{rhodes1}
\end{eqnarray}
Moreover, from the DE continuity equation given by Eq. (\ref{12deprime}), the general relation for the DE pressure $p_{D,higher,1}$ as a function of $\rho_{D,higher,1}$ and its first derivative with respect to $x$ is:
\begin{eqnarray}
p_{D,higher,1} = -\rho_{D,higher,1} - \frac{\rho_{D,higher,1}'}{3} \label{}.
\end{eqnarray}
Differentiating the expression of $\rho_{D,higher,1} $ obtained in Eq. (\ref{rhodes1}) with respect to the variable $x$, we can write the following expression for $\rho_{D,higher,1}'$:
\begin{eqnarray}
\rho_{D,higher,1}' &=& -3H_0^2 \left[ \left(\frac{\beta-\sqrt{\beta^2-8\alpha\left(\gamma-1 \right)}}{2\alpha} \right) f_0 e^{-\left(\frac{\beta-\sqrt{\beta^2-8\alpha\left(\gamma-1 \right)}}{2\alpha}\right)x}\right.\nonumber \\
&&\left.+ 2\left(\frac{2\alpha-\beta+\gamma}{1-2\alpha +\beta -\gamma}\right)\Omega_{k0}e^{-2x} +3\left(\frac{9\alpha - 3\beta +2\gamma}{2-9\alpha + 3\beta -2\gamma}\right) \Omega_{m0}e^{-3x}\right]. \label{rhodes1prime}
\end{eqnarray}
Using the continuity equation for DE given in Eq. (\ref{12de}), the pressure of DE $p_{D,higher,1}$ can be written in the following form:
\begin{eqnarray}
p_{D,higher} &=& -\frac{H_0^2}{2\alpha} \left[ \left(6\alpha - \beta + \sqrt{\beta^2 - 8\alpha (\gamma - 1)} \right) f_0 e^{-\left(\frac{\beta - \sqrt{\beta^2 - 8\alpha (\gamma - 1)}}{2\alpha}\right) x} \right. \nonumber \\
&& \left. + 2 \alpha \left(\frac{2\alpha - \beta + \gamma}{1 - 2\alpha + \beta - \gamma}\right) \Omega_{k0} e^{-2x} \right]. \label{pdnon}
\end{eqnarray}

The general expression for the EoS parameter of dark energy $\omega_{D,higher,1}$ is given by:
\begin{eqnarray}
\omega_{D,higher,1} =-1-  \frac{\rho_{D,higher,1}' }{3\rho_{D,higher,1}}  . \label{eosnon1}
\end{eqnarray}
Using the expressions of $\rho_{D,higher,1}$ and $\rho_{D,higher,1}'$ obtained in Eqs. (\ref{rhodes1}) and (\ref{rhodes1prime}), we obtain:
\begin{eqnarray}
\omega_{D,higher,1} &=& -1 + \frac{1}{3}\cdot \left\{   \left(\frac{\beta - \sqrt{\beta^2 - 8\alpha (\gamma - 1)}}{2\alpha}\right) f_0 e^{-\left(\frac{\beta - \sqrt{\beta^2 - 8\alpha (\gamma - 1)}}{2\alpha}\right) x} \right. \nonumber \\
&& \left. + 2 \left(\frac{2\alpha - \beta + \gamma}{1 - 2\alpha + \beta - \gamma}\right) \Omega_{k0} e^{-2x} + 3 \left(\frac{9\alpha - 3\beta + 2\gamma}{2 - 9\alpha + 3\beta - 2\gamma}\right) \Omega_{m0} e^{-3x}    \right\}\times\nonumber \\
&&\left\{   f_0 e^{-\left(\frac{\beta - \sqrt{\beta^2 - 8\alpha (\gamma - 1)}}{2\alpha}\right) x} \right. \nonumber \\
&& \left. + \left(\frac{2\alpha - \beta + \gamma}{1 - 2\alpha + \beta - \gamma}\right) \Omega_{k0} e^{-2x} + \left(\frac{9\alpha - 3\beta + 2\gamma}{2 - 9\alpha + 3\beta - 2\gamma}\right) \Omega_{m0} e^{-3x}     \right\}^{-1}   \label{eosnongen}.
\end{eqnarray}

Next, we derive the expression for the deceleration parameter $q_{higher,1}$, generally defined as:
\begin{eqnarray}
q_{higher,1} &=& -1 - \frac{1}{2 h_{higher,1}^2} \frac{d h_{higher,1}^2}{dx}. \label{}
\end{eqnarray}
Differentiating the expression of $h_{higher,1}^2$ obtained in Eq. (\ref{accanones1}) with respect to $x$, we can write:
\begin{eqnarray}
\frac{d h_{higher,1}^2}{dx} &=& -2\Omega_{k0}e^{-2x}-3\Omega_{m0}e^{-3x}  \nonumber \\
&&-\left(\frac{\beta - \sqrt{\beta^2 -8\alpha\left(\gamma -1\right)}}{2\alpha}\right) f_0e^{-\left(\frac{\beta - \sqrt{\beta^2 -8\alpha\left(\gamma -1\right)}}{2\alpha}\right)x} + \nonumber \\
&& -2 \left(\frac{2\alpha-\beta+\gamma}{1-2\alpha +\beta -\gamma}\right)\Omega_{k0}e^{-2x}-3 \left(\frac{9\alpha - 3\beta +2\gamma}{2-9\alpha + 3\beta -2\gamma}\right)\Omega_{m0}e^{-3x}. \label{}
\end{eqnarray}

Therefore, we obtain the following expression for $q_{higher,1}$:
\begin{eqnarray}
q_{higher,1} &=& -1 + \frac{1}{2}\cdot \left\{2\Omega_{k0} e^{-2x} +3 \Omega_{m0} e^{-3x} \right.
\nonumber \\
&&\left.+\left(\frac{\beta - \sqrt{\beta^2 - 8\alpha (\gamma - 1)}}{2\alpha}\right) f_0 e^{-\left(\frac{\beta - \sqrt{\beta^2 - 8\alpha (\gamma - 1)}}{2\alpha}\right) x} \right.\nonumber \\
&&\left. +2 \left(\frac{2\alpha - \beta + \gamma}{1 - 2\alpha + \beta - \gamma}\right) \Omega_{k0} e^{-2x} +3 \left(\frac{9\alpha - 3\beta + 2\gamma}{2 - 9\alpha + 3\beta - 2\gamma}\right) \Omega_{m0} e^{-3x}  \right\} \times\nonumber \\
&&\left\{\Omega_{k0} e^{-2x} + \Omega_{m0} e^{-3x} + f_0 e^{-\left(\frac{\beta - \sqrt{\beta^2 - 8\alpha (\gamma - 1)}}{2\alpha}\right) x} \nonumber \right.\\
&&\left. + \left(\frac{2\alpha - \beta + \gamma}{1 - 2\alpha + \beta - \gamma}\right) \Omega_{k0} e^{-2x} + \left(\frac{9\alpha - 3\beta + 2\gamma}{2 - 9\alpha + 3\beta - 2\gamma}\right) \Omega_{m0} e^{-3x}  \right\}^{-1}, \label{decelerationnon}
\end{eqnarray}

\subsubsection{Case with $f_0=0$}
We now consider the second limiting case, i.e. the one corresponding to $f_0=0$. Using this condition in Eq. (\ref{condition}), we find that $f_1$ is given by:
\begin{eqnarray}
f_1 = 1-  \frac{\Omega_{k0}}{1-2\alpha + \beta -\gamma}-\frac{2\Omega_{m0}}{2-9\alpha + 3\beta -2\gamma}. \label{esempio2}
\end{eqnarray}
The general expression of the Hubble parameter squared in this case can be written as:
\begin{eqnarray}
h^2_{higher,2} &=&\Omega_{k0}e^{-2x}+ \Omega_{m0}e^{-3x}    + f_1e^{-\left(\frac{\beta + \sqrt{\beta^2 - 8\alpha\left(\gamma -1\right)}}{2\alpha}\right)x}\nonumber \\
&&+ \left(\frac{2\alpha-\beta+\gamma}{1-2\alpha +\beta -\gamma}\right)\Omega_{k0}e^{-2x}+\left(\frac{9\alpha - 3\beta +2\gamma}{2-9\alpha + 3\beta -2\gamma}\right)\Omega_{m0}e^{-3x} . \label{aga4}
\end{eqnarray}
Moreover, the density $\rho_{D,higher,2}$ is given by the following relation:
\begin{eqnarray}
\rho_{D,higher,2} &=& 3H_0^2\left[ f_1 e^{-\left(\frac{\beta+\sqrt{\beta^2-8\alpha\left(\gamma-1 \right)}}{2\alpha}\right)x}  +\left(\frac{  2 \alpha -\beta +\gamma }{1-2 \alpha +\beta-\gamma }\right)\Omega_{k0}e^{-2x} \right. \nonumber \\
&&\left.+\left(\frac{9\alpha - 3\beta +2\gamma}{2-9\alpha + 3\beta -2\gamma}\right) \Omega_{m0}e^{-3x}\right], \label{rhoesempio2} 
\end{eqnarray}

We now want to derive the final expression of $p_{D,higher,2}$.\\
We start from the general definition:
\begin{eqnarray}
p_{D,higher,2} = -\rho_{D,higher,2} - \frac{\rho_{D,higher,2}'}{3} \label{PICU3}.
\end{eqnarray}
Differentiating with respect to $x$ the expression of $\rho_{D,higher,2}$, we can write:
\begin{eqnarray}
\rho_{D,higher,2}' &=& -3 H_0^2 \left[ \left(\frac{\beta + \sqrt{\beta^2 - 8\alpha (\gamma - 1)}}{2\alpha}\right) f_1 e^{-\left(\frac{\beta + \sqrt{\beta^2 - 8\alpha (\gamma - 1)}}{2\alpha}\right) x} \right. \nonumber \\
&& \left. + 2 \left(\frac{2\alpha - \beta + \gamma}{1 - 2\alpha + \beta - \gamma}\right) \Omega_{k0} e^{-2x} + 3 \left(\frac{9\alpha - 3\beta + 2\gamma}{2 - 9\alpha + 3\beta - 2\gamma}\right) \Omega_{m0} e^{-3x} \right]. \label{PICU4}
\end{eqnarray}
Therefore, the final expression of $p_{D,higher,2}$ is given by:
\begin{eqnarray}\
p_{D,higher,2} &=& -\left(\frac{6\alpha-\beta-\sqrt{\beta^2-8\alpha\left(\gamma-1 \right)}}{2\alpha}\right)f_1H_0^2e^{-\left(\frac{\beta+\sqrt{\beta^2-8\alpha\left(\gamma-1 \right)}}{2\alpha}\right)x}\nonumber \\
&&-H_0^2\left(\frac{  2 \alpha -\beta +\gamma }{1-2 \alpha +\beta-\gamma }\right)\Omega_{k0}e^{-2x}. \label{pressureesempio2}
\end{eqnarray}

The general expression of the EoS parameter for this case is given by:
\begin{eqnarray}
\omega_{D,higher,2} =-1-  \frac{\rho_{D,higher,2}' }{3\rho_{D,higher,2}} . \label{eosnon1}
\end{eqnarray}

Using the expressions of $\rho_{D,higher,2}$ and $\rho'_{D,higher,2}$ given in Eqs. (\ref{rhoesempio2}) and (\ref{PICU4}), we can write:
\begin{eqnarray}
\omega_{D,higher,2} &=& -1 + \frac{1}{3}\cdot \left\{  \left(\frac{\beta + \sqrt{\beta^2 - 8\alpha (\gamma - 1)}}{2\alpha}\right) f_1 e^{-\left(\frac{\beta + \sqrt{\beta^2 - 8\alpha (\gamma - 1)}}{2\alpha}\right) x} \right. \nonumber \\
&& \left. + 2 \left(\frac{2\alpha - \beta + \gamma}{1 - 2\alpha + \beta - \gamma}\right) \Omega_{k0} e^{-2x} + 3 \left(\frac{9\alpha - 3\beta + 2\gamma}{2 - 9\alpha + 3\beta - 2\gamma}\right) \Omega_{m0} e^{-3x}    \right\}\times\nonumber \\
&&\left\{  f_1 e^{-\left(\frac{\beta + \sqrt{\beta^2 - 8\alpha (\gamma - 1)}}{2\alpha}\right) x} \right. \nonumber \\
&& \left. + \left(\frac{2\alpha - \beta + \gamma}{1 - 2\alpha + \beta - \gamma}\right) \Omega_{k0} e^{-2x} + \left(\frac{9\alpha - 3\beta + 2\gamma}{2 - 9\alpha + 3\beta - 2\gamma}\right) \Omega_{m0} e^{-3x}     \right\}^{-1}   \label{eosnongen}.
\end{eqnarray}

We now want to obtain the final expression of the deceleration parameter.\\
we start from the general definition:
\begin{eqnarray}
q_{higher,2} &=& -1 - \frac{1}{2 h_{higher,2}^2} \frac{d h_{higher,2}^2}{dx}. \label{}
\end{eqnarray}
Differentiating the expression of $h_{higher,2}^2$ with respect to $x$, we have the following expression:
\begin{eqnarray}
\frac{d h_{higher,2}^2}{dx} &=&-2\Omega_{k0}e^{-2x}-3 \Omega_{m0}e^{-3x}  \nonumber \\
&&-\left(\frac{\beta + \sqrt{\beta^2 - 8\alpha\left(\gamma -1\right)}}{2\alpha}\right) f_1e^{-\left(\frac{\beta + \sqrt{\beta^2 - 8\alpha\left(\gamma -1\right)}}{2\alpha}\right)x}\nonumber \\
&&-2 \left(\frac{2\alpha-\beta+\gamma}{1-2\alpha +\beta -\gamma}\right)\Omega_{k0}e^{-2x}-3\left(\frac{9\alpha - 3\beta +2\gamma}{2-9\alpha + 3\beta -2\gamma}\right)\Omega_{m0}e^{-3x} . \label{aga4}
\end{eqnarray}
Therefore, we obtain that $q_{higher,2}$ is given by:
\begin{eqnarray}
q_{higher,2} &=& -1 + \frac{1}{2}\cdot \left\{2\Omega_{k0} e^{-2x} +3 \Omega_{m0} e^{-3x} \right.
\nonumber \\
&&\left.+\left(\frac{\beta + \sqrt{\beta^2 - 8\alpha (\gamma - 1)}}{2\alpha}\right)f_1 e^{-\left(\frac{\beta + \sqrt{\beta^2 - 8\alpha (\gamma - 1)}}{2\alpha}\right) x}\right. \nonumber \\
&&\left. +2 \left(\frac{2\alpha - \beta + \gamma}{1 - 2\alpha + \beta - \gamma}\right) \Omega_{k0} e^{-2x} +3 \left(\frac{9\alpha - 3\beta + 2\gamma}{2 - 9\alpha + 3\beta - 2\gamma}\right) \Omega_{m0} e^{-3x}  \right\} \times\nonumber \\
&&\left\{\Omega_{k0} e^{-2x} + \Omega_{m0} e^{-3x} + f_1 e^{-\left(\frac{\beta + \sqrt{\beta^2 - 8\alpha (\gamma - 1)}}{2\alpha}\right) x} \nonumber \right.\\
&&\left. + \left(\frac{2\alpha - \beta + \gamma}{1 - 2\alpha + \beta - \gamma}\right) \Omega_{k0} e^{-2x} + \left(\frac{9\alpha - 3\beta + 2\gamma}{2 - 9\alpha + 3\beta - 2\gamma}\right) \Omega_{m0} e^{-3x}  \right\}^{-1}, \label{decelerationnon}
\end{eqnarray}

\subsubsection{Case with $f_0=f_1\equiv f_{01}$}
From Eq. (\ref{condition}), we obtain that $f_{01}$ assumes the following general expression:
\begin{eqnarray}
f_{01}= \frac{1}{2}\left(1-\frac{\Omega_{k0}}{1-2\alpha + \beta -\gamma}-\frac{2\Omega_{m0}}{2-9\alpha + 3\beta -2\gamma}\right)  , \label{condition3}
\end{eqnarray}
For this limiting case, we derive that the final expression of the Hubble parameter squared is given by:
\begin{eqnarray}
h^2_{higher,3} &=& \Omega_{k0}e^{-2x}+\Omega_{m0}e^{-3x}\nonumber \\
&&+f_{01} e^{-\left(\frac{\beta - \sqrt{\beta^2 -8\alpha\left(\gamma -1\right)}}{2\alpha}\right)x} + f_{01}e^{-\left(\frac{\beta + \sqrt{\beta^2 - 8\alpha\left(\gamma -1\right)}}{2\alpha}\right)x} \nonumber \\
&&+ \left(\frac{2\alpha-\beta+\gamma}{1-2\alpha +\beta -\gamma}\right)\Omega_{k0}e^{-2x}+\left(\frac{9\alpha - 3\beta +2\gamma}{2-9\alpha + 3\beta -2\gamma}\right)\Omega_{m0}e^{-3x} , \label{accanon3}
\end{eqnarray}
Using  Eq. (\ref{accanon3}), we also obtain that the  density of  DE $\rho_{D,higher,3}$ is given by:
\begin{eqnarray}
\rho_{D,higher,3} &=&3H_0^2\left[ f_{01}e^{-\left(\frac{\beta - \sqrt{\beta^2 -8\alpha\left(\gamma -1\right)}}{2\alpha}\right)x} + f_{01}e^{-\left(\frac{\beta + \sqrt{\beta^2 - 8\alpha\left(\gamma -1\right)}}{2\alpha}\right)x}\right. \nonumber \\
&&\left. + \left(\frac{2\alpha -\beta+\gamma }{1-2\alpha +\beta -\gamma}\right)\Omega_{k0}e^{-2x}+\left(\frac{9\alpha - 3\beta +2\gamma}{2-9\alpha + 3\beta -2\gamma}\right)\Omega_{m0}e^{-3x}\right]. \label{rhodnon}
\end{eqnarray}
From the continuity equation for DE given in Eq. (\ref{12deprime}), we obtain the following general relation for the pressure of DE as function of the density of DE and its first derivative with respect to $x$:
\begin{eqnarray}
p_{D,higher,3} = - \rho_{D,higher,3} - \frac{\rho_{D,higher,3}'}{3} \label{pdnongen}.
\end{eqnarray}
 Differentiating the expression of $\rho_{D,higher,3}$ given in Eq. (\ref{rhodnon}) with respect to $x$, we obtain the following result for $\rho_{D,higher,3}$:
\begin{eqnarray}
\rho_{D,higher,3}'&=& -3 H_0^2 \left[\left(\frac{ \beta -\sqrt{\beta ^2-8 \alpha
 (\gamma-1 )} }{2 \alpha }\right)f_{01}e^{-\left(\frac{\beta -\sqrt{\beta ^2-8 \alpha  (\gamma-1 )}}{2 \alpha }\right)x} \right.\nonumber \\
&&\left. +\left(\frac{  \beta +\sqrt{\beta
^2-8 \alpha  (\gamma-1 )} }{2 \alpha }\right)f_{01}e^{-\left(\frac{\beta +\sqrt{\beta ^2-8 \alpha  (\gamma-1 )}}{2 \alpha }\right)x}\right.\nonumber \\
&&\left. +2\left(\frac{ 2 \alpha -\beta +\gamma  }{1-2 \alpha +\beta -\gamma }\right)\Omega _{k0}e^{-2 x}+3\left(\frac{
9 \alpha -3 \beta +2 \gamma   }{2-9 \alpha +3 \beta -2 \gamma }\right)\Omega _{m0}e^{-3 x}\right]. \label{rhodnonprime}
\end{eqnarray}
The final expression of $p_{D,higher,3}$ is then given by:
\begin{eqnarray}
p_{D,higher,3} &=& -\frac{H_0^2}{2\alpha} \left[ \left(6\alpha - \beta + \sqrt{\beta^2 - 8\alpha (\gamma - 1)} \right) f_{01} e^{-\left(\frac{\beta - \sqrt{\beta^2 - 8\alpha (\gamma - 1)}}{2\alpha}\right) x} \right. \nonumber \\
&& + \left(6\alpha - \beta - \sqrt{\beta^2 - 8\alpha (\gamma - 1)} \right) f_{01} e^{-\left(\frac{\beta + \sqrt{\beta^2 - 8\alpha (\gamma - 1)}}{2\alpha}\right) x} \nonumber \\
&& \left. + 2 \alpha \left(\frac{2\alpha - \beta + \gamma}{1 - 2\alpha + \beta - \gamma}\right) \Omega_{k0} e^{-2x} \right]. \label{pdnon}
\end{eqnarray}

Using the expression of the continuity equation for DE given in Eqs. (\ref{12deoldprime}), we obtain the following general relation for the EoS parameter of DE $\omega_{D,higher,3}$:
\begin{eqnarray}
\omega_{D,higher,3} = - 1 - \frac{\rho_{D,higher,3}'}{3\rho_{D,higher,3}} \label{eosnongen}.
\end{eqnarray}
Using the expressions of $\rho_{D,higher,3}'$ and $\rho_{D,higher,3}'$ derived above, we can write:

\begin{eqnarray}
\omega_{D,higher,3} &=& -1 + \frac{1}{3}\cdot \left\{   \left(\frac{\beta - \sqrt{\beta^2 - 8\alpha (\gamma - 1)}}{2\alpha}\right) f_{01} e^{-\left(\frac{\beta - \sqrt{\beta^2 - 8\alpha (\gamma - 1)}}{2\alpha}\right) x} \right. \nonumber \\
&& \left. + \left(\frac{\beta + \sqrt{\beta^2 - 8\alpha (\gamma - 1)}}{2\alpha}\right) f_{01}e^{-\left(\frac{\beta + \sqrt{\beta^2 - 8\alpha (\gamma - 1)}}{2\alpha}\right) x} \right. \nonumber \\
&& \left. + 2 \left(\frac{2\alpha - \beta + \gamma}{1 - 2\alpha + \beta - \gamma}\right) \Omega_{k0} e^{-2x} + 3 \left(\frac{9\alpha - 3\beta + 2\gamma}{2 - 9\alpha + 3\beta - 2\gamma}\right) \Omega_{m0} e^{-3x}    \right\}\times\nonumber \\
&&\left\{   f_{01} e^{-\left(\frac{\beta - \sqrt{\beta^2 - 8\alpha (\gamma - 1)}}{2\alpha}\right) x} + f_{01} e^{-\left(\frac{\beta + \sqrt{\beta^2 - 8\alpha (\gamma - 1)}}{2\alpha}\right) x} \right. \nonumber \\
&& \left. + \left(\frac{2\alpha - \beta + \gamma}{1 - 2\alpha + \beta - \gamma}\right) \Omega_{k0} e^{-2x} + \left(\frac{9\alpha - 3\beta + 2\gamma}{2 - 9\alpha + 3\beta - 2\gamma}\right) \Omega_{m0} e^{-3x}     \right\}^{-1}   \label{eosnongen}.
\end{eqnarray}

For this limiting case, we can write the deceleration parameter as:
\begin{eqnarray}
q_{higher,3} = -1 -\frac{1}{2h^2_{higher,3}}\left(\frac{dh^2_{higher,3}}{dx}\right). \label{deceleration}
\end{eqnarray}
Differentiating the expression of $h^2_{higher,3}$ with respect to the variable $x$, we can write:
\begin{eqnarray}
\frac{dh^2_{higher,3}}{dx} &=& -2\Omega_{k0}e^{-2x}-2\Omega_{m0}e^{-3x}\nonumber \\
&&- \left(\frac{\beta - \sqrt{\beta^2 -8\alpha\left(\gamma -1\right)}}{2\alpha}\right)f_{01}e^{-\left(\frac{\beta - \sqrt{\beta^2 -8\alpha\left(\gamma -1\right)}}{2\alpha}\right)x} \nonumber \\
&&-\left(\frac{\beta + \sqrt{\beta^2 -8\alpha\left(\gamma -1\right)}}{2\alpha}\right) f_{01}e^{-\left(\frac{\beta + \sqrt{\beta^2 - 8\alpha\left(\gamma -1\right)}}{2\alpha}\right)x} \nonumber \\
&&-2 \left(\frac{2\alpha-\beta+\gamma}{1-2\alpha +\beta -\gamma}\right)\Omega_{k0}e^{-2x}-3\left(\frac{9\alpha - 3\beta +2\gamma}{2-9\alpha + 3\beta -2\gamma}\right)\Omega_{m0}e^{-3x} \label{deceleration}
\end{eqnarray}

Therefore, the final expression of $q_{higher,3}$ is given by:
\begin{eqnarray}
q_{higher,3} &=& -1 + \frac{1}{2}\cdot \left\{2\Omega_{k0} e^{-2x} +3 \Omega_{m0} e^{-3x} \right.
\nonumber \\
&&\left.+\left(\frac{\beta - \sqrt{\beta^2 - 8\alpha (\gamma - 1)}}{2\alpha}\right) f_{01} e^{-\left(\frac{\beta - \sqrt{\beta^2 - 8\alpha (\gamma - 1)}}{2\alpha}\right) x} \right.\nonumber \\
&&\left.+\left(\frac{\beta + \sqrt{\beta^2 - 8\alpha (\gamma - 1)}}{2\alpha}\right)f_{01} e^{-\left(\frac{\beta + \sqrt{\beta^2 - 8\alpha (\gamma - 1)}}{2\alpha}\right) x}\right. \nonumber \\
&&\left. +2 \left(\frac{2\alpha - \beta + \gamma}{1 - 2\alpha + \beta - \gamma}\right) \Omega_{k0} e^{-2x} +3 \left(\frac{9\alpha - 3\beta + 2\gamma}{2 - 9\alpha + 3\beta - 2\gamma}\right) \Omega_{m0} e^{-3x}  \right\} \times\nonumber \\
&&\left\{\Omega_{k0} e^{-2x} + \Omega_{m0} e^{-3x} + f_{01} e^{-\left(\frac{\beta - \sqrt{\beta^2 - 8\alpha (\gamma - 1)}}{2\alpha}\right) x} + f_{01} e^{-\left(\frac{\beta + \sqrt{\beta^2 - 8\alpha (\gamma - 1)}}{2\alpha}\right) x} \nonumber \right.\\
&&\left. + \left(\frac{2\alpha - \beta + \gamma}{1 - 2\alpha + \beta - \gamma}\right) \Omega_{k0} e^{-2x} + \left(\frac{9\alpha - 3\beta + 2\gamma}{2 - 9\alpha + 3\beta - 2\gamma}\right) \Omega_{m0} e^{-3x}  \right\}^{-1}, \label{decelerationnon}
\end{eqnarray}

\subsection{Interacting Case}
We now consider the presence of interaction between DE and DM. \\
From the continuity equation for DM given in Eq. (\ref{12dmprimeI}), using the definition of $Q$ we have chosen, we obtain that:
\begin{eqnarray}\label{10}
\rho_{m}=\rho_{m_{0}}e^{-3(1-d^2)x},
\end{eqnarray}
which leads to:
\begin{eqnarray}
\Omega_{m}=\Omega_{m0}e^{-3(1-d^2)x},\label{bbb1}
\end{eqnarray}

Substituting the dark energy (DE) density $\rho_D$ expressed in Eq. (\ref{model}) along with the result of Eq. (\ref{bbb1}) into the Friedmann equation given by Eq. (\ref{7}), we obtain the following second-order differential equation for $h^2$:
\begin{eqnarray}
\left(\frac{\alpha}{2}\right)\frac{d^2h^2}{dx^2} + \left(\frac{\beta}{2}\right)\frac{dh^2}{dx} + \left(\gamma - 1\right) h^2 + \Omega_{k0} e^{-2x} + \Omega_{m0} e^{-3(1-d^2)x} = 0 , \label{}
\end{eqnarray}
which general solution is given by:
\begin{eqnarray}
h^2_{higher,I} &=& C_1 \, e^{\frac{-\beta - \sqrt{\beta^2 - 8 \alpha (\gamma-1)}}{2\alpha} x} + C_2 \, e^{\frac{-\beta + \sqrt{\beta^2 - 8 \alpha (\gamma-1)}}{2\alpha} x}\nonumber  \\
&& + \frac{ \, \Omega_{k0}e^{-2x} }{1-2\alpha + \beta - \gamma }  \nonumber \\
&& + \frac{2 \, \Omega_{m0}e^{-3(1-d^2)x}}{2-9\alpha (1-d^2)^2 + 3\beta (1-d^2) - 2 \gamma } \label{king1}
\end{eqnarray}
where $C_1$ and $C_2$ are two integration constants which can be set by the initial conditions.\\ 
Eq. (\ref{king1}) can be also written as:
\begin{eqnarray}
h^2_{higher,I} &=& \Omega_{k0} e^{-2x} + \Omega_{m0} e^{-3(1-d^2)x}\nonumber \\
&&+ C_1 e^{-\left(\frac{\beta - \sqrt{\beta^2 - 8\alpha (\gamma - 1)}}{2\alpha}\right) x} + C_2 e^{-\left(\frac{\beta + \sqrt{\beta^2 - 8\alpha (\gamma - 1)}}{2\alpha}\right) x} \nonumber \\
&& + \left(\frac{2\alpha - \beta + \gamma}{1 - 2\alpha + \beta - \gamma}\right) \Omega_{k0} e^{-2x} \nonumber \\
&&+ \left[\frac{9\alpha (1-d^2)^2 - 3\beta (1-d^2) + 2 \gamma}{2-9\alpha (1-d^2)^2 + 3\beta (1-d^2) - 2 \gamma }\right] \,\Omega_{m0}e^{-3(1-d^2)x} , \label{king2}
\end{eqnarray}

Therefore, from Eq. (\ref{king2}), we obtain that the expression for the dark energy density $\rho_{D,\text{higher},I}$ can be written as follows:
\begin{eqnarray}
\rho_{D,higher,I}&=& 3H_0^2\left\{C_1 e^{-\left(\frac{\beta - \sqrt{\beta^2 - 8\alpha (\gamma - 1)}}{2\alpha}\right) x} + C_2 e^{-\left(\frac{\beta + \sqrt{\beta^2 - 8\alpha (\gamma - 1)}}{2\alpha}\right) x} \right.\nonumber \\
&&\left. + \left(\frac{2\alpha - \beta + \gamma}{1 - 2\alpha + \beta - \gamma}\right) \Omega_{k0} e^{-2x} + \left[\frac{9\alpha (1-d^2)^2 - 3\beta (1-d^2) + 2 \gamma}{2-9\alpha (1-d^2)^2 + 3\beta (1-d^2) - 2 \gamma }\right] \, \Omega_{m0}e^{-3(1-d^2)x} \right\}, \label{}
\end{eqnarray}
We now want to derive the final expression of the pressure.\\
From the continuity equation for DE given in Eq. (\ref{12deprimeI}), we obtain the following general relation for the pressure of DE $p_{D,higher,I}$ as function of the density of DE and its first derivative with respect to $x$:
\begin{eqnarray}
p_{D,higher,I} = - \rho_{D,higher,I} - \frac{\rho_{D,higher,I}'}{3} -\frac{Q}{3H}\label{king3}.
\end{eqnarray}
Differentiating the expression of $\rho_{D,higher,I}$ with respect to $x$, we obtain that $\rho'_{D,higher,I}$ is given by: 
\begin{eqnarray}
\rho'_{D,higher,I} &=& -3H_0^2\left\{\left(\frac{\beta - \sqrt{\beta^2 - 8\alpha (\gamma - 1)}}{2\alpha}\right)C_1 e^{-\left(\frac{\beta - \sqrt{\beta^2 - 8\alpha (\gamma - 1)}}{2\alpha}\right) x}  \nonumber \right.\\
&&\left.+\left(\frac{\beta + \sqrt{\beta^2 - 8\alpha (\gamma - 1)}}{2\alpha}\right)C_2 e^{-\left(\frac{\beta + \sqrt{\beta^2 - 8\alpha (\gamma - 1)}}{2\alpha}\right) x} \right.\nonumber \\
&&\left. +2 \left(\frac{2\alpha - \beta + \gamma}{1 - 2\alpha + \beta - \gamma}\right) \Omega_{k0} e^{-2x}\nonumber \right.\\
&&\left.+3(1-d^2) \left[\frac{9\alpha (1-d^2)^2 - 3\beta (1-d^2) + 2 \gamma}{2-9\alpha (1-d^2)^2 + 3\beta (1-d^2) - 2 \gamma }\right] \, \Omega_{m0}e^{-3(1-d^2)x} \right\}, \label{}
\end{eqnarray}
Moreover, we have that:
\begin{eqnarray}
    \frac{Q}{3H} = d^2\rho_m = 3H_0^2d^2\Omega_{m0}e^{-3(1-d^2)x}
\end{eqnarray}
Inserting in Eq. (\ref{king3}) the expression of $\rho_{D,higher,I}$, $\rho'_{D,higher,I}$ and $\frac{Q}{3H}$ we derived and defined, we obtained the following expression for $p_{D,higher,I} $:
\begin{eqnarray}
p_{D,higher,I} &=& -\frac{H_0^2}{2\alpha} \Bigg[
\left( 6\alpha - \beta +\sqrt{\beta^2 - 8\alpha(\gamma - 1)} \right)
C_1 e^{-\left(\frac{\beta - \sqrt{\beta^2 - 8\alpha(\gamma - 1)}}{2\alpha}\,\right)x} \nonumber \\
&& + \left( 6\alpha - \beta - \sqrt{\beta^2 - 8\alpha(\gamma - 1)} \right)
C_2 e^{-\left(\frac{\beta + \sqrt{\beta^2 - 8\alpha(\gamma - 1)}}{2\alpha}\,\right)x} \nonumber \\
&& + 2\alpha\left(\frac{2\alpha - \beta + \gamma}{1 - 2\alpha + \beta - \gamma}\right)\Omega_{k0}\,e^{-2x} \nonumber \\
&& + 6\alpha d^2 \left(\frac{9\alpha(1-d^2)^2 - 3\beta(1-d^2) + 2\gamma}{2 - 9\alpha(1-d^2)^2 + 3\beta(1-d^2) - 2\gamma}\right)
\Omega_{m0}e^{-3(1-d^2)x}
\Bigg].
\end{eqnarray}
We now want to derive the final expression of the EoS parameter.\\
From the DE continuity equation given in Eq. (\ref{12deoldprimeI}), the general relation for the DE equation of state (EoS) parameter $\omega_D$ is:
\begin{eqnarray}
\omega_{D,higher,I} = -1 - \frac{\rho'_{D,higher,I}}{3 \rho_{D,higher,I} } -\frac{Q}{3H\rho_{D,higher,I}} \label{eosnongen}.
\end{eqnarray}
The term $\frac{Q}{3H\rho_{D,higher,I}}$ can be written as:
\begin{eqnarray}
    \frac{Q}{3H\rho_{D,higher,I}} = d^2\left(\frac{\rho_m}{\rho_{D,higher,I}}\right)
\end{eqnarray}

Therefore, using the expression of $\rho_{D,higher,I}$, $\rho'_{D,higher,I}$ and $\frac{Q}{3H\rho_{D,higher,I}}$ defined above, we obtain the following result:
\begin{eqnarray}
    \omega_{D,higher,I}&=& -1 +\frac{1}{3}\cdot \left\{\left(\frac{\beta - \sqrt{\beta^2 - 8\alpha (\gamma - 1)}}{2\alpha}\right)C_1 e^{-\left(\frac{\beta - \sqrt{\beta^2 - 8\alpha (\gamma - 1)}}{2\alpha}\right) x}  \nonumber \right.\\
&&\left.+\left(\frac{\beta + \sqrt{\beta^2 - 8\alpha (\gamma - 1)}}{2\alpha}\right)C_2 e^{-\left(\frac{\beta + \sqrt{\beta^2 - 8\alpha (\gamma - 1)}}{2\alpha}\right) x} \right.\nonumber \\
&&\left. +2 \left(\frac{2\alpha - \beta + \gamma}{1 - 2\alpha + \beta - \gamma}\right) \Omega_{k0} e^{-2x}\nonumber \right.\\
&&\left.+\left\{3(1-d^2) \left[\frac{9\alpha (1-d^2)^2 - 3\beta (1-d^2) + 2 \gamma}{2-9\alpha (1-d^2)^2 + 3\beta (1-d^2) - 2 \gamma }\right]-3d^2\right\} \, \Omega_{m0}e^{-3(1-d^2)x} \right\}\times \nonumber \\
&&\left\{C_1 e^{-\left(\frac{\beta - \sqrt{\beta^2 - 8\alpha (\gamma - 1)}}{2\alpha}\right) x} + C_2 e^{-\left(\frac{\beta + \sqrt{\beta^2 - 8\alpha (\gamma - 1)}}{2\alpha}\right) x} \right.\nonumber \\
&&\left. + \left(\frac{2\alpha - \beta + \gamma}{1 - 2\alpha + \beta - \gamma}\right) \Omega_{k0} e^{-2x} \nonumber \right.\\
&&\left.+ \left[\frac{9\alpha (1-d^2)^2 - 3\beta (1-d^2) + 2 \gamma}{2-9\alpha (1-d^2)^2 + 3\beta (1-d^2) - 2 \gamma }\right] \, \Omega_{m0}e^{-3(1-d^2)x} \right\}^{-1}
\end{eqnarray}

We now want to derive the final expression of the deceleration parameter $q_{higher,I}$, generally defined as:
\begin{eqnarray}
q_{higher,I} &=& -1 - \frac{1}{2 h_{higher,I}^2} \frac{d h_{higher,I}^2}{dx}. \label{}
\end{eqnarray}
Differentiating the expression of $h_{higher,I}^2$ with respect to the variable $x$, we can write

\begin{eqnarray}
\frac{d h_{higher,I}^2}{dx} &=& -2\Omega_{k0} e^{-2x} -3(1-d^2) \Omega_{m0} e^{-3(1-d^2)x} \nonumber \\
&&-\left(\frac{\beta - \sqrt{\beta^2 - 8\alpha (\gamma - 1)}}{2\alpha}\right) C_1 e^{-\left(\frac{\beta - \sqrt{\beta^2 - 8\alpha (\gamma - 1)}}{2\alpha}\right) x} \nonumber \\
&&-\left(\frac{\beta + \sqrt{\beta^2 - 8\alpha (\gamma - 1)}}{2\alpha}\right) C_2 e^{-\left(\frac{\beta + \sqrt{\beta^2 - 8\alpha (\gamma - 1)}}{2\alpha}\right) x} \nonumber \\
&& -2 \left(\frac{2\alpha - \beta + \gamma}{1 - 2\alpha + \beta - \gamma}\right) \Omega_{k0} e^{-2x} \nonumber \\
&&-3(1-d^2) \left[\frac{9\alpha (1-d^2)^2 - 3\beta (1-d^2) + 2 \gamma}{2-9\alpha (1-d^2)^2 + 3\beta (1-d^2) - 2 \gamma }\right] \, \Omega_{m0}e^{-3(1-d^2)x} , \label{}
\end{eqnarray}
Therefore, the final expression of $q_{higher,I}$ is given by:
\begin{eqnarray}
    q_{higher,I} &=& -1 + \frac{1}{2}\cdot \left\{  2\Omega_{k0} e^{-2x} +3(1-d^2) \Omega_{m0} e^{-3(1-d^2)x} \right.\nonumber \\
&&\left.+\left(\frac{\beta - \sqrt{\beta^2 - 8\alpha (\gamma - 1)}}{2\alpha}\right) C_1 e^{-\left(\frac{\beta - \sqrt{\beta^2 - 8\alpha (\gamma - 1)}}{2\alpha}\right) x}\right. \nonumber \\
&&\left.+\left(\frac{\beta + \sqrt{\beta^2 - 8\alpha (\gamma - 1)}}{2\alpha}\right) C_2 e^{-\left(\frac{\beta + \sqrt{\beta^2 - 8\alpha (\gamma - 1)}}{2\alpha}\right) x}\right. \nonumber \\
&&\left. +2 \left(\frac{2\alpha - \beta + \gamma}{1 - 2\alpha + \beta - \gamma}\right) \Omega_{k0} e^{-2x}\right. \nonumber \\
&&\left.+3(1-d^2) \left[\frac{9\alpha (1-d^2)^2 - 3\beta (1-d^2) + 2 \gamma}{2-9\alpha (1-d^2)^2 + 3\beta (1-d^2) - 2 \gamma }\right] \, \Omega_{m0} e^{-3(1-d^2)x}  \right\}\times \nonumber \\
&& \left\{  \Omega_{k0} e^{-2x} + \Omega_{m0} e^{-3(1-d^2)x} + C_1 e^{-\left(\frac{\beta - \sqrt{\beta^2 - 8\alpha (\gamma - 1)}}{2\alpha}\right) x} + C_2 e^{-\left(\frac{\beta + \sqrt{\beta^2 - 8\alpha (\gamma - 1)}}{2\alpha}\right) x} \right.\nonumber \\
&&\left. + \left(\frac{2\alpha - \beta + \gamma}{1 - 2\alpha + \beta - \gamma}\right) \Omega_{k0} e^{-2x} \nonumber \right.\\
&&\left.+ \left[\frac{9\alpha (1-d^2)^2 - 3\beta (1-d^2) + 2 \gamma}{2-9\alpha (1-d^2)^2 + 3\beta (1-d^2) - 2 \gamma }\right] \,\Omega_{m0} e^{-3(1-d^2)x}    \right\}^{-1}
\end{eqnarray}
From the initial constraint, namely that the sum of the current density parameters equals one, the constants $C_1$ and $C_2$ must satisfy the following relation:
\begin{eqnarray}
 C_1 + C_2 + \frac{\Omega_{k0}}{1 - 2\alpha + \beta - \gamma} + \frac{2 \, \Omega_{m0}}{2 - 9\alpha (1-d^2)^2 + 3\beta (1-d^2) - 2\gamma} =1 \label{king5}
\end{eqnarray}
In the following subsections, we will consider three limiting cases, corresponding to $C_2=0$, $C_1=0$ e $C_1=C_2\equiv C_{12}$

\subsubsection{Case with $C_2=0$}
We now start considering the first case, i.e. $C_2=0$. \\
From Eq. (\ref{king5}), we obtain that:
\begin{equation}
 C_1 =1- \frac{\Omega_{k0}}{1 - 2\alpha + \beta - \gamma} - \frac{2 \, \Omega_{m0}}{2 - 9\alpha (1-d^2)^2 + 3\beta (1-d^2) - 2\gamma} 
\end{equation}
The general expression of the Hubble parameter squared for this case is given by:
\begin{eqnarray}
h^2_{higher,I,1} &=& \Omega_{k0} e^{-2x} + \Omega_{m0} e^{-3(1-d^2)x} + C_1 e^{-\left(\frac{\beta - \sqrt{\beta^2 - 8\alpha (\gamma - 1)}}{2\alpha}\right) x} \nonumber \\
&& + \left(\frac{2\alpha - \beta + \gamma}{1 - 2\alpha + \beta - \gamma}\right) \Omega_{k0} e^{-2x} + \left[\frac{9\alpha (1-d^2)^2 - 3\beta (1-d^2) + 2 \gamma}{2-9\alpha (1-d^2)^2 + 3\beta (1-d^2) - 2 \gamma }\right] \, \Omega_{m0}e^{-3(1-d^2)x} , \label{}
\end{eqnarray}

The energy density of DE for this case can be written as:
\begin{eqnarray}
\rho_{D,higher,I,1} &=& 3H_0^2\left\{C_1 e^{-\left(\frac{\beta - \sqrt{\beta^2 - 8\alpha (\gamma - 1)}}{2\alpha}\right) x} + \left(\frac{2\alpha - \beta + \gamma}{1 - 2\alpha + \beta - \gamma}\right) \Omega_{k0} e^{-2x}\right.\nonumber \\
&&\left.  + \left[\frac{9\alpha (1-d^2)^2 - 3\beta (1-d^2) + 2 \gamma}{2-9\alpha (1-d^2)^2 + 3\beta (1-d^2) - 2 \gamma }\right] \, \Omega_{m0}e^{-3(1-d^2)x} \right\}, \label{}
\end{eqnarray}
The final expression of the pressure is given by:
\begin{eqnarray}
p_{D,higher,I,1} &=& -\frac{H_0^2}{2\alpha} \Bigg[
\left( 6\alpha - \beta +\sqrt{\beta^2 - 8\alpha(\gamma - 1)} \right)
C_1 e^{-\left(\frac{\beta - \sqrt{\beta^2 - 8\alpha(\gamma - 1)}}{2\alpha}\,\right)x} \nonumber \\
&& + 2\alpha\left(\frac{2\alpha - \beta + \gamma}{1 - 2\alpha + \beta - \gamma}\right)\Omega_{k0}\,e^{-2x} \nonumber \\
&& + 6\alpha d^2 \left(\frac{9\alpha(1-d^2)^2 - 3\beta(1-d^2) + 2\gamma}{2 - 9\alpha(1-d^2)^2 + 3\beta(1-d^2) - 2\gamma}\right)\Omega_{m0}
e^{-3(1-d^2)x}
\Bigg].
\end{eqnarray}
where we used the fact that
\begin{eqnarray}
\rho'_{D,higher,I,1} &=& -3H_0^2\left\{\left(\frac{\beta - \sqrt{\beta^2 - 8\alpha (\gamma - 1)}}{2\alpha}\right)C_1 e^{-\left(\frac{\beta - \sqrt{\beta^2 - 8\alpha (\gamma - 1)}}{2\alpha}\right) x}  \nonumber \right.\\
&&\left. +2 \left(\frac{2\alpha - \beta + \gamma}{1 - 2\alpha + \beta - \gamma}\right) \Omega_{k0} e^{-2x}\nonumber \right.\\
&&\left.+3(1-d^2) \left[\frac{9\alpha (1-d^2)^2 - 3\beta (1-d^2) + 2 \gamma}{2-9\alpha (1-d^2)^2 + 3\beta (1-d^2) - 2 \gamma }\right] \, \Omega_{m0}e^{-3(1-d^2)x} \right\}, \label{}
\end{eqnarray}
Moroever, using the expressions of $\rho_{D,higher,I,1}$ and $\rho'_{D,higher,I,1}$ along with the term $\frac{Q}{3H\rho_{D,higher,I,1}}$, we obtained the following expression for $\omega_{D,higher,I,1}$:
\begin{eqnarray}
    \omega_{D,higher,I,1}&=& -1 +\frac{1}{3}\cdot \left\{\left(\frac{\beta - \sqrt{\beta^2 - 8\alpha (\gamma - 1)}}{2\alpha}\right)C_1 e^{-\left(\frac{\beta - \sqrt{\beta^2 - 8\alpha (\gamma - 1)}}{2\alpha}\right) x}  \nonumber \right.\\
&&\left. +2 \left(\frac{2\alpha - \beta + \gamma}{1 - 2\alpha + \beta - \gamma}\right) \Omega_{k0} e^{-2x}\nonumber \right.\\
&&\left.+\left\{3(1-d^2) \left[\frac{9\alpha (1-d^2)^2 - 3\beta (1-d^2) + 2 \gamma}{2-9\alpha (1-d^2)^2 + 3\beta (1-d^2) - 2 \gamma }\right]-3d^2\right\} \, \Omega_{m0}e^{-3(1-d^2)x} \right\}\times \nonumber \\
&&\left\{C_1 e^{-\left(\frac{\beta - \sqrt{\beta^2 - 8\alpha (\gamma - 1)}}{2\alpha}\right) x}  + \left(\frac{2\alpha - \beta + \gamma}{1 - 2\alpha + \beta - \gamma}\right) \Omega_{k0} e^{-2x} \nonumber \right.\\
&&\left.+ \left[\frac{9\alpha (1-d^2)^2 - 3\beta (1-d^2) + 2 \gamma}{2-9\alpha (1-d^2)^2 + 3\beta (1-d^2) - 2 \gamma }\right] \, \Omega_{m0}e^{-3(1-d^2)x} \right\}^{-1}
\end{eqnarray}
Finally, we obtain the following relation for the deceleration parameter:
\begin{eqnarray}
    q_{higher,I,1} &=& -1 + \frac{1}{2}\cdot \left\{  2\Omega_{k0} e^{-2x} +3(1-d^2) \Omega_{m0} e^{-3(1-d^2)x} \right.\nonumber \\
&&\left.+\left(\frac{\beta - \sqrt{\beta^2 - 8\alpha (\gamma - 1)}}{2\alpha}\right) C_1 e^{-\left(\frac{\beta - \sqrt{\beta^2 - 8\alpha (\gamma - 1)}}{2\alpha}\right) x}\right. \nonumber \\
&&\left. +2 \left(\frac{2\alpha - \beta + \gamma}{1 - 2\alpha + \beta - \gamma}\right) \Omega_{k0} e^{-2x}\right. \nonumber \\
&&\left.+3(1-d^2) \left[\frac{9\alpha (1-d^2)^2 - 3\beta (1-d^2) + 2 \gamma}{2-9\alpha (1-d^2)^2 + 3\beta (1-d^2) - 2 \gamma }\right] \, \Omega_{m0}e^{-3(1-d^2)x}  \right\}\times \nonumber \\
&& \left\{  \Omega_{k0} e^{-2x} + \Omega_{m0} e^{-3(1-d^2)x} + C_1 e^{-\left(\frac{\beta - \sqrt{\beta^2 - 8\alpha (\gamma - 1)}}{2\alpha}\right) x} \right.\nonumber \\
&&\left. + \left(\frac{2\alpha - \beta + \gamma}{1 - 2\alpha + \beta - \gamma}\right) \Omega_{k0} e^{-2x} \nonumber \right.\\
&&\left.+ \left[\frac{9\alpha (1-d^2)^2 - 3\beta (1-d^2) + 2 \gamma}{2-9\alpha (1-d^2)^2 + 3\beta (1-d^2) - 2 \gamma }\right] \, \Omega_{m0}e^{-3(1-d^2)x}    \right\}^{-1}
\end{eqnarray}
where we used the fact that:
\begin{eqnarray}
\frac{dh^2_{higher,I,1}}{dx} &=& -2\Omega_{k0} e^{-2x} -3(1-d^2) \Omega_{m0} e^{-3(1-d^2)x} \nonumber \\
&&-\left(\frac{\beta - \sqrt{\beta^2 - 8\alpha (\gamma - 1)}}{2\alpha}\right) C_1 e^{-\left(\frac{\beta - \sqrt{\beta^2 - 8\alpha (\gamma - 1)}}{2\alpha}\right) x} \nonumber \\
&& -2 \left(\frac{2\alpha - \beta + \gamma}{1 - 2\alpha + \beta - \gamma}\right) \Omega_{k0} e^{-2x} \nonumber \\
&&-3(1-d^2) \left[\frac{9\alpha (1-d^2)^2 - 3\beta (1-d^2) + 2 \gamma}{2-9\alpha (1-d^2)^2 + 3\beta (1-d^2) - 2 \gamma }\right] \, \Omega_{m0}e^{-3(1-d^2)x} , \label{}
\end{eqnarray}

\subsubsection{Case with $C_1=0$}
We how consider the second limiting case, i.e. $C_1=0$.\\
In this case, we derive the following expression for $C_2$ from Eq. (\ref{king5}):
\begin{equation}
 C_2 =1- \frac{\Omega_{k0}}{1 - 2\alpha + \beta - \gamma} - \frac{2 \, \Omega_{m0}}{2 - 9\alpha (1-d^2)^2 + 3\beta (1-d^2) - 2\gamma} 
\end{equation}
The final expression of the Hubble parameter squared for this case is given by:
\begin{eqnarray}
h^2_{higher,I,2} &=& \Omega_{k0} e^{-2x} + \Omega_{m0} e^{-3(1-d^2)x}  + C_2 e^{-\left(\frac{\beta + \sqrt{\beta^2 - 8\alpha (\gamma - 1)}}{2\alpha}\right) x} \nonumber \\
&& + \left(\frac{2\alpha - \beta + \gamma}{1 - 2\alpha + \beta - \gamma}\right) \Omega_{k0} e^{-2x} \nonumber \\
&&+ \left[\frac{9\alpha (1-d^2)^2 - 3\beta (1-d^2) + 2 \gamma}{2-9\alpha (1-d^2)^2 + 3\beta (1-d^2) - 2 \gamma }\right] \, \Omega_{m0}e^{-3(1-d^2)x} , \label{}
\end{eqnarray}

Moreover, we have that the energy density of DE can be expressed as:
\begin{eqnarray}
\rho_{D,higher,I,2} &=& 3H_0^2\left\{C_2 e^{-\left(\frac{\beta + \sqrt{\beta^2 - 8\alpha (\gamma - 1)}}{2\alpha}\right) x}+ \left(\frac{2\alpha - \beta + \gamma}{1 - 2\alpha + \beta - \gamma}\right) \Omega_{k0} e^{-2x} \right.\nonumber \\
&&\left.  + \left[\frac{9\alpha (1-d^2)^2 - 3\beta (1-d^2) + 2 \gamma}{2-9\alpha (1-d^2)^2 + 3\beta (1-d^2) - 2 \gamma }\right] \, \Omega_{m0}e^{-3(1-d^2)x} \right\}, \label{}
\end{eqnarray}
The final expression of the pressure is given by:
\begin{eqnarray}
p_{D,higher,I,2}  &=& -\frac{H_0^2}{2\alpha} \Bigg[  \left( 6\alpha - \beta - \sqrt{\beta^2 - 8\alpha(\gamma - 1)} \right)
C_2 e^{-\left(\frac{\beta + \sqrt{\beta^2 - 8\alpha(\gamma - 1)}}{2\alpha}\,\right)x} \nonumber \\
&& + 2\alpha\left(\frac{2\alpha - \beta + \gamma}{1 - 2\alpha + \beta - \gamma}\right)\Omega_{k0}\,e^{-2x} \nonumber \\
&& + 6\alpha d^2 \left(\frac{9\alpha(1-d^2)^2 - 3\beta(1-d^2) + 2\gamma}{2 - 9\alpha(1-d^2)^2 + 3\beta(1-d^2) - 2\gamma}\right)\Omega_{m0}
e^{-3(1-d^2)x}
\Bigg].
\end{eqnarray}
where we used the fact that:
\begin{eqnarray}
\rho'_{D,higher,I,2} &=& -3H_0^2\left\{\left(\frac{\beta + \sqrt{\beta^2 - 8\alpha (\gamma - 1)}}{2\alpha}\right)C_2 e^{-\left(\frac{\beta + \sqrt{\beta^2 - 8\alpha (\gamma - 1)}}{2\alpha}\right) x} \right.\nonumber \\
&&\left. +2 \left(\frac{2\alpha - \beta + \gamma}{1 - 2\alpha + \beta - \gamma}\right) \Omega_{k0} e^{-2x}\nonumber \right.\\
&&\left.+3(1-d^2) \left[\frac{9\alpha (1-d^2)^2 - 3\beta (1-d^2) + 2 \gamma}{2-9\alpha (1-d^2)^2 + 3\beta (1-d^2) - 2 \gamma }\right] \, \Omega_{m0}e^{-3(1-d^2)x} \right\}, \label{}
\end{eqnarray}
Using the expression of $\rho_{D,higher,I,2}$ and $\rho'_{D,higher,I,2}$ along with the term $\frac{Q}{3H\rho_{D,higher,I,2}}=d^2\left( \frac{\rho_m}{\rho_{D,higher,I,2}} \right)$, we obtain the following expression for the EoS parameter:
\begin{eqnarray}
    \omega_{D,higher,I,2} &=& -1 +\frac{1}{3}\cdot \left\{\left(\frac{\beta + \sqrt{\beta^2 - 8\alpha (\gamma - 1)}}{2\alpha}\right)C_2 e^{-\left(\frac{\beta + \sqrt{\beta^2 - 8\alpha (\gamma - 1)}}{2\alpha}\right) x} \right.\nonumber \\
&&\left. +2 \left(\frac{2\alpha - \beta + \gamma}{1 - 2\alpha + \beta - \gamma}\right) \Omega_{k0} e^{-2x}\nonumber \right.\\
&&\left.+\left\{3(1-d^2) \left[\frac{9\alpha (1-d^2)^2 - 3\beta (1-d^2) + 2 \gamma}{2-9\alpha (1-d^2)^2 + 3\beta (1-d^2) - 2 \gamma }\right]-3d^2\right\} \, \Omega_{m0}e^{-3(1-d^2)x} \right\}\times \nonumber \\
&&\left\{C_1 e^{-\left(\frac{\beta - \sqrt{\beta^2 - 8\alpha (\gamma - 1)}}{2\alpha}\right) x} + C_2 e^{-\left(\frac{\beta + \sqrt{\beta^2 - 8\alpha (\gamma - 1)}}{2\alpha}\right) x} \right.\nonumber \\
&&\left. + \left(\frac{2\alpha - \beta + \gamma}{1 - 2\alpha + \beta - \gamma}\right) \Omega_{k0} e^{-2x} \nonumber \right.\\
&&\left.+ \left[\frac{9\alpha (1-d^2)^2 - 3\beta (1-d^2) + 2 \gamma}{2-9\alpha (1-d^2)^2 + 3\beta (1-d^2) - 2 \gamma }\right] \, \Omega_{m0}e^{-3(1-d^2)x} \right\}^{-1}
\end{eqnarray}
Finally, we obtain the following expression for the deceleration parameter:
\begin{eqnarray}
    q_{higher,I,2}  &=& -1 + \frac{1}{2}\cdot \left\{  2\Omega_{k0} e^{-2x} +3(1-d^2) \Omega_{m0} e^{-3(1-d^2)x} \right.\nonumber \\
&&\left.+\left(\frac{\beta + \sqrt{\beta^2 - 8\alpha (\gamma - 1)}}{2\alpha}\right) C_2 e^{-\left(\frac{\beta + \sqrt{\beta^2 - 8\alpha (\gamma - 1)}}{2\alpha}\right) x}\right. \nonumber \\
&&\left. +2 \left(\frac{2\alpha - \beta + \gamma}{1 - 2\alpha + \beta - \gamma}\right) \Omega_{k0} e^{-2x}\right. \nonumber \\
&&\left.+3(1-d^2) \left[\frac{9\alpha (1-d^2)^2 - 3\beta (1-d^2) + 2 \gamma}{2-9\alpha (1-d^2)^2 + 3\beta (1-d^2) - 2 \gamma }\right] \, \Omega_{m0}e^{-3(1-d^2)x}  \right\}\times \nonumber \\
&& \left\{  \Omega_{k0} e^{-2x} + \Omega_{m0} e^{-3(1-d^2)x}  + C_2 e^{-\left(\frac{\beta + \sqrt{\beta^2 - 8\alpha (\gamma - 1)}}{2\alpha}\right) x} \right.\nonumber \\
&&\left. + \left(\frac{2\alpha - \beta + \gamma}{1 - 2\alpha + \beta - \gamma}\right) \Omega_{k0} e^{-2x} \nonumber \right.\\
&&\left.+ \left[\frac{9\alpha (1-d^2)^2 - 3\beta (1-d^2) + 2 \gamma}{2-9\alpha (1-d^2)^2 + 3\beta (1-d^2) - 2 \gamma }\right] \, \Omega_{m0}e^{-3(1-d^2)x}    \right\}^{-1}
\end{eqnarray}
where we used the fact that:
\begin{eqnarray}
h'^2_{higher,I,2}  &=& -2\Omega_{k0} e^{-2x} -3(1-d^2) \Omega_{m0} e^{-3(1-d^2)x} \nonumber \\
&&-\left(\frac{\beta + \sqrt{\beta^2 - 8\alpha (\gamma - 1)}}{2\alpha}\right) C_2 e^{-\left(\frac{\beta + \sqrt{\beta^2 - 8\alpha (\gamma - 1)}}{2\alpha}\right) x} \nonumber \\
&& -2 \left(\frac{2\alpha - \beta + \gamma}{1 - 2\alpha + \beta - \gamma}\right) \Omega_{k0} e^{-2x} \nonumber \\
&&-3(1-d^2) \left[\frac{9\alpha (1-d^2)^2 - 3\beta (1-d^2) + 2 \gamma}{2-9\alpha (1-d^2)^2 + 3\beta (1-d^2) - 2 \gamma }\right] \, \Omega_{m0}e^{-3(1-d^2)x} , \label{}
\end{eqnarray}

\subsubsection{Case with $C_1=C_2\equiv C_{12}$}
We now consider the third and last limiting case considered for this model, i.e. $C_1=C_2\equiv C_{12}$.\\
From Eq. (\ref{king5}), we obtain the following expression for $ C_{12}$
\begin{equation}
 C_{12} = \frac{1}{2}\left[1- \frac{\Omega_{k0}}{1 - 2\alpha + \beta - \gamma} - \frac{2 \, \Omega_{m0}}{2 - 9\alpha (1-d^2)^2 + 3\beta (1-d^2) - 2\gamma}\right]
\end{equation}
The expression of the Hubble parameter squared for this case is given by:
\begin{eqnarray}
h^2_{higher,I,3}  &=& \Omega_{k0} e^{-2x} + \Omega_{m0} e^{-3(1-d^2)x} + C_{12} e^{-\left(\frac{\beta - \sqrt{\beta^2 - 8\alpha (\gamma - 1)}}{2\alpha}\right) x} + C_ {12}e^{-\left(\frac{\beta + \sqrt{\beta^2 - 8\alpha (\gamma - 1)}}{2\alpha}\right) x} \nonumber \\
&& + \left(\frac{2\alpha - \beta + \gamma}{1 - 2\alpha + \beta - \gamma}\right) \Omega_{k0} e^{-2x} \nonumber \\
&&+ \left[\frac{9\alpha (1-d^2)^2 - 3\beta (1-d^2) + 2 \gamma}{2-9\alpha (1-d^2)^2 + 3\beta (1-d^2) - 2 \gamma }\right] \, \Omega_{m0}e^{-3(1-d^2)x} , \label{}
\end{eqnarray}
Instead, the energy density of DE can be written in the following way:
\begin{eqnarray}
\rho_{D,higher,I,3}  &=& 3H_0^2\left\{C_{12} e^{-\left(\frac{\beta - \sqrt{\beta^2 - 8\alpha (\gamma - 1)}}{2\alpha}\right) x} + C_{12} e^{-\left(\frac{\beta + \sqrt{\beta^2 - 8\alpha (\gamma - 1)}}{2\alpha}\right) x} \right.\nonumber \\
&&\left. + \left(\frac{2\alpha - \beta + \gamma}{1 - 2\alpha + \beta - \gamma}\right) \Omega_{k0} e^{-2x} \right. \nonumber \\
&&\left.+ \left[\frac{9\alpha (1-d^2)^2 - 3\beta (1-d^2) + 2 \gamma}{2-9\alpha (1-d^2)^2 + 3\beta (1-d^2) - 2 \gamma }\right] \, \Omega_{m0}e^{-3(1-d^2)x} \right\}, \label{}
\end{eqnarray}
The final expression of the pressure of DE is given by:
\begin{eqnarray}
p_{D,higher,I,3}  &=& -\frac{H_0^2}{2\alpha} \Bigg[
\left( 6\alpha - \beta +\sqrt{\beta^2 - 8\alpha(\gamma - 1)} \right)
C_{12} e^{-\left(\frac{\beta - \sqrt{\beta^2 - 8\alpha(\gamma - 1)}}{2\alpha}\,\right)x} \nonumber \\
&& + \left( 6\alpha - \beta - \sqrt{\beta^2 - 8\alpha(\gamma - 1)} \right)
C_{12} e^{-\left(\frac{\beta + \sqrt{\beta^2 - 8\alpha(\gamma - 1)}}{2\alpha}\,\right)x} \nonumber \\
&& + 2\alpha\left(\frac{2\alpha - \beta + \gamma}{1 - 2\alpha + \beta - \gamma}\right)\Omega_{k0}\,e^{-2x} \nonumber \\
&& + 6\alpha d^2 \left(\frac{9\alpha(1-d^2)^2 - 3\beta(1-d^2) + 2\gamma}{2 - 9\alpha(1-d^2)^2 + 3\beta(1-d^2) - 2\gamma}\right)\Omega_{m0}
e^{-3(1-d^2)x}
\Bigg].
\end{eqnarray}

where we used the fact that:
\begin{eqnarray}
\rho'_{D,higher,I,3}  &=& -3H_0^2\left\{\left(\frac{\beta - \sqrt{\beta^2 - 8\alpha (\gamma - 1)}}{2\alpha}\right)C_{12} e^{-\left(\frac{\beta - \sqrt{\beta^2 - 8\alpha (\gamma - 1)}}{2\alpha}\right) x}  \nonumber \right.\\
&&\left.+\left(\frac{\beta + \sqrt{\beta^2 - 8\alpha (\gamma - 1)}}{2\alpha}\right)C_{12} e^{-\left(\frac{\beta + \sqrt{\beta^2 - 8\alpha (\gamma - 1)}}{2\alpha}\right) x} \right.\nonumber \\
&&\left. +2 \left(\frac{2\alpha - \beta + \gamma}{1 - 2\alpha + \beta - \gamma}\right) \Omega_{k0} e^{-2x}\nonumber \right.\\
&&\left.+3(1-d^2) \left[\frac{9\alpha (1-d^2)^2 - 3\beta (1-d^2) + 2 \gamma}{2-9\alpha (1-d^2)^2 + 3\beta (1-d^2) - 2 \gamma }\right] \, \Omega_{m0}e^{-3(1-d^2)x} \right\}, \label{}
\end{eqnarray}
Using the expression of $\rho_{D,higher,I,3} $ and $\rho'_{D,higher,I,3} $ along with the term $\frac{Q}{3H\rho_{D,higher,I,3}}=d^2\left( \frac{\rho_m}{\rho_{D,higher,I,3}} \right)$, we obtain the following expression for the EoS parameter:
\begin{eqnarray}
    \omega_{D,higher,I,3} &=& -1 +\frac{1}{3}\cdot \left\{\left(\frac{\beta - \sqrt{\beta^2 - 8\alpha (\gamma - 1)}}{2\alpha}\right)C_{12} e^{-\left(\frac{\beta - \sqrt{\beta^2 - 8\alpha (\gamma - 1)}}{2\alpha}\right) x}  \nonumber \right.\\
&&\left.+\left(\frac{\beta + \sqrt{\beta^2 - 8\alpha (\gamma - 1)}}{2\alpha}\right)C_{12} e^{-\left(\frac{\beta + \sqrt{\beta^2 - 8\alpha (\gamma - 1)}}{2\alpha}\right) x} \right.\nonumber \\
&&\left. +2 \left(\frac{2\alpha - \beta + \gamma}{1 - 2\alpha + \beta - \gamma}\right) \Omega_{k0} e^{-2x}\nonumber \right.\\
&&\left.+\left\{3(1-d^2) \left[\frac{9\alpha (1-d^2)^2 - 3\beta (1-d^2) + 2 \gamma}{2-9\alpha (1-d^2)^2 + 3\beta (1-d^2) - 2 \gamma }\right]-3d^2\right\} \, \Omega_{m0}e^{-3(1-d^2)x} \right\}\times \nonumber \\
&&\left\{C_{12} e^{-\left(\frac{\beta - \sqrt{\beta^2 - 8\alpha (\gamma - 1)}}{2\alpha}\right) x} + C_{12} e^{-\left(\frac{\beta + \sqrt{\beta^2 - 8\alpha (\gamma - 1)}}{2\alpha}\right) x} \right.\nonumber \\
&&\left. + \left(\frac{2\alpha - \beta + \gamma}{1 - 2\alpha + \beta - \gamma}\right) \Omega_{k0} e^{-2x} \nonumber \right.\\
&&\left.+ \left[\frac{9\alpha (1-d^2)^2 - 3\beta (1-d^2) + 2 \gamma}{2-9\alpha (1-d^2)^2 + 3\beta (1-d^2) - 2 \gamma }\right] \, \Omega_{m0}e^{-3(1-d^2)x} \right\}^{-1}
\end{eqnarray}
Finally, the deceleration parameter can be written as:
\begin{eqnarray}
    q_{higher,I,3}  &=& -1 + \frac{1}{2}\cdot \left\{  2\Omega_{k0} e^{-2x} +3(1-d^2) \Omega_{m0} e^{-3(1-d^2)x} \right.\nonumber \\
&&\left.+\left(\frac{\beta - \sqrt{\beta^2 - 8\alpha (\gamma - 1)}}{2\alpha}\right) C_{12} e^{-\left(\frac{\beta - \sqrt{\beta^2 - 8\alpha (\gamma - 1)}}{2\alpha}\right) x}\right. \nonumber \\
&&\left.+\left(\frac{\beta + \sqrt{\beta^2 - 8\alpha (\gamma - 1)}}{2\alpha}\right) C_{12} e^{-\left(\frac{\beta + \sqrt{\beta^2 - 8\alpha (\gamma - 1)}}{2\alpha}\right) x}\right. \nonumber \\
&&\left. +2 \left(\frac{2\alpha - \beta + \gamma}{1 - 2\alpha + \beta - \gamma}\right) \Omega_{k0} e^{-2x}\right. \nonumber \\
&&\left.+3(1-d^2) \left[\frac{9\alpha (1-d^2)^2 - 3\beta (1-d^2) + 2 \gamma}{2-9\alpha (1-d^2)^2 + 3\beta (1-d^2) - 2 \gamma }\right] \, \Omega_{m0}e^{-3(1-d^2)x}  \right\}\times \nonumber \\
&& \left\{  \Omega_{k0} e^{-2x} + \Omega_{m0} e^{-3(1-d^2)x} + C_{12} e^{-\left(\frac{\beta - \sqrt{\beta^2 - 8\alpha (\gamma - 1)}}{2\alpha}\right) x} + C_{12} e^{-\left(\frac{\beta + \sqrt{\beta^2 - 8\alpha (\gamma - 1)}}{2\alpha}\right) x} \right.\nonumber \\
&&\left. + \left(\frac{2\alpha - \beta + \gamma}{1 - 2\alpha + \beta - \gamma}\right) \Omega_{k0} e^{-2x} \nonumber \right.\\
&&\left.+ \left[\frac{9\alpha (1-d^2)^2 - 3\beta (1-d^2) + 2 \gamma}{2-9\alpha (1-d^2)^2 + 3\beta (1-d^2) - 2 \gamma }\right] \, \Omega_{m0}e^{-3(1-d^2)x}    \right\}^{-1}
\end{eqnarray}
where we used the fact that:
\begin{eqnarray}
\frac{h'^2_{higher,I,3} }{dx} &=& -2\Omega_{k0} e^{-2x} -3(1-d^2) \Omega_{m0} e^{-3(1-d^2)x} \nonumber \\
&&-\left(\frac{\beta - \sqrt{\beta^2 - 8\alpha (\gamma - 1)}}{2\alpha}\right) C_{12} e^{-\left(\frac{\beta - \sqrt{\beta^2 - 8\alpha (\gamma - 1)}}{2\alpha}\right) x} \nonumber \\
&&-\left(\frac{\beta + \sqrt{\beta^2 - 8\alpha (\gamma - 1)}}{2\alpha}\right) C_{12} e^{-\left(\frac{\beta + \sqrt{\beta^2 - 8\alpha (\gamma - 1)}}{2\alpha}\right) x} \nonumber \\
&& -2 \left(\frac{2\alpha - \beta + \gamma}{1 - 2\alpha + \beta - \gamma}\right) \Omega_{k0} e^{-2x} \nonumber \\
&&-3(1-d^2) \left[\frac{9\alpha (1-d^2)^2 - 3\beta (1-d^2) + 2 \gamma}{2-9\alpha (1-d^2)^2 + 3\beta (1-d^2) - 2 \gamma }\right] \, \Omega_{m0}e^{-3(1-d^2)x} , \label{}
\end{eqnarray}

It is worth noting that, in the limiting case $d^2 = 0$, which corresponds to the absence of interaction, the results obtained for the three scenarios $C_1 = 0$, $C_2 = 0$, and $C_1 = C_2$ consistently reduce to those derived in the subsection devoted to the non-interacting case.

\subsection{Dark Dominated Universe}
We now aim to find explicit expressions for the scale factor $a(t)$, the Hubble parameter $H$, the  density of DE $\rho_D$, the pressure of DE $p_D$, the equation of state (EoS) parameter $\omega_D$ and finally the deceleration parameter $q$, all as functions of cosmic time $t$.\\
Considering a flat Dark Dominated Universe, we have that Eq. (\ref{accanon}) reduces to:
\begin{eqnarray}
h^2_{DD,higher}=&  f_0e^{-\left(\frac{\beta - \sqrt{\beta^2 -8\alpha\left(\gamma -1\right)}}{2\alpha}\right)x} + f_1e^{-\left(\frac{\beta + \sqrt{\beta^2 - 8\alpha\left(\gamma -1\right)}}{2\alpha}\right)x}. \label{hquadro3}
\end{eqnarray}

From Eq.~(\ref{condition}), we obtain the following relation between $f_0$ and $f_1$:
\begin{eqnarray}
f_0 + f_1  = 1 \rightarrow f_1 = 1-f_0 \label{condition-DD}
\end{eqnarray}

Therefore, we can write:
\begin{eqnarray}
h^2_{DD,higher} &=&  f_0e^{-\left(\frac{\beta - \sqrt{\beta^2 -8\alpha\left(\gamma -1\right)}}{2\alpha}\right)x} + (1-f_0)e^{-\left(\frac{\beta + \sqrt{\beta^2 - 8\alpha\left(\gamma -1\right)}}{2\alpha}\right)x}. \label{hquadro3-2}
\end{eqnarray}
If we consider the case with $f_0=f_1$, we obtain:
\begin{eqnarray}
h_{DD,higher}^2 &=&  \frac{1}{2}\left[e^{-\left(\frac{\beta - \sqrt{\beta^2 -8\alpha\left(\gamma -1\right)}}{2\alpha}\right)x} + e^{-\left(\frac{\beta + \sqrt{\beta^2 - 8\alpha\left(\gamma -1\right)}}{2\alpha}\right)x}\right], \label{hquadro3-3}
\end{eqnarray}
Since Eq.~(\ref{hquadro3-3}) does not admit a simple analytical solution in general, we consider the limiting case in which $\beta^2 - 8\alpha(\gamma - 1) = 0$. Under this condition, Eq.~(\ref{hquadro3-3}) simplifies to the following expression for $h^2$:
\begin{eqnarray}
h^2_{DD,higher} &=&  e^{-\frac{\beta x}{2\alpha}}. \label{}
\end{eqnarray}

By changing variables from $x$ to the scale factor $a$, we obtain a second-order differential equation for $a(t)$, whose solution reads:
\begin{eqnarray}
a_{DD,higher} (t)= \left[\frac{\beta(t + C_3)}{4\alpha}\right]^{\frac{4\alpha}{\beta}}, \label{scale3}
\end{eqnarray}
where $C_3$ is an integration constant.

Using the result from Eq.~(\ref{scale3}), we can straightforwardly compute the Hubble parameter as:
\begin{eqnarray}
H_{DD,higher}(t) &=& \frac{4\alpha}{\beta(t + C_3)}. \label{hubble3!!!}
\end{eqnarray}

Inserting this expression for $H$ into Eq.~(\ref{model}), we derive the following expression for the energy density of dark energy:
\begin{eqnarray}
\rho_{DD,higher}(t) = \frac{6\alpha(8\alpha\gamma - \beta^2)}{\left[\beta(t + C_3)\right]^2} = \frac{48\alpha^2}{\left[\beta(t + C_3)\right]^2}, \label{rhot3}
\end{eqnarray}
where in the last step we used the condition $8\alpha\gamma - \beta^2 = 8\alpha$ derived from $\beta^2 - 8\alpha(\gamma - 1) = 0$.

Similarly, we find the pressure of dark energy:
\begin{eqnarray}
p_{DD,higher}(t) = \frac{(6\alpha - \beta)(\beta^2 - 8\alpha\gamma)}{\left[\beta(t + C_3)\right]^2} = -\frac{8(6\alpha - \beta)\alpha}{\left[\beta(t + C_3)\right]^2},
\end{eqnarray}
where we used:
\begin{eqnarray}
\dot{\rho}_{DD,higher}(t) = -\frac{12\alpha(8\alpha\gamma - \beta^2)}{\beta^2(t + C_3)^3} = -\frac{96\alpha^2}{\beta^2(t + C_3)^3}.
\end{eqnarray}

Furthermore, we obtain the following expression for the EoS parameter of dark energy:
\begin{eqnarray}
\omega_{DD,higher}(t) = -1 + \frac{\beta}{6\alpha}, \label{nicomu4}
\end{eqnarray}
and for the deceleration parameter:
\begin{eqnarray}
q(t)_{DD,higher} = -1 + \frac{\beta}{4\alpha}.
\end{eqnarray}

In the limiting case corresponding to $C_3 = 0$, the above expressions simplify as follows:
\begin{eqnarray}
a_{DD,,higher,lim}(t) &=& \left(\frac{\beta t}{4\alpha}\right)^{\frac{4\alpha}{\beta}}, \label{scale30}\\
H_{DD,,higher,lim}(t) &=& \left(\frac{4\alpha}{\beta}\right)\left(\frac{1}{t}\right), \label{hubble3}\\
\rho_{DD,lim}(t) &=& \frac{6\alpha(8\alpha\gamma - \beta^2)}{(\beta t)^2} = \frac{48\alpha^2}{(\beta t)^2}, \label{fonza1}\\
p_{DD,,higher,lim}(t) &=& \frac{(6\alpha - \beta)(\beta^2 - 8\alpha\gamma)}{(\beta t)^2} = -\frac{8(6\alpha - \beta)\alpha}{(\beta t)^2}.
\end{eqnarray}

\section{New Holographic Dark Energy (NHDE) model}
We now consider the NHDE model. \\
As we have seen above, the DE energy density of the NHDE model can be written as:
\begin{eqnarray}
\rho_{NHDE}=3(\mu H^2+\lambda\dot{H}) \label{king6}
\end{eqnarray}
In the following Sections, as we have done for the first model, we derive expressions for several important cosmological quantities as functions of the variable $x = \ln a$:  the reduced Hubble parameter squared $h^2$, the DE energy density $\rho_D$, the DE pressure $p_D$, the DE equation of state (EoS) parameter $\omega_D$, and the deceleration parameter $q$. We also consider some limiting cases of the integration constants we obtain during calculations. Moreover, we calculate the same quantities for the limiting case of a flat Dark Dominated Universe, i.e. for $\Omega_m=\Omega_k=0$ and $\Omega_D=1$.

\subsection{Non Interacting Case}
We start considering the non interacting case.\\
Inserting the expression of $\rho_{NHDE}$ obtained in Eq. (\ref{king6}) in the Friedmann equation, we obtain the following differential equation for $h^2$:
\begin{eqnarray}\label{11}
\frac{dh^2}{dx}+\frac{2(\mu-1)}{\lambda}h^2-\frac{2\Omega_{k0}}{\lambda}e^{-2x}
+\frac{2\Omega_{m0}}{\lambda}e^{-3x} =0, \label{king7}
\end{eqnarray}
The general solution of Eq. (\ref{king7}) is
\begin{eqnarray}
h^2_{NHDE} = 
\frac{\Omega_{k0}e^{-2x}}{\mu - 1 - \lambda} \, 
- \frac{2\Omega_{m0}e^{-3x}}{2(\mu-1) - 3\lambda} \, 
+ C \, e^{-\tfrac{2(\mu-1)}{\lambda}x}, \label{king8--}
\end{eqnarray}
where $C$ is an integration constant which can be determined from the initial conditions.\\
Eq. (\ref{king8--}) can be also rewritten as:
\begin{eqnarray}
 h^2_{NHDE}&=&  \Omega_{k0} e^{-2x} + \Omega_{m0} e^{-3x} \nonumber \\
&&+ \, \left(\frac{2-\mu+\lambda}{\mu-\lambda -1}\right)\, \Omega_{k0}e^{-2x} 
+ \, \left[\frac{3\lambda-2\mu}{2(\mu-1)-3\lambda}\right]\, \Omega_{m0}e^{-3x} 
+ C \, e^{-\frac{2(\mu-1)}{\lambda} x}.
\end{eqnarray}
The initial condition tell us that $h^2=1$ for $x=0$ (considering also $a_0=1$), and hence we have:
\begin{eqnarray}
C  = 1 - \frac{\Omega_{k0}}{\mu - 1 - \lambda} + \frac{2 \Omega_{m0}}{2(\mu-1) - 3\lambda}.
\end{eqnarray}
Therefore, we can write:
\begin{eqnarray}
 h^2_{NHDE}&=&  \Omega_{k0} e^{-2x} + \Omega_{m0} e^{-3x}\nonumber  \\
&&+ \, \left(\frac{2-\mu+\lambda}{\mu-\lambda -1}\right)\, \Omega_{k0}e^{-2x} 
+ \, \left[\frac{3\lambda-2\mu}{2(\mu-1)-3\lambda}\right]\, \Omega_{m0}e^{-3x}\nonumber \\ 
&&+ \left[  1 - \frac{\Omega_{k0}}{\mu - 1 - \lambda} + \frac{2 \Omega_{m0}}{2(\mu-1) - 3\lambda} \right] \, e^{-\frac{2(\mu-1)}{\lambda} x}.\label{king8}
\end{eqnarray}
From the expression of $h^2_{NHDE}$ given in Eq. (\ref{king8}), we obtain that the energy density of the HNDE model can be also written as:
\begin{eqnarray}
 \rho_{D,NHDE}&=& 3H_0^2\left\{ \left(\frac{2-\mu+\lambda}{\mu-\lambda -1}\right)\, \Omega_{k0}e^{-2x} 
+ \, \left[\frac{3\lambda-2\mu}{2(\mu-1)-3\lambda}\right]\, \Omega_{m0}e^{-3x}\right.\nonumber \\ 
&&\left.+ \left[  1 - \frac{\Omega_{k0}}{\mu - 1 - \lambda} + \frac{2 \Omega_{m0}}{2(\mu-1) - 3\lambda} \right] \, e^{-\frac{2(\mu-1)}{\lambda} x}\right\}.
\end{eqnarray}
Moreover, from the DE continuity equation given by Eq. (\ref{12deprime}), the general relation for the DE pressure $p_{D,NHDE}$ as a function of $\rho_{D,NHDE}$ and its derivative with respect to $x$ is:
\begin{eqnarray}
p_{D,NHDE} = -\rho_{D,NHDE} - \frac{\rho_{D,NHDE}'}{3} \label{pdnongen}.
\end{eqnarray}
Differentiating with respect to $x$ the expression of $\rho_{D,NHDE}$, we obtain:
\begin{eqnarray}
 \rho'_{D,NHDE}&=& -3H_0^2\left\{ 2\left(\frac{2-\mu+\lambda}{\mu-\lambda -1}\right)\, \Omega_{k0}e^{-2x} 
+3 \, \left[\frac{3\lambda-2\mu}{2(\mu-1)-3\lambda}\right]\, \Omega_{m0}e^{-3x}\right.\nonumber \\ 
&&\left.+\frac{2(\mu-1)}{\lambda} \left[  1 - \frac{\Omega_{k0}}{\mu - 1 - \lambda} + \frac{2 \Omega_{m0}}{2(\mu-1) - 3\lambda} \right] \, e^{-\frac{2(\mu-1)}{\lambda} x}\right\}.
\end{eqnarray}

Therefore we obtain the following final expression of $p$:
\begin{eqnarray}
p_{D,NHDE} &=&
- H_0^2 \left\{\left(\frac{2-\mu+\lambda}{\mu-\lambda -1}\right) \, \Omega_{k0} \, e^{-2x} \nonumber \right.\\
&&\left.- \left[ \frac{2(\mu-1)-3\lambda}{\lambda}  \right] 
\left[ 1 - \frac{\Omega_{k0}}{\mu-1-\lambda} + \frac{2 \, \Omega_{m0}}{2(\mu-1)-3\lambda} \right] e^{- \frac{2(\mu-1)}{\lambda} x}\right\}.
\end{eqnarray}

We now want to derive the final expression of the EoS parameter. 
From the DE continuity equation given in Eq. (\ref{12deoldprime}), the general relation for the DE equation of state (EoS) parameter $\omega_{D,NHDE}$ is:
\begin{eqnarray}
\omega_{D,NHDE} = -1 - \frac{\rho_{D,NHDE}'}{3 \rho_{D,NHDE}}  \label{eosnongen}.
\end{eqnarray}
Using the relations of $\rho_{D,NHDE}$ and $\rho_{D,NHDE}'$ we already derived, we obtain the following expression for $\omega_D$:
\begin{eqnarray}
\omega_{D,NHDE} &=& -1 + \frac{1}{3}\cdot \left\{ 2\left(\frac{2-\mu+\lambda}{\mu-\lambda -1}\right)\, \Omega_{k0}e^{-2x} 
+3 \, \left[\frac{3\lambda-2\mu}{2(\mu-1)-3\lambda}\right]\, \Omega_{m0}e^{-3x}\right.\nonumber \\ 
&&\,\,\,\,\,\,\,\,\,\,\,\,\,\,\,\,\,\,\,\,\,\,\,\,\,\,\,\,\,\,\,\left.+\frac{2(\mu-1)}{\lambda} \left[  1 - \frac{\Omega_{k0}}{\mu - 1 - \lambda} + \frac{2 \Omega_{m0}}{2(\mu-1) - 3\lambda} \right] \, e^{-\frac{2(\mu-1)}{\lambda} x}  \right\}\times \nonumber\\
&& \left\{  \left(\frac{2-\mu+\lambda}{\mu-\lambda -1}\right)\, \Omega_{k0}e^{-2x} 
+ \, \left[\frac{3\lambda-2\mu}{2(\mu-1)-3\lambda}\right]\, \Omega_{m0}e^{-3x}\right.\nonumber \\ 
&&\left.+ \left[  1 - \frac{\Omega_{k0}}{\mu - 1 - \lambda} + \frac{2 \Omega_{m0}}{2(\mu-1) - 3\lambda} \right] \, e^{-\frac{2(\mu-1)}{\lambda} x}  \right\}^{-1}
\end{eqnarray}
The deceleration parameter is defined as:
\begin{eqnarray}
q_{NHDE} &=&  -1 - \frac{1}{2 h_{NHDE}^2} \frac{d h^2_{NHDE}2}{dx}. \label{}
\end{eqnarray}
Differentiating the expression of $h_{NHDE}^2$ with respect to the variable $x$, we can write
\begin{eqnarray}\label{12}
\frac{d h_{NHDE}^2}{dx}&=& -2\Omega_{k0} e^{-2x} -3 \Omega_{m0} e^{-3x} \nonumber \\
&&- \, 2\left(\frac{2-\mu+\lambda}{\mu-\lambda -1}\right)\, \Omega_{k0}e^{-2x} 
-3 \, \left[\frac{3\lambda-2\mu}{2(\mu-1)-3\lambda}\right]\, \Omega_{m0}e^{-3x}\nonumber \\ 
&&-\frac{2(\mu-1)}{\lambda} \left[  1 - \frac{\Omega_{k0}}{\mu - 1 - \lambda} + \frac{2 \Omega_{m0}}{2(\mu-1) - 3\lambda} \right] \, e^{-\frac{2(\mu-1)}{\lambda} x}.
\end{eqnarray}
Therefore, we obtain the following final expression for $q_{NHDE}$
\begin{eqnarray}
    q_{NHDE}&=&-1+\frac{1}{2}\cdot \left\{ 2\Omega_{k0} e^{-2x} +3 \Omega_{m0} e^{-3x}\right. \nonumber \\
&&\left.+ \, 2\left(\frac{2-\mu+\lambda}{\mu-\lambda -1}\right)\, \Omega_{k0}e^{-2x} 
+3 \, \left[\frac{3\lambda-2\mu}{2(\mu-1)-3\lambda}\right]\, \Omega_{m0}e^{-3x}\nonumber \right.\\ 
&&\left.+\frac{2(\mu-1)}{\lambda} \left[  1 - \frac{\Omega_{k0}}{\mu - 1 - \lambda} + \frac{2 \Omega_{m0}}{2(\mu-1) - 3\lambda} \right] \, e^{-\frac{2(\mu-1)}{\lambda} x}\right\} \times\nonumber \\
&&\left\{ \Omega_{k0} e^{-2x} + \Omega_{m0} e^{-3x} \right.\nonumber \\
&&\left.+ \, \left(\frac{2-\mu+\lambda}{\mu-\lambda -1}\right)\, \Omega_{k0}e^{-2x} 
+ \, \left[\frac{3\lambda-2\mu}{2(\mu-1)-3\lambda}\right]\, \Omega_{m0}e^{-3x}\nonumber\right. \\ 
&&\left.+ \left[  1 - \frac{\Omega_{k0}}{\mu - 1 - \lambda} + \frac{2 \Omega_{m0}}{2(\mu-1) - 3\lambda} \right] \, e^{-\frac{2(\mu-1)}{\lambda} x}\right\} ^{-1}
\end{eqnarray}

\subsection{Interacting Case}
We now consider the NHDE model in presence of interaction between DM and DE.\\
Also in this case, we consider the same expression of $Q$ we have considered for the previous DE model. \\
We remember that, for the interacting case, $\rho_m$ can be expressed as:
\begin{eqnarray}\label{10}
\rho_{m}=\rho_{m_{0}}e^{-3(1-d^2)x},
\end{eqnarray}
which leads to:
\begin{eqnarray}\label{10}
\Omega_{m}=\Omega_{m{0}}e^{-3(1-d^2)x},
\end{eqnarray}

In this case, the differential equation for $h^2$ is given by:
\begin{eqnarray}\label{11}
\frac{dh^2}{dx}+\frac{2(\mu-1)}{\lambda}h^2-\frac{2\Omega_{k0}}{\lambda}e^{-2x}
+\frac{2\Omega_{m0}}{\lambda}e^{-3(1-d^2)x} =0,
\end{eqnarray}
which general solution is given by:
\begin{eqnarray}
h^2_{NHDE,I}=\frac{\Omega_{k0}e^{-2x}}{\mu-\lambda-1}
-\frac{2\Omega_{m0}e^{-3(1-d^2)x}}{2(\mu-1)-3(1-d^2)\lambda}+B
e^{-\frac{2}{\lambda}{(\mu-1)x}}.\label{king10}
\end{eqnarray}
where $B$ is an integration constant which can be set from the initial conditions. \\
Eq. (\ref{king10}) can be also written as:
\begin{eqnarray}
h^2_{NHDE,I}&=& \Omega_{k0} e^{-2x} + \Omega_{m0} e^{-3x} \nonumber \\
&&+\left( \frac{2 - \mu + \lambda}{\mu - \lambda - 1}\right)\Omega_{k0} \, e^{-2x}\nonumber \\
&&+ \left[\frac{ 3(1-d^2)\lambda-2\mu}{2(\mu-1) - 3(1-d^2)\lambda}\right]\Omega_{m0} \, e^{-3(1-d^2)x}
+ B \, e^{-\frac{2(\mu-1)}{\lambda} x}\label{king11}
\end{eqnarray}

The initial condition, i.e. that $h^2(0)=1$  $a_0=1$, we obtain the following expression for B:
\begin{eqnarray}\label{13}
B =  1 - \frac{\Omega_{k0}}{\mu - \lambda - 1} + \frac{2 \, \Omega_{m0}}{2(\mu-1) - 3(1-d^2)\lambda}.
\end{eqnarray}

Therefore, we have that Eq. (\ref{king11}) can be rewritten as:
\begin{eqnarray}
h^2_{NHDE,I}&=& \Omega_{k0} e^{-2x} + \Omega_{m0} e^{-3(1-d^2)x}\nonumber  \\
&&  \left( \frac{2 - \mu + \lambda}{\mu - \lambda - 1}\right)\Omega_{k0} \, e^{-2x}
+ \left[\frac{ 3(1-d^2)\lambda-2\mu}{2(\mu-1) - 3(1-d^2)\lambda}\right]\Omega_{m0} \, e^{-3(1-d^2)x}\nonumber \\
&&+ \left\{  1 - \frac{\Omega_{k0}}{\mu - \lambda - 1} + \frac{2 \, \Omega_{m0}}{2(\mu-1) - 3(1-d^2)\lambda}. \right\}\, e^{-\frac{2(\mu-1)}{\lambda} x}.
\end{eqnarray}

From the expression of $h^2$, we obtain that the energy density of the HNDE model can be also written as:
\begin{eqnarray}
\rho_{D,NHDE,I} &=& 3H_0^2 \left\{   \left( \frac{2 - \mu + \lambda}{\mu - \lambda - 1}\right)\Omega_{k0} \, e^{-2x}
+ \left[\frac{ 3(1-d^2)\lambda-2\mu}{2(\mu-1) - 3(1-d^2)\lambda}\right]\Omega_{m0} \, e^{-3(1-d^2)x}\right.\nonumber \\
&&\left.+ \left\{  1 - \frac{\Omega_{k0}}{\mu - \lambda - 1} + \frac{2 \, \Omega_{m0}}{2(\mu-1) - 3(1-d^2)\lambda}. \right\}\, e^{-\frac{2(\mu-1)}{\lambda} x}.  \right\}
\end{eqnarray}
Moreover, from the DE continuity equation given by Eq. (\ref{12deprimeI}), the general relation for the DE pressure $p_{D,NHDE}$ as a function of $\rho_{D,NHDE}$ and its derivative with respect to $x$ is:
\begin{eqnarray}
p_{D,NHDE,I} = -\rho_{D,NHDE,I} - \frac{\rho_{D,NHDE,I}'}{3} - \frac{Q}{3H} \label{pdnongen}.
\end{eqnarray}
Differentiating with respect to $x$ the expression of $\rho_{D,NHDE}$, we obtain:
\begin{eqnarray}
\rho'_{D,NHDE} &=& -3H_0^2 \left\{ 2  \left( \frac{2 - \mu + \lambda}{\mu - \lambda - 1}\right)\Omega_{k0} \, e^{-2x}
+3(1-d^2) \left[\frac{ 3(1-d^2)\lambda-2\mu}{2(\mu-1) - 3(1-d^2)\lambda}\right]\Omega_{m0} \, e^{-3(1-d^2)x}\right.\nonumber \\
&&\left.+\frac{2(\mu-1)}{\lambda} \left\{  1 - \frac{\Omega_{k0}}{\mu - \lambda - 1} + \frac{2 \, \Omega_{m0}}{2(\mu-1) - 3(1-d^2)\lambda}. \right\}\, e^{-\frac{2(\mu-1)}{\lambda} x}.  \right\}
\end{eqnarray}
Moreover, we have that:
\begin{eqnarray}
    \frac{Q}{3H} = d^2\rho_m = 3H_0^2d^2\Omega_{m0}e^{-3(1-d^2)x}
\end{eqnarray}
Therefore we obtain the following final expression of $p_{D,NHDE}$:
\begin{eqnarray}
p_{D,NHDE} &=&-H_0^2 \left\{
\left( \frac{2 - \mu + \lambda}{\mu - \lambda - 1}\right)\Omega_{k0} \, e^{-2x} 
\nonumber \right.\\
&&\left. + 3 d^2 \left[\frac{ 3(1-d^2)\lambda-2\mu}{2(\mu-1) - 3(1-d^2)\lambda}+1\right]\Omega_{m0} \, e^{-3(1-d^2)x} \right.\nonumber \\
&& \left.-\left[\frac{2(\mu-1)-3\lambda}{\lambda}  \right] 
\left[ 1 - \frac{\Omega_{k0}}{\mu - \lambda - 1} + \frac{2 \, \Omega_{m0}}{2(\mu-1) - 3(1-d^2)\lambda} \right] e^{-\frac{2(\mu-1)}{\lambda} x} 
\right\}.
\end{eqnarray}
From the DE continuity equation given in Eq. (\ref{12deoldprimeI}), the general relation for the DE equation of state (EoS) parameter is:
\begin{eqnarray}
\omega_{D,NHDE,I} = -1 - \frac{\rho'_{D,NHDE,I}}{3 \rho_{D,NHDE,I} } -\frac{Q}{3H\rho_{D,NHDE,I}} \label{eosnongen}.
\end{eqnarray}
Using the expressions of $\rho_{D,NHDE,I}$, $\rho'_{D,NHDE,I}$ and $\frac{Q}{3H\rho_{D,NHDE,I}}$ we described above, we obtain the EoS parameter of the interacting NHDE model is given by:
\begin{eqnarray}\nonumber
\omega_{D,HDNE,I}&=&-1 + \frac{1}{3}\left\{   2  \left( \frac{2 - \mu + \lambda}{\mu - \lambda - 1}\right)\Omega_{k0} \, e^{-2x}\nonumber \right.\\
&&\left.
+\left\{3(1-d^2) \left[\frac{ 3(1-d^2)\lambda-2\mu}{2(\mu-1) - 3(1-d^2)\lambda}\right]-3d^2\right\}\Omega_{m0} \, e^{-3(1-d^2)x}\right.\nonumber \\
&&\left.+\frac{2(\mu-1)}{\lambda} \left\{  1 - \frac{\Omega_{k0}}{\mu - \lambda - 1} + \frac{2 \, \Omega_{m0}}{2(\mu-1) - 3(1-d^2)\lambda}. \right\}\, e^{-\frac{2(\mu-1)}{\lambda} x} \right\} \times\nonumber \\
&&\left\{   \left( \frac{2 - \mu + \lambda}{\mu - \lambda - 1}\right)\Omega_{k0} \, e^{-2x}
+ \left[\frac{ 3(1-d^2)\lambda-2\mu}{2(\mu-1) - 3(1-d^2)\lambda}\right]\Omega_{m0} \, e^{-3(1-d^2)x}\right.\nonumber \\
&&\left.+ \left\{  1 - \frac{\Omega_{k0}}{\mu - \lambda - 1} + \frac{2 \, \Omega_{m0}}{2(\mu-1) - 3(1-d^2)\lambda}. \right\}\, e^{-\frac{2(\mu-1)}{\lambda} x}  \right\}^{-1}
\end{eqnarray}

In this case, the deceleration parameter is defined as:
\begin{eqnarray}
    q_{NHDE,I} &=& -1 - \frac{1}{2h^2_{NHDE,I}}\frac{d h^2_{NHDE,I}}{dx}
\end{eqnarray}
Differentiating the expression of $h^2_{NHDE,I}$ with respect to the variable $x$, we obtain:
\begin{eqnarray}\label{}
\frac{d h^2_{NHDE,I}}{dx}&=&-2\Omega_{k0} e^{-2x} -3(1-d^2) \Omega_{m0} e^{-3(1-d^2)x}\nonumber  \\
&&  -2\left( \frac{2 - \mu + \lambda}{\mu - \lambda - 1}\right)\Omega_{k0} \, e^{-2x}
-3(1-d^2) \left[\frac{ 3(1-d^2)\lambda-2\mu}{2(\mu-1) - 3(1-d^2)\lambda}\right]\Omega_{m0} \, e^{-3(1-d^2)x}\nonumber \\
&&-\frac{2(\mu-1)}{\lambda} \left\{  1 - \frac{\Omega_{k0}}{\mu - \lambda - 1} + \frac{2 \, \Omega_{m0}}{2(\mu-1) - 3(1-d^2)\lambda}. \right\}\, e^{-\frac{2(\mu-1)}{\lambda} x}.
\end{eqnarray}
Therefore, the final expression of the deceleration parameter is given by:
\begin{eqnarray}
q_{NHDE,I} &=& -1 + \frac{1}{2} \left\{2\Omega_{k0} e^{-2x} +3(1-d^2) \Omega_{m0} e^{-3(1-d^2)x}\right.\nonumber  \\
&&\left.  +2\left( \frac{2 - \mu + \lambda}{\mu - \lambda - 1}\right)\Omega_{k0} \, e^{-2x}
+3(1-d^2) \left[\frac{ 3(1-d^2)\lambda-2\mu}{2(\mu-1) - 3(1-d^2)\lambda}\right]\Omega_{m0} \, e^{-3(1-d^2)x}\right.\nonumber \\
&&\left.+\frac{2(\mu-1)}{\lambda} \left\{  1 - \frac{\Omega_{k0}}{\mu - \lambda - 1} + \frac{2 \, \Omega_{m0}}{2(\mu-1) - 3(1-d^2)\lambda}. \right\}\, e^{-\frac{2(\mu-1)}{\lambda} x}
\right\}\times\nonumber \\
&&\left\{\Omega_{k0} e^{-2x} + \Omega_{m0} e^{-3(1-d^2)x}\nonumber\right.  \\
&&\left.  \left( \frac{2 - \mu + \lambda}{\mu - \lambda - 1}\right)\Omega_{k0} \, e^{-2x}
+ \left[\frac{ 3(1-d^2)\lambda-2\mu}{2(\mu-1) - 3(1-d^2)\lambda}\right]\Omega_{m0} \, e^{-3(1-d^2)x}\right.\nonumber \\
&&\left.+ \left\{  1 - \frac{\Omega_{k0}}{\mu - \lambda - 1} + \frac{2 \, \Omega_{m0}}{2(\mu-1) - 3(1-d^2)\lambda}. \right\}\, e^{-\frac{2(\mu-1)}{\lambda} x}
\right\}^{-1}
\end{eqnarray}

\subsection{Dark Dominated Case}
We now want to study the limiting case of a Dark Dominated Universe. In this case, we can write:
\begin{eqnarray}
h^2_{DD,NHDE}&=&e^{-\frac{2(\mu-1)}{\lambda}x  }\\
\rho_{D,DD,NHDE} &=& 3H_0^2e^{-\frac{2(\mu-1)}{\lambda}x}\\
\omega_{D,DD,NHDE}&=& \frac{2(\mu - 1)}{3\lambda} - 1=-\frac{3\lambda-2\mu+2}{3\lambda}\\
q_{DD,NHDE} &=&   -1 + \frac{\mu-1}{\lambda} = \frac{\mu - 1 - \lambda}{\lambda}. \label{deceleration}
\end{eqnarray}
In order to have an accelerated phase of the Universe, we must have $q<0$, therefore we must have $\mu < 1 + \frac{\lambda}{2}$ with $\lambda >0$.\\
In the limiting case corresponding to the Ricci DE, i.e. for $\mu=2$ and $\lambda=1$, we obtain:
\begin{eqnarray}
h^2_{DD,NHDE,Ricci}&=&e^{-2x  }\\
\rho_{D,DD,NHDE,Ricci} &=& 3H_0^2e^{- 2x}\\
\omega_{D,DD,NHDE,Ricci}&=&-\frac{1}{3}\\
q_{DD,NHDE,Ricci} &=&  0
\end{eqnarray}
For the chosen model parameters $\mu = 2$ and $\lambda = 1$, we obtain a deceleration parameter $q = 0$. A vanishing deceleration parameter indicates that the Universe expands at a constant rate, without acceleration or deceleration. This corresponds to a coasting cosmology, characterized by a linearly growing scale factor:
\begin{eqnarray}
a(t) \propto t + C.
\end{eqnarray}
In this scenario the scale factor $\dot a(t)$ remains constant in time, the Hubble parameter decreases as $H(t) = 1/(t+C)$, consistent with the linear growth of $a(t)$ and the energy density evolves as $\rho(t) \propto (t+C)^{-2}$.\\
Such a model represents a Universe where gravitational deceleration and dark energy–driven acceleration exactly cancel, producing a linear expansion history.\\
Instead, in the limiting cases corresponding to the values found by Wang \& Xu, i.e. $\mu= 0.8502$ and $\lambda = 0.4817$, we obtain:
\begin{eqnarray}
h^2_{DD,NHDE,WX} &&\approx e^{0.6220 \, x}, \\
\rho_{D,DD,NHDE,WX} &&\approx  3 H_0^2 \, e^{0.6220 \, x}, \\
\omega_{D,DD,NHDE,WX} &&\approx -1.2073, \\
q_{DD,NHDE} &&\approx -1.311.
\end{eqnarray}
Therefore, in this case, we obtain a negative value for $q$, which is consistent with current observational data.

We now want to find the expressions of some parameters as a function of time $t$. Solving the expression of $h^2_{DD}$, we have:
\begin{eqnarray}
a_{DD,NHDE}(t) = \left[ \left(\frac{\mu - 1}{\lambda}\right) (t + C) \right]^{\frac{\lambda}{\mu - 1}}
\end{eqnarray}
where $C$ is an integration constant.\\
Therefore, we obtain:
\begin{eqnarray}
    H_{DD,NHDE}(t) = \left(\frac{\lambda}{\mu - 1}\right) \cdot \frac{1}{t + C}
\end{eqnarray}
The results we obtained here for $a(t)$ and $H(t)$ are in agreement with the results we obtain from the results $q=0$ we obtained above.\\
Inserting the expression of $H(t)$ in the general expression of the energy density, we have:
\begin{eqnarray}
    \rho_{D,DD,NHDE}(t) = \frac{3 \lambda^2}{(\mu - 1)^2 (t + C)^2}
\end{eqnarray}
In the limiting case corresponding to the Ricci DE, i.e. for $\mu=2$ and $\lambda=1$, we obtain:
\begin{eqnarray}
a_{DD,NHDE,Ricci}(t) &=& (t + C),\\
H_{DD,NHDE.Ricci}(t) &=& \frac{1}{t + C},\\
\rho_{D,DD,NHDE,Ricci}(t) &=& \frac{3}{(t + C)^2},
\end{eqnarray}
Instead, in the limiting cases corresponding to the values found by Wang \& Xu, we obtain:
\begin{eqnarray}
 a_{DD,NHDE,WX}(t) &&\approx \left[-0.3110 \,(t+C)\right]^{-3.21495}\\
H_{DD,NHDE,WX}(t) &&\approx \frac{-3.21495}{t+C}\\
\rho_{D,DD,NHDE,WX}(t) &&\approx \frac{31.0202}{(t+C)^2}
\end{eqnarray}
In order to have positive values of $a_{DD,NHDE,WX}$ and $H_{DD,NHDE,WX}$ we need to have $t+C<0$.

 In the limiting case of $C=0$, we have:
\begin{eqnarray}
a_{DD,NHDE,lim}(t) &=& \left[\left( \frac{\mu - 1}{\lambda}\right) t  \right]^{\frac{\lambda}{\mu - 1}}\\
    H_{DD,NHDE,lim}(t) &=& \left(\frac{\lambda}{\mu - 1}\right) \cdot \frac{1}{t }\\
    \rho_{D,DD,NHDE,im}(t) &=& \frac{3 \lambda^2}{(\mu - 1)^2 t ^2}
\end{eqnarray}

In the limiting case corresponding to the Ricci DE, i.e. for $\mu=2$ and $\lambda=1$, we obtain:
\begin{eqnarray}
a_{DD,NHDE,Ricci,lim}(t) &=& t ,\\
H_{DD,NHDE,Ricci,lim}(t) &=& \frac{1}{t },\\
\rho_{D,DD,NHDE,Ricci,lim}(t) &=& \frac{3}{t^2},
\end{eqnarray}

Instead, in the limiting cases corresponding to the values found by Wang \& Xu, we obtain:
\begin{eqnarray}
 a_{DD,NHDE,WX,lim}(t) &&\approx \left[-0.3110 \,t\right]^{-3.21495}\\
H_{DD,NHDE,WX,lim}(t) &&\approx \frac{-3.21495}{t}\\
\rho_{D,DD,NHDE,WX,lim}(t) &&\approx \frac{31.0202}{t^2}
\end{eqnarray}
In this case, the computed values of $a_{DD,NHDE,WX,lim}$ and $H_{DD,NHDE,WX,lim}$ are problematic, as they lead to unrealistic or unphysical results. For this reason, we will not consider for furhter calculations.

\section{CORRESPONDENCE BETWEEN THE  MODEL  AND SCALAR FIELDS MODELS}
In this Section, we aim to establish a correspondence between the dark energy (DE) model investigated in this work and several well-known scalar field models, namely: the Generalized Chaplygin Gas (GCG), the Modified Chaplygin Gas (MCG), the Modified Variable Chaplygin Gas (MVCG), the Viscous Generalized Chaplygin Gas (VGCG), the Dirac-Born-Infeld (DBI) model, the Yang-Mills (YM) model and finally the Non-Linear Electrodynamics (NLED) model.

This choice is motivated by the fact that scalar field models are widely regarded as effective frameworks for describing dark energy phenomena. Therefore, we proceed by first comparing the energy density of the DE model proposed in this paper with the energy densities of the corresponding scalar field models. Subsequently, we match the equation of state (EoS) parameter of each scalar field model under consideration with the EoS parameter of our DE model.

\subsection{Generalized Chaplygin Gas (GCG) Model}
We start by establishing a correspondence between the two DE models analyzed in this work and the first scalar field model under consideration, namely the Generalized Chaplygin Gas (GCG) model. In a seminal work, Kamenshchik \emph{et al.}~\cite{gcg1} proposed a homogeneous model known as the Chaplygin Gas (CG) model, which is based on a single fluid obeying the equation of state (EoS) $p = -\frac{A_0}{\rho}$, where $p$ and $\rho$ denote the pressure and energy density of the fluid, respectively, and $A_0$ is a positive constant parameter. The authors subsequently introduced a generalization of this framework, referred to as the Generalized Chaplygin Gas (GCG) model.

An appealing feature of the GCG model is its ability to interpolate between a dust-dominated Universe at early times and a late-time accelerated expansion phase. As such, it provides a viable fit to current observational cosmological data~\cite{gcg6}.
The EoS of the GCG model is defined as follows~\cite{gcg9,gcg16,gcg17}:
\begin{eqnarray}
    p_D = -\frac{D}{\rho_D^{\theta}}, \label{gcg1}
\end{eqnarray}
where $D$ and $\theta$ are two free constant parameters, with $D > 0$ and $\theta$ constrained to the range $0 < \theta < 1$. The standard Chaplygin Gas (CG) model is recovered for $\theta = 1$. The EoS in Eq.~(\ref{gcg1}) with $\theta = 1$ was first studied in 1904 by Chaplygin in the context of adiabatic processes~\cite{gcg1}. The more general case with $\theta \neq 1$ was explored in~\cite{cgas2}.

The idea that a cosmological model based on the Chaplygin gas could provide a unified description of dark energy (DE) and dark matter (DM) was initially proposed for $\theta = 1$ in~\cite{gcg4,gcg5}, and later extended to the case $\theta \neq 1$ in~\cite{cgas2}.

It was shown by Gorini \emph{et al.}~\cite{gcg15} that the matter power spectrum predicted by the GCG model is consistent with the observed one only if $\theta < 10^{-5}$. This constraint implies that the GCG model becomes practically indistinguishable from the standard $\Lambda$CDM model. In~\cite{gcg7}, Chaplygin inflation was studied within the framework of Loop Quantum Cosmology.

The evolution of the energy density $\rho_D$ in the GCG model is given by:
\begin{eqnarray}
    \rho_D = \left[D + \frac{B}{a^{3\left(\theta+1\right)}}\right]^{\frac{1}{\theta+1}}, \label{gcg2}
\end{eqnarray}
where $B$ is an integration constant.

In principle, Eq.~(\ref{gcg2}) allows for a wide range of positive values of the parameter $\theta$. However, we must ensure that $\theta$ satisfies the condition that the sound speed (given by $c_s^2 = \frac{D\theta}{\rho^{\theta+1}}$) does not exceed the speed of light, $c$. Moreover, as discussed in Bento \emph{et al.}~\cite{cgas2}, only values in the range $0 < \theta < 1$ lead to a physically meaningful analysis of the evolution of energy density fluctuations.

We now aim to reconstruct the potential and dynamics of this scalar field model within the framework of the DE model considered in this work. The energy density $\rho_D$ and the pressure $p_D$ of a homogeneous and time-dependent scalar field $\phi$ are given, respectively, by:
\begin{eqnarray}
\rho_D &=& \frac{1}{2}\dot{\phi}^2 + V\left(\phi\right), \label{gcg3}\\
p_D &=& \frac{1}{2}\dot{\phi}^2 - V\left(\phi\right). \label{gcg4}
\end{eqnarray}
Using Eqs.~(\ref{gcg3}) and (\ref{gcg4}), we can express the equation of state (EoS) parameter $\omega_D$ of the GCG model as:
\begin{eqnarray}
\omega_D &=& \frac{\frac{1}{2} \dot{\phi}^2 - V\left(\phi\right)}{\frac{1}{2} \dot{\phi}^2 + V\left(\phi\right)} = \frac{\dot{\phi}^2 - 2 V\left(\phi\right)}{\dot{\phi}^2 + 2 V\left(\phi\right)}. \label{gcg5}
\end{eqnarray}

Adding Eqs.~(\ref{gcg3}) and (\ref{gcg4}), we easily obtain the kinetic energy term $\dot{\phi}^2$ as
\begin{eqnarray}
    \dot{\phi}^2 &=& \rho_D + p_D. \label{gcg6-1}
\end{eqnarray}
Substituting the definition of $p_D$ from Eq.~(\ref{gcg1}) into Eq.~(\ref{gcg6-1}), we get
\begin{eqnarray}
    \dot{\phi}^2 &=& \rho_D - \frac{D}{\rho_D^\theta} = \rho_D \left(1 - \frac{D}{\rho_D^{\theta+1}}\right). \label{gcg6-11}
\end{eqnarray}
Considering the relation $\dot{\phi} = H \phi'$, and using the expression of $\rho_D$ given in Eq.~(\ref{gcg2}), the evolution of the scalar field $\phi$ can be written as
\begin{eqnarray}
    \phi &=& \int_{a_0}^a \sqrt{3\Omega_D} \sqrt{1 - \frac{D}{D + \frac{B}{a^{3(\theta+1)}}}} \frac{da}{a}, \label{gcg6-1newint}
\end{eqnarray}
whose solution, in the case of a flat, dark energy dominated Universe, reads
\begin{eqnarray}
 \phi(a) &=& \frac{2 \sqrt{3}}{3(1+\theta)} \times \nonumber \\
 && \left[\log\left(a^{\frac{3}{2}(1+\theta)}\right) - \log\left(B + \sqrt{B\left(B + a^{3(1+\theta)} D\right)}\right)\right]. \label{murano4}
\end{eqnarray}

Subtracting Eqs.~(\ref{gcg4}) from (\ref{gcg3}), we find the scalar potential $V(\phi)$ as
\begin{eqnarray}
 V(\phi) &=& \frac{1}{2} (\rho_D - p_D) = \frac{\rho_D}{2} \left(1 + \frac{D}{\rho_D^{\theta+1}}\right), \label{gcg7-1}
\end{eqnarray}
where we have used the definition of $p_D$ from Eq.~(\ref{gcg1}). Using the general form of $\rho_D$ in Eq.~(\ref{gcg2}), we write $V(\phi)$ explicitly as
\begin{eqnarray}
V(\phi) &=& \frac{1}{2}\left[D + \frac{B}{a^{3(\theta+1)}}\right]^{\frac{1}{\theta+1}} + \frac{1}{2} \frac{D}{\left[D + \frac{B}{a^{3(\theta+1)}}\right]^{\frac{\theta}{\theta+1}}}. \label{gcg7-1new}
\end{eqnarray}

Next, we derive general expressions for the parameters $D$ and $B$ in terms of cosmological quantities. Dividing Eq.~(\ref{gcg1}) by $\rho_D$, the EoS parameter $\omega_D$ reads
\begin{eqnarray}
    \omega_D = - \frac{D}{\rho_D^{\theta+1}}, \label{gcg8}
\end{eqnarray}
which leads to the expression
\begin{eqnarray}
    D = - \omega_D \rho_D^{\theta+1}. \label{gcg9}
\end{eqnarray}
From Eq.~(\ref{gcg2}), the integration constant $B$ can be expressed as
\begin{eqnarray}
B = a^{3(\theta+1)} \left(\rho_D^{\theta+1} - D \right), \label{murano5}
\end{eqnarray}
which, substituting Eq.~(\ref{gcg9}) for $D$, becomes
\begin{eqnarray}
B = \left(a^3 \rho_D\right)^{\theta+1} (1 + \omega_D). \label{BBB}
\end{eqnarray}
For the model we studied in this paper, we obtain the followin results for $D$ and $B$:
\begin{eqnarray}
D_{1,higher}&=& -\rho_{D,higher,1}^{\theta + 1}\cdot \omega_{D,higher,1}, \label{murano6}\\
D_{2,higher}&=& -\rho_{D,higher,2}^{\theta + 1}\cdot \omega_{D,higher,2}, \label{murano6-2}\\
D_{3,higher}&=& -\rho_{D,higher,3}^{\theta + 1}\cdot \omega_{D,higherm3}, \label{murano6-3}\\
B_{1,higher}&=&\left(a^3\rho_{D,higher,1}\right)^{\theta+1}\cdot \left( 1+\omega_{D,higher,1}\right)\label{murano7} \\
B_{2,higher}&=&\left(a^3\rho_{D,higher,2}\right)^{\theta+1}\cdot \left( 1+\omega_{D,higher,2}\right)\label{murano7-2} \\
B_{3,higher}&=&\left(a^3\rho_{D,higher,3}\right)^{\theta+1}\cdot \left( 1+\omega_{D,higher,3}\right)\label{murano7-3} 
\end{eqnarray}

For the interacting case, we obtain:
\begin{eqnarray}
D_{1,higher,I}&=& -\rho_{D,higher,I,1}^{\theta + 1}\cdot \omega_{D,higher,I,1}, \label{murano6}\\
D_{2,higher,I}&=& -\rho_{D,higher,I,2}^{\theta + 1}\cdot \omega_{D,higher,I,2}, \label{murano6-2}\\
D_{3,higher,I}&=& -\rho_{D,higher,I,3}^{\theta + 1}\cdot \omega_{D,higher,I,3}, \label{murano6-3}\\
B_{1,higher,I}&=&\left(a^3\rho_{D,higher,I,1}\right)^{\theta+1}\cdot \left( 1+\omega_{D,higher,I,1}\right)\label{murano7} \\
B_{2,higher,I}&=&\left(a^3\rho_{D,higher,I,2}\right)^{\theta+1}\cdot \left( 1+\omega_{D,higher,I,2}\right)\label{murano7-2} \\
B_{3,higher,I}&=&\left(a^3\rho_{D,higher,I,3}\right)^{\theta+1}\cdot \left( 1+\omega_{D,higher,I,3}\right)\label{murano7-3} 
\end{eqnarray}

In the limiting case of a flat, Dark Dominated Universe, we find that $D_{\rm Dark}$ and $B_{\rm Dark}$ are given by the following relations:
\begin{eqnarray}
D_{ Dark,higher} &=& -\left\{ \frac{48\alpha^2}{\left[\beta \left( t+C_3  \right)\right]^2}\right\}^{\theta+1}\left( \frac{\beta}{6\alpha}  -1\right), \label{murano10}\\
B_{ Dark,higher} &=&   3^{\theta+1}\left[\frac{\beta\left( t+C_3 \right)}{4\alpha}\right]^{\frac{\left(12\alpha -2\beta\right)\left(\theta+1\right)}{\beta}}   \left( \frac{\beta}{6\alpha}  -1\right), \label{murano11}
\end{eqnarray}
In the limiting case of $C_3=0$, we obtain:
\begin{eqnarray}
D_{ Dark,higher,lim} &=& -\left\{ \frac{48\alpha^2}{\left(\beta  t\right)^2}\right\}^{\theta+1}\left( \frac{\beta}{6\alpha}  -1\right), \label{murano10}\\
B_{ Dark,higher,lim} &=&   3^{\theta+1}\left[\frac{\beta t}{4\alpha}\right]^{\frac{\left(12\alpha -2\beta\right)\left(\theta+1\right)}{\beta}}   \left( \frac{\beta}{6\alpha}  -1\right), \label{murano11}
\end{eqnarray}

For the HNDE model, we obtain the following results for $D$ and $B$:
\begin{eqnarray}
D_{NHDE}&=& -\rho_{D,NHDE}^{\theta + 1}\cdot \omega_{D,NHDE}, \label{murano6}\\
B_{NHDE}&=&\left(a^3\rho_{D,NHDE}\right)^{\theta+1}\cdot \left( 1+\omega_{D,NHDE}\right)\label{murano7}  
\end{eqnarray}
For the interacting NHDE model, we obtain:
\begin{eqnarray}
D_{NHDE,I}&=& -\rho_{D,NHDE,I}^{\theta + 1}\cdot \omega_{D,NHDE,I}, \label{murano6}\\
B_{NHDE,I}&=&\left(a^3\rho_{D,NHDE,I}\right)^{\theta+1}\cdot \left( 1+\omega_{D,NHDE,I}\right)\label{murano7}  
\end{eqnarray}

In the limiting case of a flat, Dark Dominated Universe, we find that $D_{\rm Dark}$ and $B_{\rm Dark}$ are given by the following relations:
\begin{eqnarray}
D_{ Dark,NHDE} &=&\left[ \frac{3 \lambda^2}{(\mu - 1)^2 (t + C)^2} \right]^{\theta+1}\left(\frac{3\lambda-2\mu+2}{3\lambda}\right) , \label{murano10}\\
B_{ Dark,NHDE} &=& \left[ \frac{\mu - 1}{\lambda} (t + C) \right]^{\frac{3(\theta+1)\lambda}{\mu - 1}} \left[ \frac{3 \lambda^2}{(\mu - 1)^2 (t + C)^2} \right]^{\theta+1}\left( \frac{2\mu-2}{3\lambda} \right), \label{murano11}
\end{eqnarray}
In the limiting case corresponding to the Ricci DE, i.e. for $\mu=2$ and $\lambda=1$, we obtain:
\begin{eqnarray}
D_{\rm Dark,NHDE,Ricci} &=& 3^{\theta} (t + C)^{-2(\theta+1)}, \\
B_{\rm Dark,NHDE,Ricci} &=& 2 \cdot 3^{\theta} (t + C)^{\theta+1}.
\end{eqnarray}

In the limiting case of $C=0$, we obtain:
\begin{eqnarray}
D_{ Dark,lim} &=&\left[ \frac{3 \lambda^2}{(\mu - 1)^2 t ^2} \right]^{\theta+1}\left(\frac{3\lambda-2\mu+2}{3\lambda}\right) , \label{murano10}\\
B_{ Dark,lim} &=& \left[ \left(\frac{\mu - 1}{\lambda}\right) t  \right]^{\frac{3(\theta+1)\lambda}{\mu - 1}} \left[ \frac{3 \lambda^2}{(\mu - 1)^2 t ^2} \right]^{\theta+1}\left( \frac{2\mu-2}{3\lambda} \right), \label{murano11}
\end{eqnarray}
In the limiting case corresponding to the Ricci DE and $C=0$,we obtain:
\begin{eqnarray}
D_{Dark,NHDE,Ricci,lim} &=& 3^{\theta} t^{-2(\theta+1)}, \\
B_{Dark,NHDE,Ricci,lim} &=& 2 \cdot 3^{\theta} t^{\theta+1}.
\end{eqnarray}
In the limiting case corresponding to the values found by Wang \& Xu, we obtain:
\begin{eqnarray}
D_{Dark,NHDE,WX} &\approx& 1.207\cdot\left[ \frac{31.020}{(t + C)^2} \right]^{\theta+1}  \\
B_{Dark,NHDE,WX} &\approx& -0.207\cdot\left[ -0.311 \, (t + C) \right]^{-9.65 (\theta+1)} \left[ \frac{30.99}{(t + C)^2} \right]^{\theta+1} 
\end{eqnarray}

Moreover, using the general definition of EoS parameter $\omega_D$,  we can rewrite Eqs. (\ref{gcg6-1}) and (\ref{gcg7-1}) in the following way:
\begin{eqnarray}
    \dot{\phi}^2 &=& \left( 1+ \omega_D \right)\rho_D, \label{gcg6mura1}\\
    V\left(  \phi \right) &=& \frac{1}{2}\left( 1 - \omega_D \right)\rho_D. \label{gcg7mura2}
\end{eqnarray}
For the DE model studied in this paper, we obtain the following expressions for  $\dot{\phi}^2$ and $ V\left(  \phi \right)$:
\begin{eqnarray}
    \dot{\phi}^2_{higher,1} &=& \rho_{D,higher,1}\left(1+\omega_{D,higher,1} \right)  \label{gcg6mura1model1}\\
    \dot{\phi}^2_{higher,2} &=& \rho_{D,higher,2}\left(1+\omega_{D,higher,2} \right)  \label{gcg6mura1model1-2}\\
    \dot{\phi}^2_{higher,3} &=& \rho_{D,higher,3}\left(1+\omega_{D,higher,3} \right)  \label{gcg6mura1mode1-3}\\
    V\left(  \phi \right)_{higher,1} &=& \frac{\rho_{D,higher,1}}{2}\left(1-\omega_{D,higher,1} \right)   \label{gcg7mura2model1}\\
    V\left(  \phi \right)_{higher,2} &=& \frac{\rho_{D,higher,2}}{2}\left(1-\omega_{D,higher,2} \right)   \label{gcg7mura2model1-2}\\
    V\left(  \phi \right)_{higher,3} &=& \frac{\rho_{D,higher,3}}{2}\left(1-\omega_{D,higher,3} \right)   \label{gcg7mura2model1-3}
\end{eqnarray}

We can obtain the evolutionary form of the GCG model integrating the expressions of $\dot{\phi}^2$ we obtained with respect to the scale factor $a\left(t\right)$
\begin{eqnarray}
	\phi\left(a\right)_{higher,1} - \phi\left(a_0\right)_{higher,1} &=& \int_{a_0}^{a}\sqrt{3\Omega_{D}} \times \left(1+\omega_{D,higher,1} \right)  \frac{da}{a}, \label{}\\
    \phi\left(a\right)_{higher,2} - \phi\left(a_0\right)_{higher,2} &=& \int_{a_0}^{a}\sqrt{3\Omega_{D}} \times \left(1+\omega_{D,higher,2} \right)  \frac{da}{a}, \label{}\\
    \phi\left(a\right)_{higher,3} - \phi\left(a_0\right)_{higher,3} &=& \int_{a_0}^{a}\sqrt{3\Omega_{D}} \times \left(1+\omega_{D,higher,3} \right)  \frac{da}{a}, \label{}
\end{eqnarray}

For the interacting case, we obtain:
\begin{eqnarray}
    \dot{\phi}^2_{higher,I,1} &=& \rho_{D,higher,I,1}\left(1+\omega_{D,higher,I,1} \right)  \label{gcg6mura1model1}\\
    \dot{\phi}^2_{higher,I,2} &=& \rho_{D,higher,I,2}\left(1+\omega_{D,higher,I,2} \right)  \label{gcg6mura1model1-2}\\
    \dot{\phi}^2_{higher,3,I} &=& \rho_{D,higher,I,3}\left(1+\omega_{D,higher,I,3} \right)  \label{gcg6mura1mode1-3}\\
    V\left(  \phi \right)_{higher,I,1} &=& \frac{\rho_{D,higher,I,1}}{2}\left(1-\omega_{D,higher,I,1} \right)   \label{gcg7mura2model1}\\
    V\left(  \phi \right)_{higher,I,2} &=& \frac{\rho_{D,higher,I,2}}{2}\left(1-\omega_{D,higher,I,2} \right)   \label{gcg7mura2model1-2}\\
    V\left(  \phi \right)_{higher,I,3} &=& \frac{\rho_{D,higher,I,3}}{2}\left(1-\omega_{D,higher,I,3} \right)   \label{gcg7mura2model1-3}
\end{eqnarray}
and the evolutionary form of the GCG model are given by:
\begin{eqnarray}
	\phi\left(a\right)_{higher,I,1} - \phi\left(a_0\right)_{higher,I,1} &=& \int_{a_0}^{a}\sqrt{3\Omega_{D}} \times \left(1+\omega_{D,higher,I,1} \right)  \frac{da}{a}, \label{}\\
    \phi\left(a\right)_{higher,I,2} - \phi\left(a_0\right)_{higher,I,2} &=& \int_{a_0}^{a}\sqrt{3\Omega_{D}} \times \left(1+\omega_{D,higher,I,2} \right)  \frac{da}{a}, \label{}\\
    \phi\left(a\right)_{higher,I,3} - \phi\left(a_0\right)_{higher,I,3} &=& \int_{a_0}^{a}\sqrt{3\Omega_{D}} \times \left(1+\omega_{D,higher,I,3} \right)  \frac{da}{a}, \label{}
\end{eqnarray}

In the limiting case for a flat Dark Dominated Universe, we obtain:
\begin{eqnarray}
\dot{\phi}_{Dark,higher}^2\left(t \right) &=&\frac{8\alpha}{\beta}\frac{1}{ \left( t+C_3  \right)^2},\label{dusan1} \\
V_{Dark,higher}\left(t\right)&=&\left(2-\frac{\beta}{6\alpha}   \right) \frac{24\alpha^2}{\left[\beta \left( t+C_3  \right)\right]^2}.\label{murano13gcg}
\end{eqnarray}
Integrating with respect to the time $t$ the expression of $\dot{\phi}_{Dark,higher}^2\left(t \right) $, we obtain:
\begin{eqnarray}
\phi_{Dark,higher}\left(t \right) &=&2\sqrt{\frac{2\alpha}{\beta}}\cdot \ln \left( t+C_3  \right)+constant,\label{dusan1}
\end{eqnarray}
In the limiting case of $C_3=0$, we obtain:
\begin{eqnarray}
\phi_{Dark,higher}\left(t \right) &=&2\sqrt{\frac{2\alpha}{\beta}}\cdot \ln t+constant,\label{dusan1}\\
V_{Darkhigher}\left(t\right)&=&\left(2-\frac{\beta}{6\alpha}   \right) \frac{24\alpha^2}{\left(\beta  t\right)^2}.\label{murano13gcg}
\end{eqnarray}

For the NHDE model, we obtain the following expressions for $\dot{\phi}^2$ and $ V\left(  \phi \right)$:
\begin{eqnarray}
    \dot{\phi}^2_{NHDE} &=& \rho_{D,NHDE}\left(1+\omega_{D,NHDE} \right)  \label{gcg6mura1model1}\\
    V\left(  \phi \right)_{NHDE} &=& \frac{\rho_{D,NHDE,1}}{2}\left(1-\omega_{D,NHDE} \right)   \label{gcg7mura2model1}
\end{eqnarray}
We can obtain the evolutionary form of the GCG model integrating the expressions of $\dot{\phi}^2$ we obtained with respect to the scale factor 
\begin{eqnarray}
	\phi\left(a\right)_{NHDE} - \phi\left(a_0\right)_{NHDE} &=& \int_{a_0}^{a}\sqrt{3\Omega_{D}} \times \left(1+\omega_{D,NHDE} \right)  \frac{da}{a}, \label{}
\end{eqnarray}
For the omteracting NHDE model, we obtain:
\begin{eqnarray}
    \dot{\phi}^2_{NHDE,I} &=& \rho_{D,NHDE,I}\left(1+\omega_{D,NHDE,I} \right)  \label{gcg6mura1model1}\\
    V\left(  \phi \right)_{NHDE,I} &=& \frac{\rho_{D,NHDE,I}}{2}\left(1-\omega_{D,NHDE,I} \right)   \label{gcg7mura2model1}
\end{eqnarray}
and the evolutionary form of the GCG model is given by:
\begin{eqnarray}
	\phi\left(a\right)_{NHDE,I} - \phi\left(a_0\right)_{NHDE,I} &=& \int_{a_0}^{a}\sqrt{3\Omega_{D}} \times \left(1+\omega_{D,NHDE,I} \right)  \frac{da}{a}, \label{}
\end{eqnarray}

In the limiting case for a flat Dark Dominated Universe, we obtain:
\begin{eqnarray}
\dot{\phi}_{Dark,NHDE}^2\left(t \right) &=& \left(\frac{ 2\lambda}{\mu - 1}\right) \frac{1}{(t+C)^2},\label{dusan1} \\
V_{Dark,NHDE}\left(t\right)&=&\frac{ \lambda (3\lambda-\mu+1)}{(\mu - 1)^2 }\cdot \frac{1}{(t + C)^2}  \label{murano13gcg}
\end{eqnarray}
Integrating with respect to the time $t$ the expression of $\dot{\phi}_{Dark,NHDE}^2\left(t \right) $, we obtain:
\begin{eqnarray}
\phi_{Dark,NHDE}\left(t \right) &=&\sqrt{ \frac{ 2\lambda}{\mu - 1}}\cdot \ln (t+C) + const,\label{dusan1}
\end{eqnarray}
In the limiting case corresponding to $\lambda =1$ e $\mu=2$, i.e. the case corresponding to the Ricci scale, we obtain:
\begin{eqnarray}
\phi_{\rm Dark,NHDE,Ricci}(t) &=& \sqrt{2} \, \ln(t + C) + \text{const}, \\
V_{\rm Dark,NHDE,Ricci}(t) &=& \frac{2}{(t + C)^2}.
\end{eqnarray}

In the limiting case of $C=0$, we obtain:
\begin{eqnarray}
\phi_{Dark,NHDE,lim}\left(t \right) &=&\sqrt{ \frac{ 2\lambda}{\mu - 1}}\cdot \ln t + const,\label{dusan1} \\
V_{Dark,NHDE,lim}\left(t\right)&=& \frac{ \lambda (3\lambda-\mu+1)}{(\mu - 1)^2 }\cdot \frac{1}{t^2}  \label{murano13gcg}
\end{eqnarray}

In the limiting case corresponding to $\lambda =1$ e $\mu=2$, i.e. the case corresponding to the Ricci scale, we obtain:
\begin{eqnarray}
\phi_{\rm Dark,NHDE,Ricci,lim}(t) &=& \sqrt{2} \, \ln t  + \text{const}, \\
V_{\rm Dark,NHDE,Ricci,lim}(t) &=& \frac{2}{t ^2}.
\end{eqnarray}

We can easily derive the following relation between $\phi_{Dark,NHDE,Ricci,lim}$ and $V_{Dark,NHDE,Ricci,lim}$:
\begin{eqnarray}
 V(\phi) = 2 \, e^{ -\sqrt{2} \phi } 
\end{eqnarray}

In the limiting case corresponding to the values obtained by Wang \& Xu, the scalar field $\phi_{Dark}$ becomes complex, while the potential reads
\begin{eqnarray}
V_{Dark,NHDE,WX}(t) &\approx& \frac{34.22}{t^2}.
\end{eqnarray}
We find that $\phi_{Dark,NHDE}$ is not a real function.

\subsection{Modified  Chaplygin Gas (MCG) Model}
We now focus on the second scalar field model considered in this work, namely the Modified Chaplygin Gas (MCG) model, which generalizes the Generalized Chaplygin Gas (GCG) model \cite{mcg1,mcg2,mcg3,mcg4} by the addition of a barotropic term.

The equation of state (EoS) for the MCG model is expressed as \cite{mcg1,mcg5}:
\begin{eqnarray}
p_D = A \rho_D - \frac{D}{\rho_D^\theta}, \label{mcg1}
\end{eqnarray}
where $A$ and $D$ are positive constants, and the parameter $\theta$ lies within the interval $0 \leq \theta \leq 1$.

A notable feature of the MCG EoS is its ability to describe a radiation-dominated era at early times, while at late times it behaves like a model consistent with the $\Lambda$CDM paradigm.

The corresponding energy density $\rho_D$ evolves according to:
\begin{eqnarray}
\rho_D = \left[ \frac{D}{A+1} + \frac{B}{a^{3(\theta+1)(A+1)}} \right]^{\frac{1}{\theta+1}}, \label{mcg2}
\end{eqnarray}
where $B$ is an integration constant.\\

We now proceed to reconstruct the potential and dynamics of the scalar field $\phi$ corresponding to the Modified Chaplygin Gas (MCG) model. The energy density $\rho_D$ and pressure $p_D$ of a homogeneous and time-dependent scalar field are given by:
\begin{eqnarray}
\rho_D &=& \frac{1}{2} \dot{\phi}^2 + V(\phi), \label{mcg3} \\
p_D &=& \frac{1}{2} \dot{\phi}^2 - V(\phi). \label{mcg4}
\end{eqnarray}

Using these relations, the equation of state (EoS) parameter $\omega_D$ can be expressed as:
\begin{eqnarray}
\omega_D = \frac{p_D}{\rho_D} = \frac{\frac{1}{2} \dot{\phi}^2 - V(\phi)}{\frac{1}{2} \dot{\phi}^2 + V(\phi)} = \frac{\dot{\phi}^2 - 2 V(\phi)}{\dot{\phi}^2 + 2 V(\phi)}. \label{mcg5}
\end{eqnarray}

Adding Eqs. \eqref{mcg3} and \eqref{mcg4}, the kinetic term of the scalar field can be written as:
\begin{eqnarray}
\dot{\phi}^2 = \rho_D + p_D. \label{mcg6-1}
\end{eqnarray}

Substituting the MCG pressure from Eq. \eqref{mcg1} into Eq. \eqref{mcg6-1}, we obtain:
\begin{eqnarray}
\dot{\phi}^2 = \rho_D + A \rho_D - \frac{D}{\rho_D^\theta} = \rho_D \left(1 + A - \frac{D}{\rho_D^{\theta + 1}}\right). \label{dotfi}
\end{eqnarray}

Inserting the expression of $\rho_D$ from Eq. \eqref{mcg2} into Eq. \eqref{dotfi}, we write the kinetic term as
\begin{eqnarray}
\dot{\phi}^2 = \rho_D \left[ 1 + A - \frac{D}{\frac{D}{A+1} + \frac{B}{a^{3(\theta+1)(A+1)}}} \right]. \label{dotphi2}
\end{eqnarray}

Using the relation $\dot{\phi} = H \frac{d\phi}{d\ln a} = H \phi'$, where the prime denotes differentiation with respect to $\ln a$, the evolutionary form of the scalar field $\phi$ can be expressed as
\begin{eqnarray}
\phi = \int_{a_0}^a \sqrt{3 \Omega_D} \sqrt{(1+A) - \frac{D}{\frac{D}{A+1} + \frac{B}{a^{3(\theta+1)(A+1)}}}} , \frac{da}{a}. \label{phi_integral}
\end{eqnarray}

For a flat Dark Dominated Universe, the solution to the integral \eqref{phi_integral} is given by
\begin{eqnarray}
\phi(a) = \frac{2}{\sqrt{3} \sqrt{1+A} , (1+\theta)} \cdot \mathrm{ArcTanh}\left[\frac{\sqrt{(1+A) B + a^{3(1+A)(1+\theta)} D}}{\sqrt{B(1+A)}}\right].
\end{eqnarray}

Subtracting Eq. \eqref{mcg4} from Eq. \eqref{mcg3}, we derive the scalar potential $V(\phi)$ as
\begin{eqnarray}
V(\phi) = \frac{1}{2} (\rho_D - p_D). \label{V_phi}
\end{eqnarray}

Using the explicit forms of $\rho_D$ and $p_D$ from Eqs. \eqref{mcg2} and \eqref{mcg1}, respectively, the potential reads
\begin{eqnarray}
V(\phi) = \frac{1 - A}{2} \left[\frac{D}{A+1} + \frac{B}{a^{3(\theta+1)(A+1)}}\right]^{\frac{1}{\theta+1}} + \frac{D}{2} \left[\frac{D}{A+1} + \frac{B}{a^{3(\theta+1)(A+1)}}\right]^{-\frac{\theta}{\theta+1}}. \label{V_phi_explicit}
\end{eqnarray}

Dividing the equation of state \eqref{mcg1} by \(\rho_D\), the equation of state parameter \(\omega_D\) can be expressed as
\begin{eqnarray}
\omega_D = A - \frac{D}{\rho_D^{\theta + 1}}. \label{mcg8}
\end{eqnarray}

From this relation, the parameter \(D\) can be isolated as
\begin{eqnarray}
D = \rho_D^{\theta + 1} (A - \omega_D). \label{mcg9}
\end{eqnarray}

Moreover, the parameter \(A\) can be written in terms of \(\omega_D\) and \(D\) as
\begin{eqnarray}
A = \omega_D + \frac{D}{\rho_D^{\theta + 1}}. \label{mcg8new}
\end{eqnarray}

Starting from the expression of the energy density \eqref{mcg2}, the integration constant \(B\) can be rewritten as
\begin{eqnarray}
B = a^{3(\theta + 1)(A+1)} \left( \rho_D^{\theta + 1} - \frac{D}{A+1} \right). \label{mcgb}
\end{eqnarray}

Substituting the expression of \(D\) from \eqref{mcg9} into \eqref{mcgb}, we obtain
\begin{eqnarray}
B = \left[a^{3(A+1)} \rho_D \right]^{\theta + 1} \left( \frac{1 + \omega_D}{1 + A} \right). \label{MMmcg}
\end{eqnarray}

For the DE model we are considering in this paper, we derive the following expressions for $D$ and $B$:
\begin{eqnarray}
D_{higher,1} &=&\left(\rho_{D,higher,1}\right)^{\theta +1}\left( A-  \omega_{D,higher,1}\right),\label{mcg11} \\
D_{higher,2} &=&\left(\rho_{D,higher,2}\right)^{\theta +1}\left( A-  \omega_{D,higher,2}\right),\label{mcg11-}\\
D_{higher,3} &=&\left(\rho_{D,higher,3}\right)^{\theta +1}\left( A-  \omega_{D,higher,3}\right),\label{mcg11--}\\
B_{higher,1} &=& \frac{[a^{3(A+1)}\rho_{D,higher,1}]^{\theta +1}}{1+A}\left( 1+\omega_{D,higher,1}\right).\label{mcg13}\\
B_{higher,2} &=& \frac{[a^{3(A+1)}\rho_{D,higher,2}]^{\theta +1}}{1+A}\left( 1+\omega_{D,higher,2}\right).\label{mcg13-}\\
B_{higher,3} &=& \frac{[a^{3(A+1)}\rho_{D,higher,3}]^{\theta +1}}{1+A}\left( 1+\omega_{D,higher,3}\right).\label{mcg13--}
\end{eqnarray}

For the interacting case, we obtain:
\begin{eqnarray}
D_{higher,I,1} &=&\left(\rho_{D,higher,I,1}\right)^{\theta +1}\left( A-  \omega_{D,higher,I,1}\right),\label{mcg11} \\
D_{higher,I,2} &=&\left(\rho_{D,higher,I,2}\right)^{\theta +1}\left( A-  \omega_{D,higher,I,2}\right),\label{mcg11-}\\
D_{higher,I,3} &=&\left(\rho_{D,higher,I,3}\right)^{\theta +1}\left( A-  \omega_{D,higher,I,3}\right),\label{mcg11--}\\
B_{higher,I,1} &=& \frac{[a^{3(A+1)}\rho_{D,higher,I,1}]^{\theta +1}}{1+A}\left( 1+\omega_{D,higher,I,1}\right).\label{mcg13}\\
B_{higher,I,2} &=& \frac{[a^{3(A+1)}\rho_{D,higher,I,2}]^{\theta +1}}{1+A}\left( 1+\omega_{D,higher,I,2}\right).\label{mcg13-}\\
B_{higher,I,3} &=& \frac{[a^{3(A+1)}\rho_{D,higher,I,3}]^{\theta +1}}{1+A}\left( 1+\omega_{D,higher,I,3}\right).\label{mcg13--}
\end{eqnarray}

In the limiting case of a flat Dark Dominated Universe, we obtain that the expressions of $D_{ Dark}$ and $B_{ Dark}$ are given, respectively, by
\begin{eqnarray}
D_{ Dark,higher} &=& \left\{ \frac{48\alpha^2}{\left[\beta \left( t+C_3  \right)\right]^2}\right\}^{\theta+1}\left( A+1 -\frac{\beta}{6\alpha}  \right),\label{mcg11-3} \\
B_{ Dark,higher} &=&\frac{3^{\theta+1}\beta}{6\alpha\left(1+A\right)} \left[\frac{\beta\left( t+C_3 \right)}{4\alpha}\right]^{\left[\frac{12\alpha(A+1)}{\beta}-2\right](\theta+1)}   \label{mcg13-3}
\end{eqnarray}
In the limiting case of $C_3=0$, we obtain:
\begin{eqnarray}
D_{ Dark,higher,lim} &=& \left\{ \frac{48\alpha^2}{\left(\beta  t\right)^2}\right\}^{\theta+1}\left( A+1 -\frac{\beta}{6\alpha}  \right),\label{mcg11-3} \\
B_{ Dark,higher,lim} &=&\frac{3^{\theta+1}\beta}{6\alpha\left(1+A\right)} \left(\frac{\beta t}{4\alpha}\right)^{\left[\frac{12\alpha(A+1)}{\beta}-2\right](\theta+1)}   \label{mcg13-3}
\end{eqnarray}

For the NHDE non-interacting model, we obtain:
\begin{eqnarray}
D_{NHDE} &=&\left(\rho_{D,NHDE}\right)^{\theta +1}\left( A-  \omega_{D,NHDE,1}\right),\label{mcg11} \\
B_{NHDE} &=& \frac{[a^{3(A+1)}\rho_{D,NHDE}]^{\theta +1}}{1+A}\left( 1+\omega_{D,NHDE}\right).\label{mcg13}
\end{eqnarray}
Instead for the  NHDE interacting model, we obtain:
\begin{eqnarray}
D_{NHDE,I} &=&\left(\rho_{D,NHDE,I,1}\right)^{\theta +1}\left( A-  \omega_{D,NHDE,I,1}\right),\label{mcg11} \\
B_{NHDE,I} &=& \frac{[a^{3(A+1)}\rho_{D,NHDE,I}]^{\theta +1}}{1+A}\left( 1+\omega_{D,NHDE,I}\right).\label{mcg13}
\end{eqnarray}
In the limiting case of a flat Dark Dominated Universe, we obtain that the expressions of $D_{ Dark}$ and $B_{ Dark}$ are given, respectively, by
\begin{eqnarray}
D_{ Dark, NHDE} &=& \left[  \frac{3 \lambda^2}{(\mu - 1)^2 (t+C) ^2}\right]^{\theta +1} \left( A +\frac{3\lambda-2\mu+2}{3\lambda} \right),\label{mcg11-3} \\
B_{ Dark,NHDE} &=& \left[\left( \frac{\mu - 1}{\lambda}\right) (t+C)  \right]^{\frac{\lambda (\theta+1)}{\mu - 1}}\left[  \frac{3 \lambda^2}{(\mu - 1)^2 (t+C) ^2}\right]^{\theta +1}\left[   \frac{2\mu-2}{3\lambda(1+A)} \right]  \label{mcg13-3}
\end{eqnarray}

In the limiting care corresponding to the Ricci scale, we find:
\begin{eqnarray}
D_{\rm Dark,NHDE,Ricci} &=&\left( A +\frac{1}{3} \right) \left[  \frac{3 }{ (t+C) ^2}\right]^{\theta +1}= \left(A + \frac{1}{3} \right)3^{\theta+1} (t + C)^{-2(\theta+1)} , \\
B_{\rm Dark,NHDE,Ricci} &=&\left( \frac{2 \cdot 3^{\theta}}{1+A}\right) (t + C)^{-(\theta+1)}.
\end{eqnarray}

In the limiting case of $C=0$, we obtain:
\begin{eqnarray}
D_{ Dark,lim} &=& \left[  \frac{3 \lambda^2}{(\mu - 1)^2 t ^2}\right]^{\theta +1} \left( A +\frac{3\lambda-2\mu+2}{3\lambda} \right),\label{mcg11-3} \\
B_{ Dark,lim} &=& \left[\left( \frac{\mu - 1}{\lambda}\right) t  \right]^{\frac{\lambda (\theta+1)}{\mu - 1}}\left[  \frac{3 \lambda^2}{(\mu - 1)^2 t ^2}\right]^{\theta +1}\left[   \frac{2\mu-2}{3\lambda(1+A)} \right]  \label{mcg13-3}
\end{eqnarray}
In the limiting case corresponding to the Ricci DE and $C=0$, we obtain:
\begin{eqnarray}
D_{\rm Dark,NHDE,Ricci,lim} &=&\left( A +\frac{1}{3} \right) \left(  \frac{3 }{ t ^2}\right)^{\theta +1}= \left(A + \frac{1}{3} \right)3^{\theta+1} t ^{-2(\theta+1)} , \\
B_{\rm Dark,NHDE,Ricci,lim} &=&\left( \frac{2 \cdot 3^{\theta}}{1+A}\right) t ^{-(\theta+1)}.
\end{eqnarray}
In the limiting case corresponding to the values found by Wang \& Xu, we can write:
\begin{eqnarray}
D_{Dark,NHDE,WX} &\approx& \left[ \frac{31.020}{(t+C)^2} \right]^{\theta+1} (A + 0.995), \\
B_{Dark,NHDE,WX} &\approx& -\frac{0.299 \, (-0.311)^{-3.216 (\theta+1)} \, 30.990^{\theta+1}}{1+A} \, (t+C)^{-5.216 (\theta+1)}
\end{eqnarray}

Moreover, using the general definition of EoS parameter $\omega_D$,  we can rewrite $ \dot{\phi}^2$ and $V\left(  \phi \right)$ in the following way:
\begin{eqnarray}
    \dot{\phi}^2 &=& \left( 1+ \omega_D \right)\rho_D, \label{gcg6mura1}\\
    V\left(  \phi \right) &=& \frac{1}{2}\left( 1 - \omega_D \right)\rho_D. \label{gcg7mura2}
\end{eqnarray}
For the DE model studied in this paper, we obtain the following expressions for  $\dot{\phi}^2$ and $ V\left(  \phi \right)$:
\begin{eqnarray}
    \dot{\phi}^2_{higher,1} &=& \rho_{D,higher,1}\left(1+\omega_{D,higher,1} \right)  \label{gcg6mura1model1}\\
    \dot{\phi}^2_{higher,2} &=& \rho_{D,higher,2}\left(1+\omega_{D,higher,2} \right)  \label{gcg6mura1model1-2}\\
    \dot{\phi}^2_{higher,3} &=& \rho_{D,higher,3}\left(1+\omega_{D,higher,3} \right)  \label{gcg6mura1mode1-3}\\
    V\left(  \phi \right)_{higher,1} &=& \frac{\rho_{D,higher,1}}{2}\left(1-\omega_{D,higher,1} \right)   \label{gcg7mura2model1}\\
    V\left(  \phi \right)_{higher,2} &=& \frac{\rho_{D,higher,2}}{2}\left(1-\omega_{D,higher,2} \right)   \label{gcg7mura2model1-2}\\
    V\left(  \phi \right)_{higher,3} &=& \frac{\rho_{D,higher,3}}{2}\left(1-\omega_{D,higher,3} \right)   \label{gcg7mura2model1-3}
\end{eqnarray}

We can obtain the evolutionary form of the GCG model integrating the expressions of $\dot{\phi}^2$ we obtained with respect to the scale factor $a\left(t\right)$
\begin{eqnarray}
	\phi\left(a\right)_{higher,1} - \phi\left(a_0\right)_{higher,1} &=& \int_{a_0}^{a}\sqrt{3\Omega_{D}} \times \left(1+\omega_{D,higher,1} \right)  \frac{da}{a}, \label{}\\
    \phi\left(a\right)_{higher,2} - \phi\left(a_0\right)_{higher,2} &=& \int_{a_0}^{a}\sqrt{3\Omega_{D}} \times \left(1+\omega_{D,higher,2} \right)  \frac{da}{a}, \label{}\\
    \phi\left(a\right)_{higher,3} - \phi\left(a_0\right)_{higher,3} &=& \int_{a_0}^{a}\sqrt{3\Omega_{D}} \times \left(1+\omega_{D,higher,3} \right)  \frac{da}{a}, \label{}
\end{eqnarray}

For the interacting case, we obtain:
\begin{eqnarray}
    \dot{\phi}^2_{higher,I,1} &=& \rho_{D,higher,I,1}\left(1+\omega_{D,higher,I,1} \right)  \label{gcg6mura1model1}\\
    \dot{\phi}^2_{higher,I,2} &=& \rho_{D,higher,I,2}\left(1+\omega_{D,higher,I,2} \right)  \label{gcg6mura1model1-2}\\
    \dot{\phi}^2_{higher,3,I} &=& \rho_{D,higher,I,3}\left(1+\omega_{D,higher,I,3} \right)  \label{gcg6mura1mode1-3}\\
    V\left(  \phi \right)_{higher,I,1} &=& \frac{\rho_{D,higher,I,1}}{2}\left(1-\omega_{D,higher,I,1} \right)   \label{gcg7mura2model1}\\
    V\left(  \phi \right)_{higher,I,2} &=& \frac{\rho_{D,higher,I,2}}{2}\left(1-\omega_{D,higher,I,2} \right)   \label{gcg7mura2model1-2}\\
    V\left(  \phi \right)_{higher,I,3} &=& \frac{\rho_{D,higher,I,3}}{2}\left(1-\omega_{D,higher,I,3} \right)   \label{gcg7mura2model1-3}
\end{eqnarray}
and the evolutionary form of the GCG model are given by:
\begin{eqnarray}
	\phi\left(a\right)_{higher,I,1} - \phi\left(a_0\right)_{higher,I,1} &=& \int_{a_0}^{a}\sqrt{3\Omega_{D}} \times \left(1+\omega_{D,higher,I,1} \right)  \frac{da}{a}, \label{}\\
    \phi\left(a\right)_{higher,I,2} - \phi\left(a_0\right)_{higher,I,2} &=& \int_{a_0}^{a}\sqrt{3\Omega_{D}} \times \left(1+\omega_{D,higher,I,2} \right)  \frac{da}{a}, \label{}\\
    \phi\left(a\right)_{higher,I,3} - \phi\left(a_0\right)_{higher,I,3} &=& \int_{a_0}^{a}\sqrt{3\Omega_{D}} \times \left(1+\omega_{D,higher,I,3} \right)  \frac{da}{a}, \label{}
\end{eqnarray}

In the limiting case for a flat Dark Dominated Universe, we obtain:
\begin{eqnarray}
\dot{\phi}_{Dark,higher}^2\left(t \right) &=&\frac{8\alpha}{\beta}\frac{1}{ \left( t+C_3  \right)^2},\label{dusan1} \\
V_{Dark,higher}\left(t\right)&=&\left(2-\frac{\beta}{6\alpha}   \right) \frac{24\alpha^2}{\left[\beta \left( t+C_3  \right)\right]^2}.\label{murano13gcg}
\end{eqnarray}
Integrating with respect to the time $t$ the expression of $\dot{\phi}_{Dark,higher}^2\left(t \right) $, we obtain:
\begin{eqnarray}
\phi_{Dark,higher}\left(t \right) &=&2\sqrt{\frac{2\alpha}{\beta}}\cdot \ln \left( t+C_3  \right)+constant,\label{dusan1}
\end{eqnarray}
In the limiting case of $C_3=0$, we obtain:
\begin{eqnarray}
\phi_{Dark,higher,lim}\left(t \right) &=&2\sqrt{\frac{2\alpha}{\beta}}\cdot \ln t+constant,\label{dusan1}\\
V_{Dark,higher,lim}\left(t\right)&=&\left(2-\frac{\beta}{6\alpha}   \right) \frac{24\alpha^2}{\left(\beta  t\right)^2}.\label{murano13gcg}
\end{eqnarray}

For the NHDE model, we obtain the following expressions for $\dot{\phi}^2$ and $ V\left(  \phi \right)$:
\begin{eqnarray}
    \dot{\phi}^2_{NHDE} &=& \rho_{D,NHDE}\left(1+\omega_{D,NHDE} \right)  \label{gcg6mura1model1}\\
    V\left(  \phi \right)_{NHDE} &=& \frac{\rho_{D,NHDE,1}}{2}\left(1-\omega_{D,NHDE} \right)   \label{gcg7mura2model1}
\end{eqnarray}
We can obtain the evolutionary form of the GCG model integrating the expressions of $\dot{\phi}^2$ we obtained with respect to the scale factor 
\begin{eqnarray}
	\phi\left(a\right)_{NHDE} - \phi\left(a_0\right)_{NHDE} &=& \int_{a_0}^{a}\sqrt{3\Omega_{D}} \times \left(1+\omega_{D,NHDE} \right)  \frac{da}{a}, \label{}
\end{eqnarray}
For the omteracting NHDE model, we obtain:
\begin{eqnarray}
    \dot{\phi}^2_{NHDE,I} &=& \rho_{D,NHDE,I}\left(1+\omega_{D,NHDE,I} \right)  \label{gcg6mura1model1}\\
    V\left(  \phi \right)_{NHDE,I} &=& \frac{\rho_{D,NHDE,I}}{2}\left(1-\omega_{D,NHDE,I} \right)   \label{gcg7mura2model1}
\end{eqnarray}
and the evolutionary form of the GCG model is given by:
\begin{eqnarray}
	\phi\left(a\right)_{NHDE,I} - \phi\left(a_0\right)_{NHDE,I} &=& \int_{a_0}^{a}\sqrt{3\Omega_{D}} \times \left(1+\omega_{D,NHDE,I} \right)  \frac{da}{a}, \label{}
\end{eqnarray}

In the limiting case for a flat Dark Dominated Universe, we obtain:
\begin{eqnarray}
\dot{\phi}_{Dark,NHDE}^2\left(t \right) &=& \left(\frac{ 2\lambda}{\mu - 1}\right) \frac{1}{(t+C)^2},\label{dusan1} \\
V_{Dark,NHDE}\left(t\right)&=&\frac{ \lambda (3\lambda-\mu+1)}{(\mu - 1)^2 }\cdot \frac{1}{(t + C)^2}  \label{murano13gcg}
\end{eqnarray}
Integrating with respect to the time $t$ the expression of $\dot{\phi}_{Dark,NHDE}^2\left(t \right) $, we obtain:
\begin{eqnarray}
\phi_{Dark,NHDE}\left(t \right) &=&\sqrt{ \frac{ 2\lambda}{\mu - 1}}\cdot \ln (t+C) + const,\label{dusan1}
\end{eqnarray}
In the limiting case corresponding to $\lambda =1$ e $\mu=2$, i.e. the case corresponding to the Ricci scale, we obtain:
\begin{eqnarray}
\phi_{\rm Dark,NHDE,Ricci}(t) &=& \sqrt{2} \, \ln(t + C) + \text{const}, \\
V_{\rm Dark,NHDE,Ricci}(t) &=& \frac{2}{(t + C)^2}.
\end{eqnarray}

In the limiting case of $C=0$, we obtain:
\begin{eqnarray}
\phi_{Dark,NHDE,lim}\left(t \right) &=&\sqrt{ \frac{ 2\lambda}{\mu - 1}}\cdot \ln t + const,\label{dusan1} \\
V_{Dark,NHDE,lim}\left(t\right)&=& \frac{ \lambda (3\lambda-\mu+1)}{(\mu - 1)^2 }\cdot \frac{1}{t^2}  \label{murano13gcg}
\end{eqnarray}

In the limiting case corresponding to $\lambda =1$ e $\mu=2$, i.e. the case corresponding to the Ricci scale, we obtain:
\begin{eqnarray}
\phi_{\rm Dark,NHDE,Ricci,lim}(t) &=& \sqrt{2} \, \ln t  + \text{const}, \\
V_{\rm Dark,NHDE,Ricci,lim}(t) &=& \frac{2}{t ^2}.
\end{eqnarray}

We can easily derive the following relation between $\phi_{Dark,NHDE,Ricci,lim}$ and $V_{Dark,NHDE,Ricci,lim}$:
\begin{eqnarray}
 V(\phi) = 2 \, e^{ -\sqrt{2} \phi } 
\end{eqnarray}

In the limiting case corresponding to the values obtained by Wang \& Xu, the scalar field $\phi_{Dark}$ becomes complex, while the potential reads
\begin{eqnarray}
V_{Dark,NHDE,WX}(t) &\approx& \frac{34.22}{t^2}.
\end{eqnarray}
We find that $\phi_{Dark,NHDE}$ is not a real function.

\subsection{Modified Variable Chaplygin Gas (MVCG) Model}
We now consider the third model of this paper, i.e. the Modified Variable Chaplygin Gas (MVCG) model. Guo \& Zhang \cite{mvcg1} recently introduced a model known as the Variable Chaplygin Gas (VCG) with the following equation of state:
\begin{eqnarray}
p_D = - \frac{B}{\rho_D}, \label{murano16}
\end{eqnarray}
where \(B\) is a function of the scale factor \(a(t)\), i.e. \(B = B(a(t))\). This assumption appears reasonable since it is related to the scalar potential if the Chaplygin Gas is interpreted through a Born-Infeld scalar field \cite{mvcg2}. In the following, for simplicity, we omit the explicit time dependence of the scale factor. The VCG model has been studied recently in \cite{mvcg3,mvcg4}. 

Debnath \cite{mvcg5} proposed the equation of state of the Modified Variable Chaplygin Gas (MVCG) model in the form
\begin{eqnarray}
p_D = A \rho_D - \frac{B(a)}{\rho_D^{\theta}}, \label{murano17}
\end{eqnarray}
where in this work we choose \(B(a) = B_0 a^{-\delta_1}\). Therefore, the pressure \(p_D\) of the MVCG model can be written as
\begin{eqnarray}
p_D = A \rho_D - \frac{B_0 a^{-\delta_1}}{\rho_D^{\theta}}. \label{murano18}
\end{eqnarray}
Here, \(A\), \(B_0\), and \(\delta_1\) are three positive constant parameters, with \(B_0\) representing the present-day value of \(B\), and \(\delta_1\) the exponent of the scale factor. Moreover, \(\theta\) is usually taken in the range \(0 \leq \theta \leq 1\).

In the limiting case corresponding to \(B_0 = 0\), Eq. (\ref{murano18}) reduces to a barotropic equation of state (EoS), or equivalently, to a barotropic fluid. In general, the barotropic EoS \(p = A \rho\) can describe different types of media. For example, the limiting case with \(A = -1\) (i.e. \(p = -\rho\)) corresponds to the Cosmological Constant; the limiting case with \(A = -\frac{2}{3}\) describes domain walls; \(A = -\frac{1}{3}\) corresponds to cosmic strings; \(A = 0\) represents dust or matter; when \(A = \frac{1}{3}\), it describes a relativistic gas; the case \(A = \frac{2}{3}\) gives a perfect gas; and finally, \(A = 1\) corresponds to ultra-stiff matter. 

If we consider \(B\) as a constant, i.e. \(B = B_0\) in Eq. (\ref{murano18}) (which corresponds to the limiting case \(\delta_1 = 0\)), we recover the EoS of the original Modified Chaplygin Gas (MCG) model. Eq. (\ref{murano18}) shows that the MVCG scenario interpolates between a radiation-dominated phase \(\left(A = \frac{1}{3}\right)\) and a quintessence-dominated phase described by a constant EoS. In the limiting case \(A = 0\) and \(\alpha = 1\), the usual Chaplygin Gas is recovered. 

Recently, it was shown using the latest Supernovae data that models with \(\alpha > 1\) are also viable \cite{mvcg8}. It is also worth emphasizing that, in the limiting case \(A = 0\), Eq. (\ref{murano18}) describes a fluid with negative pressure, which is generally characteristic of the quintessence regime.

This modified form of the Chaplygin Gas also has a phenomenological motivation, as it can explain the flat rotation curves of galaxies \cite{mvcg9}. The galactic rotational velocity \(V_c\) is related to the MVCG parameter \(A\) through the relation \(V_c = \sqrt{2A}\), while the density parameter \(\rho\) is related to the radial size of the galaxy \(r\) by
\begin{eqnarray}
\rho = \frac{A}{2 \pi G r^2}.
\end{eqnarray}
At high densities, the first term of the MVCG model dominates and produces the flat rotation curve consistent with current observations. The parameter \(A\) varies from galaxy to galaxy due to the variations in \(V_c\).

The  density \(\rho_D\) of the MVCG model is given by the following general relation:
\begin{eqnarray}
\rho_D = \left\{ \frac{3(\theta + 1) B_0}{3(\theta + 1)(A + 1) - \delta_1} \frac{1}{a^{\delta_1}} - \frac{C}{a^{3(\theta + 1)(A + 1)}} \right\}^{\frac{1}{1 + \theta}}, \label{murano19}
\end{eqnarray}
where \(C\) is a positive integration constant, and the condition
\[
3(\theta + 1)(A + 1) > \delta_1
\]
ensures that the first term inside the braces is positive. The parameter \(\delta_1\) must be positive; otherwise, the scale factor \(a\) would tend to infinity, which would imply that the energy density \(\rho_D\) also diverges, contradicting the behavior expected for an expanding Universe.

We can now reconstruct the expressions for both the potential and the dynamics of the scalar field model. To this end, we consider a time-dependent scalar field \(\phi(t)\) with potential \(V(\phi)\), which are directly related to the energy density and pressure of the MVCG model as follows:
\begin{eqnarray}
\rho_D &=& \frac{1}{2} \dot{\phi}^2 + V(\phi), \label{murano20} \\
p_D &=& \frac{1}{2} \dot{\phi}^2 - V(\phi). \label{murano21}
\end{eqnarray}
Since we have that the kinetic term is positive, the MVCG corresponds to a quintessence-type scalar field.

It is known that the deceleration parameter \(q\) can be expressed as:
\begin{eqnarray}
q = -\frac{\ddot{a}}{a H^2}. \label{murano22}
\end{eqnarray}
In order for the Universe to accelerate, the deceleration parameter must be negative, i.e. \(\ddot{a} > 0\), since the scale factor \(a\) is positive and \(H^2\) is always positive. The condition \(\ddot{a} > 0\) implies the following inequality:
\begin{eqnarray}
\left[\frac{2(1+\theta) - \delta_1}{3(1+\theta)(1+A) - \delta_1}\right] a^{3(1+\theta)(1+A) - \delta_1} > \frac{C(1+3A)}{3 B_0}. \label{murano23}
\end{eqnarray}
This result requires that \(\delta_1 < 2(1+\theta)\). Since \(0 \leq \theta \leq 1\), it follows that \(\delta_1\) must lie in the range \(0 < \delta_1 < 4\).

This expression shows that for small values of the scale factor, the Universe is decelerating, while for large values of the scale factor, it is accelerating. The transition occurs at the scale factor value
\begin{eqnarray}
a = \left\{ \frac{C(1+3A) \left[3(1+\theta)(1+A) - \delta_1 \right]}{3 B_0 \left[2(1+\theta) - \delta_1 \right]} \right\}^{\frac{1}{3(1+\theta)(1+A) - \delta_1}}. \label{murano24}
\end{eqnarray}

For small values of the scale factor \(a(t)\), the energy density \(\rho\) approximately satisfies
\begin{eqnarray}
\rho \cong \frac{C^{\frac{1}{1+\theta}}}{a^{3(1+A)}}, \label{murano25}
\end{eqnarray}
which corresponds to a Universe dominated by a fluid with equation of state \(p = A \rho\).

For large values of the scale factor $a$, the density \(\rho\) approximately satisfies the relation:
\begin{eqnarray}
\rho \cong \left[ \frac{3(1 + \theta) B_0}{3(1 + \theta)(1 + A) - \delta_1} \right]^{\frac{1}{1+\theta}} 
a^{-\frac{\delta_1}{1+\theta}}, \label{murano26}
\end{eqnarray}
which corresponds to the equation of state:
\begin{eqnarray}
p = \left[ \frac{\delta_1}{3(1 + \theta)} - 1 \right] \rho, \label{murano27}
\end{eqnarray}
describing a quintessence model \cite{mvcg6}.

In the limiting case \(\delta_1 = 0\), Eq. \eqref{murano27} reduces to the original modified Chaplygin gas scenario. More generally, Eq. \eqref{murano27} shows that in the variable modified Chaplygin gas model, the cosmic evolution interpolates between a radiation-dominated phase (corresponding to \(A = \frac{1}{3}\)) and a quintessence-dominated phase characterized by the constant equation of state parameter \(\gamma = -1 + \frac{\delta_1}{3(1 + \theta)} < -\frac{1}{3}\).

We must also require that the density given in Eq. \eqref{murano19} remains positive, which implies that the scale factor \(a(t)\) must satisfy the following condition:
\begin{eqnarray}
a\left( t \right) > \left\{ -\frac{C\left[3\left( \alpha +1 \right)\left(A+1\right)-\delta_1 \right]}{3\left( \alpha +1 \right)B_0}  \right\}^{\frac{1}{3\left( \alpha +1 \right)\left(A+1\right)-\delta_1}}. \label{murano28}
\end{eqnarray}

Therefore, the minimum value of the scale factor $a\left( t\right)$ is given by the following expression:
\begin{eqnarray}
a_{min}\left( t\right) = \left\{ -\frac{C\left[3\left( \theta +1 \right)\left(A+1\right)-\delta_1 \right]}{3\left( \theta+1 \right)B_0}  \right\}^{\frac{1}{3\left( \theta +1 \right)\left(A+1\right)-\delta_1}}. \label{murano29}
\end{eqnarray}
Adding the results of  Eqs. (\ref{murano20}) and (\ref{murano21}), we easily derive the kinetic energy $\dot{\phi}^2$  term as follows:
\begin{eqnarray}
    \dot{\phi}^2 &=& \rho_D + p_D. \label{murano30}
\end{eqnarray}
Alternatively, subtracting Eqs. \eqref{murano21} from \eqref{murano20}, we can straightforwardly derive the scalar potential \( V(\phi) \) as follows:
\begin{eqnarray}
 V\left(  \phi \right) &=& \frac{\left( \rho_D - p_D\right)}{2}. \label{murano31}
\end{eqnarray}
Inserting in Eq. (\ref{murano30}) the expression of $p_D$ and $\rho_D$ given, respectively, in Eqs. (\ref{murano18}) and (\ref{murano19}), we obtain the following expression for the kinetic term $\dot{\phi}^2$:
\begin{eqnarray}
\dot{\phi}^2 &=& \left( 1+A\right) \left\{\frac{3\left( \theta +1 \right)B_0}{\left[3\left( \theta +1 \right)\left(A+1\right)-\delta_1 \right]}\left(\frac{1}{a^{\delta_1}}\right)  - \frac{C}{a^{3\left( \theta +1 \right)\left(A+1\right)}} \right\}^{\frac{1}{1+\theta}}\nonumber\\
 &-& \frac{B_0 a^{-\delta_1}}{\left\{ \frac{3\left( \theta +1 \right)B_0}{\left[3\left( \theta +1 \right)\left(A+1\right)-\delta_1 \right]}\left(\frac{1}{a^{\delta_1}}\right)  - \frac{C}{a^{3\left( \theta +1 \right)\left(A+1\right)}}  \right\}^{\frac{\theta}{1+\theta}}}.\label{murano32}
\end{eqnarray}
Using the general relation $\dot{\phi} = H \phi'$, we obtain:
\begin{eqnarray}
\phi &=& \int_{t_0}^t  \left( 1+A\right)^{1/2}  \left\{\frac{3\left( \theta +1 \right)B_0}{\left[3\left( \theta +1 \right)\left(A+1\right)-\delta_1 \right]}\left(\frac{1}{a^{\delta_1}}\right)  - \frac{C}{a^{3\left( \theta +1 \right)\left(A+1\right)}} \right\}^{\frac{1}{2\left(1+\theta\right)}}dt \nonumber \\
&-&\int_{t_0}^t  \frac{\left(B_0 a^{-\delta_1}\right)^{1/2}}{\left\{ \frac{3\left( \theta +1 \right)B_0}{\left[3\left( \theta +1 \right)\left(A+1\right)-\delta_1 \right]}\left(\frac{1}{a^{\delta_1}}\right)  - \frac{C}{a^{3\left( \theta +1 \right)\left(A+1\right)}}  \right\}^{\frac{\theta}{2\left(1+\theta\right)}}}dt.   \label{murano33}
\end{eqnarray}
From the Friedmann equation given in Eq.~(\ref{7}), for \( k = 0 \) (i.e., a flat Universe), we obtain the explicit expression of the cosmic time \( t \) as a function of the scale factor \( a(t) \) as follows:
\begin{eqnarray}
t=Ka^{\frac{\delta_1}{2\left(1+\theta   \right)}}\,_2F_1\left[ \frac{1}{2\left(1+\theta \right)}, -z,1-z,-\left(\frac{C}{K}\right)a^{-\frac{\delta_1}{2\left(1+\theta   \right)z}} \right],\label{murano34}
\end{eqnarray}
where the two quantities $K$ and $z$ are defined as follows:
\begin{eqnarray}
K&=&\frac{2}{\delta_1}\left[\left( 1+\theta \right)^{\theta} \sqrt{\frac{\delta_1}{6B_0z}}   \right]^{\frac{1}{1+\theta}}, \label{murano35}\\
z &=& \frac{\delta_1}{2\left( 1+\theta \right)\left[3\left( 1+A  \right)\left( 1+\theta \right)-\delta_1   \right]}.\label{murano36}
\end{eqnarray}
Moreover, the term $_2F_1$ represents the hypergeomtric function of second type.

For a flat Universe, considering the expressions of $t$ expressed in Eq. (\ref{murano34}), we obtain the following expression of $\phi$:
\begin{eqnarray}
\phi &=& \frac{\sqrt{1+A}}{3\left( 1+A  \right)\left( 1+\theta \right)-\delta_1  }\times \nonumber \\
&&\left\{2\log \left( \sqrt{u+x} + \sqrt{u+y}  \right) - \sqrt{\frac{y}{x}}\log \left[ \frac{\left( \sqrt{x\left(u+x\right)} + \sqrt{y\left(u+y\right)}  \right)^2}{x^{3/2}\sqrt{y}u}  \right]     \right\},\label{murano37}
\end{eqnarray}
where the three parameters $x$, $y$ and $u$ are defined, respectively, in the following way:
\begin{eqnarray}
x &=& \frac{\delta_1}{1+A}, \label{murano38} \\
y &=&3\left(1+\theta \right), \label{murano39}\\
u &=& \left(\frac{\delta_1 C}{B_0}\right)a^{\delta_1\left( 1-\frac{y}{x}  \right)}.\label{murano40}
\end{eqnarray}

Moreover, inserting in Eq. (\ref{murano31}) the expression of $p_D$ and $\rho_D$ given, respectively, in Eqs. (\ref{murano18}) and (\ref{murano19}), we obtain the following expression for $V\left(\phi\right)$:
\begin{eqnarray}
V\left(\phi\right) &=& \frac{\left( 1-A\right)}{2}\left\{ \frac{3\left( \theta +1 \right)B_0}{\left[3\left( \theta +1 \right)\left(A+1\right)-\delta_1 \right]}\left(\frac{1}{a^{\delta_1}}\right)  - \frac{C}{a^{3\left( \theta +1 \right)\left(A+1\right)}}  \right\}^{\frac{1}{1+\theta}} \nonumber\\
 && +\frac{B_0 a^{-\delta_1}}{2\left\{  \frac{3\left( \theta +1 \right)B_0}{\left[3\left( \theta +1 \right)\left(A+1\right)-\delta_1 \right]}\left(\frac{1}{a^{\delta_1}}\right)  - \frac{C}{a^{3\left( \theta +1 \right)\left(A+1\right)}}  \right\}^{\frac{\theta}{1+\theta}}}.\label{murano40}
\end{eqnarray}
We now aim to derive the general expressions for the two parameters \( B_0 \) and \( C \).
.
Dividing the expression of \( p_D \) written in Eq. (\ref{murano18}) by the density \( \rho_D \), we easily derive the following relation for the equation of state parameter \( \omega_D \):
\begin{eqnarray}
\omega_D = A - \frac{B_0 a^{-\delta_1}}{\rho_D^{\theta + 1}}, \label{murano41}
\end{eqnarray}
which leads to the following expression for \( B_0 \):
\begin{eqnarray}
B_0 = a^{\delta_1} \left( A - \omega_D \right) \rho_D^{\theta + 1}. \label{murano42}
\end{eqnarray}
On the other hand, from the expression of \( \rho_D \) given in Eq. (\ref{murano19}), we find that
\begin{eqnarray}
C = \left\{ \frac{3(\theta + 1) B_0}{3(\theta + 1)(A + 1) - \delta_1} \frac{1}{a^{\delta_1}} - \rho_D^{1 + \theta} \right\} a^{-3(\theta + 1)(A + 1)}. \label{murano43}
\end{eqnarray}
Substituting the expression of \( B_0 \) from Eq. (\ref{murano42}) into Eq. (\ref{murano43}), we obtain the following expression for \( C \):
\begin{eqnarray}
C = \left[ \rho_D a^{-3(A + 1)} \right]^{\theta + 1} \left[ \frac{3(\theta + 1)(A - \omega_D)}{3(\theta + 1)(A + 1) - \delta_1} - 1 \right]. \label{murano44}
\end{eqnarray}

For the DE model we are considering in this paper, we derive the following expressions for  $B_0$ and $C$:
\begin{eqnarray}
B_{0-higher,1} &=&a^{\delta_1}\left[ A-\omega_{D,higher,1} \right]  \rho_{D,higher,1} ^{\theta +1},\label{murano45}\\
B_{0-higher,2} &=&a^{\delta_1}\left[ A-\omega_{D,higher,2} \right]  \rho_{D,higher,2} ^{\theta +1},\label{murano452}\\
B_{0-higher,3} &=&a^{\delta_1}\left[ A-\omega_{D,higher,3} \right]  \rho_{D,higher,3} ^{\theta +1},\label{murano453}\\
C_{higher,1}&=& \left[\rho_{D,higher,1}  a^{-3\left(A+1\right)}\right]^{\theta +1}\left\{\frac{3\left(\theta+1 \right)\left[A  -\omega_{D,higher,1}  \right]}{3\left(\theta+1 \right)\left( A+1  \right)-\delta_1}-1     \right\} \label{murano46}\\
C_{higher,2}&=& \left[\rho_{D,higher,2}  a^{-3\left(A+1\right)}\right]^{\theta +1}\left\{\frac{3\left(\theta+1 \right)\left[A  -\omega_{D,higher,2}  \right]}{3\left(\theta+1 \right)\left( A+1  \right)-\delta_1}-1     \right\} \label{murano46-2}\\
C_{higher,3}&=& \left[\rho_{D,higher,3}  a^{-3\left(A+1\right)}\right]^{\theta +1}\left\{\frac{3\left(\theta+1 \right)\left[A  -\omega_{D,higher,3}  \right]}{3\left(\theta+1 \right)\left( A+1  \right)-\delta_1}-1     \right\} \label{murano463}
\end{eqnarray}

For the interacting case, we obtain:
\begin{eqnarray}
B_{0-higher,I,1} &=&a^{\delta_1}\left[ A-\omega_{D,higher,I,1} \right]  \rho_{D,higher,I,1} ^{\theta +1},\label{murano45}\\
B_{0-higher,I,2} &=&a^{\delta_1}\left[ A-\omega_{D,higher,I,2} \right]  \rho_{D,higher,I,2} ^{\theta +1},\label{murano452}\\
B_{0-higher,I,3} &=&a^{\delta_1}\left[ A-\omega_{D,higher,I,3} \right]  \rho_{D,higher,I,3} ^{\theta +1},\label{murano453}\\
C_{higher,I,1}&=& \left[\rho_{D,higher,I,1}  a^{-3\left(A+1\right)}\right]^{\theta +1}\left\{\frac{3\left(\theta+1 \right)\left[A  -\omega_{D,higher,I,1}  \right]}{3\left(\theta+1 \right)\left( A+1  \right)-\delta_1}-1     \right\} \label{murano46}\\
C_{higher,I,2}&=& \left[\rho_{D,higher,I,2}  a^{-3\left(A+1\right)}\right]^{\theta +1}\left\{\frac{3\left(\theta+1 \right)\left[A  -\omega_{D,higher,I,2}  \right]}{3\left(\theta+1 \right)\left( A+1  \right)-\delta_1}-1     \right\} \label{murano46-2}\\
C_{higher,I,3}&=& \left[\rho_{D,higher,I,3}  a^{-3\left(A+1\right)}\right]^{\theta +1}\left\{\frac{3\left(\theta+1 \right)\left[A  -\omega_{D,higher,I,3}  \right]}{3\left(\theta+1 \right)\left( A+1  \right)-\delta_1}-1     \right\} \label{murano463}
\end{eqnarray}

In the limiting case of a flat Dark Dominated Universe, we obtain the following expressions of $B_{0,Dark}$ and $C_{Dark}$:
\begin{eqnarray}
B_{0, Dark,higher} &=&  3^{\theta+1} \left[\frac{\beta\left( t+C_3 \right)}{4\alpha}\right]^{\frac{4\alpha \delta_1}{\beta}-2(\theta+1)}\left( A+1- \frac{\beta}{6\alpha}\right),\label{murano49}\\
C_{Dark,higher}&=& 3^{\theta+1} \left[\frac{\beta\left( t+C_3 \right)}{4\alpha}\right]^{\frac{-12\alpha(A+1)(\theta+1)-2}{\beta}}\times \nonumber \\
&&\left\{\frac{3\left(\theta+1 \right)\left[A  +1-\beta/(6\alpha)  \right]}{3\left(\theta+1 \right)\left( A+1  \right)-\delta_1}-1     \right\},
 \label{murano50}
\end{eqnarray}
In the limiting case of $C_3=0$, we can write:
\begin{eqnarray}
B_{0, Dark,higher,lim} &=&  3^{\theta+1} \left[\frac{\beta t}{4\alpha}\right]^{\frac{4\alpha \delta_1}{\beta}-2(\theta+1)}\left( A+1- \frac{\beta}{6\alpha}\right),\label{murano49}\\
C_{Dark,higher,lim}&=& 3^{\theta+1} \left[\frac{\beta t}{4\alpha}\right]^{\frac{-12\alpha(A+1)(\theta+1)-2}{\beta}}\times \nonumber \\
&&\left\{\frac{3\left(\theta+1 \right)\left[A  +1-\beta/(6\alpha)  \right]}{3\left(\theta+1 \right)\left( A+1  \right)-\delta_1}-1     \right\},
 \label{murano50}
\end{eqnarray}

For the non interacting NHDE model, we obtain the following expressions of $B_{0}$ and $C$:
\begin{eqnarray}
B_{0-NHDE} &=&a^{\delta_1}\left[ A-\omega_{D,NHDE} \right]  \rho_{D,NHDE} ^{\theta +1},\label{murano45}\\
C_{NHDE}&=& \left[\rho_{D,NHDE}  a^{-3\left(A+1\right)}\right]^{\theta +1}\left\{\frac{3\left(\theta+1 \right)\left[A  -\omega_{D,NHDE}  \right]}{3\left(\theta+1 \right)\left( A+1  \right)-\delta_1}-1     \right\} \label{murano46}
\end{eqnarray}

Instead, for the interacting NHDE model, we obtain we obtain the following expressions of $B_{0}$ and $C$:
\begin{eqnarray}
B_{0-NHDE,I} &=&a^{\delta_1}\left[ A-\omega_{D,NHDE,I} \right]  \rho_{D,NHDE,I} ^{\theta +1},\label{murano45}\\
C_{NHDE,I}&=& \left[\rho_{D,NHDE,I}  a^{-3\left(A+1\right)}\right]^{\theta +1}\left\{\frac{3\left(\theta+1 \right)\left[A  -\omega_{D,NHDE,I}  \right]}{3\left(\theta+1 \right)\left( A+1  \right)-\delta_1}-1     \right\} \label{murano46}
\end{eqnarray}
In the limiting case of a flat Dark Dominated Universe, we obtain the following expressions of $B_{0,Dark}$ and $C_{Dark}$:
\begin{eqnarray}
B_{0, Dark,NHDE} &=& \left(A + \frac{3\lambda-2\mu+2}{3\lambda}\right) \left[\left( \frac{\mu - 1}{\lambda}\right) (t+C)  \right]^{\frac{\lambda \delta_1}{\mu - 1}}\left[  \frac{3 \lambda^2}{(\mu - 1)^2 (t +C)^2} \right]^{\theta+1} \nonumber \\
&=& 3^{\theta+1} \left(A + \frac{3\lambda - 2\mu + 2}{3\lambda}\right)
\left[\left(\frac{\mu - 1}{\lambda}\right) (t + C)\right]^{\frac{\lambda \delta_1}{\mu - 1} - 2(\theta + 1)} \label{murano49}\\\
C_{Dark,NHDE}&=&  \left\{\frac{3 \lambda^2}{(\mu - 1)^2 (t + C)^2} \left[ \left(\frac{\mu - 1}{\lambda}\right) (t + C)\right]^{\frac{-3\lambda (A+1)}{\mu - 1}}\right\}^{\theta +1}\times \nonumber \\
&&\left\{\frac{3\left(\theta+1 \right)\left[A  -\frac{3\lambda-2\mu+2}{3\lambda}  \right]}{3\left(\theta+1 \right)\left( A+1  \right)-\delta_1}-1     \right\}, \label{murano50}
\end{eqnarray}
In the limiting case corresponding to the Ricci scale, we obtain:
\begin{eqnarray}
B_{0, Dark,NHDE, Ricci}(t) &=& 3^{\theta}\left(3A+1\right) \,  (t + C)^{\delta_1 - 2(\theta+1)}, \\
C_{Dark,NHDE,Ricci}(t) &=& 
\left[ \frac{(\theta+1)(3A + 1)}{3(\theta+1)(A+1) - \delta_1} - 1 \right]3^{\theta+1} (t + C)^{-(3A+5)(\theta+1)} .
\end{eqnarray}

In the limiting case of $C=0$, we obtain:
\begin{eqnarray}
B_{0, Dark, HNDE,lim} &=& \left(A + \frac{3\lambda-2\mu+2}{3\lambda}\right) \left[\left( \frac{\mu - 1}{\lambda}\right) t  \right]^{\frac{\lambda \delta_1}{\mu - 1}}\left[  \frac{3 \lambda^2}{(\mu - 1)^2 t^2} \right]^{\theta+1} \nonumber \\
&=& 3^{\theta+1} \left(A + \frac{3\lambda - 2\mu + 2}{3\lambda}\right)
\left[\left(\frac{\mu - 1}{\lambda}\right) t \right]^{\frac{\lambda \delta_1}{\mu - 1} - 2(\theta + 1)},\label{murano49}\\
C_{Dark,HNDE,lim}&=& \left\{\frac{3 \lambda^2}{(\mu - 1)^2 t ^2} \left[ \left(\frac{\mu - 1}{\lambda}\right) t \right]^{\frac{-3\lambda (A+1)}{\mu - 1}}\right\}^{\theta +1}\times \nonumber \\
&&\left\{\frac{3\left(\theta+1 \right)\left[A  -\frac{3\lambda-2\mu+2}{3\lambda}  \right]}{3\left(\theta+1 \right)\left( A+1  \right)-\delta_1}-1     \right\},\label{murano50}
\end{eqnarray}
In the limiting case corresponding to the Ricci scale and $C=0$, we obtain:
\begin{eqnarray}
B_{0, Dark,NHDE, Ricci,lim}(t) &=&  3^{\theta}\left( 3A+1\right) \, t ^{\delta_1 - 2(\theta+1)}, \\
C_{Dark,NHDE,Ricci,lim}(t) &=& 
\left[ \frac{(\theta+1)(3A + 1)}{3(\theta+1)(A+1) - \delta_1} - 1 \right]3^{\theta+1} t ^{-(3A+5)(\theta+1)} .
\end{eqnarray}

In the limiting case corresponding to the values found by Wang \& Xu, we obtain:
\begin{eqnarray}
B_{0,Dark,NHDE,WX} &=& 3^{\theta+1}\,(A+1.207)\,
\left[-0.311\,(t+C)\right]^{-3.216\,\delta_1 - 2(\theta+1)}, \\
C_{Dark,NHDE,WX} &=& 
\left\{\frac{31.020}{(t+C)^2}\,\Big[-0.311\,(t+C)\Big]^{9.644\,(A+1)}\right\}^{\theta+1}\times\nonumber \\
&&\left\{\frac{3(\theta+1)(A-1.207)}{3(\theta+1)(A+1)-\delta_1}-1\right\}.
\end{eqnarray}

Moreover, using the general definition of EoS parameter $\omega_D$,  we can rewrite $ \dot{\phi}^2$ and $V\left(  \phi \right)$ in the following way:
\begin{eqnarray}
    \dot{\phi}^2 &=& \left( 1+ \omega_D \right)\rho_D, \label{gcg6mura1}\\
    V\left(  \phi \right) &=& \frac{1}{2}\left( 1 - \omega_D \right)\rho_D. \label{gcg7mura2}
\end{eqnarray}
For the DE model studied in this paper, we obtain the following expressions for  $\dot{\phi}^2$ and $ V\left(  \phi \right)$:
\begin{eqnarray}
    \dot{\phi}^2_{higher,1} &=& \rho_{D,higher,1}\left(1+\omega_{D,higher,1} \right)  \label{gcg6mura1model1}\\
    \dot{\phi}^2_{higher,2} &=& \rho_{D,higher,2}\left(1+\omega_{D,higher,2} \right)  \label{gcg6mura1model1-2}\\
    \dot{\phi}^2_{higher,3} &=& \rho_{D,higher,3}\left(1+\omega_{D,higher,3} \right)  \label{gcg6mura1mode1-3}\\
    V\left(  \phi \right)_{higher,1} &=& \frac{\rho_{D,higher,1}}{2}\left(1-\omega_{D,higher,1} \right)   \label{gcg7mura2model1}\\
    V\left(  \phi \right)_{higher,2} &=& \frac{\rho_{D,higher,2}}{2}\left(1-\omega_{D,higher,2} \right)   \label{gcg7mura2model1-2}\\
    V\left(  \phi \right)_{higher,3} &=& \frac{\rho_{D,higher,3}}{2}\left(1-\omega_{D,higher,3} \right)   \label{gcg7mura2model1-3}
\end{eqnarray}

We can obtain the evolutionary form of the GCG model integrating the expressions of $\dot{\phi}^2$ we obtained with respect to the scale factor $a\left(t\right)$
\begin{eqnarray}
	\phi\left(a\right)_{higher,1} - \phi\left(a_0\right)_{higher,1} &=& \int_{a_0}^{a}\sqrt{3\Omega_{D}} \times \left(1+\omega_{D,higher,1} \right)  \frac{da}{a}, \label{}\\
    \phi\left(a\right)_{higher,2} - \phi\left(a_0\right)_{higher,2} &=& \int_{a_0}^{a}\sqrt{3\Omega_{D}} \times \left(1+\omega_{D,higher,2} \right)  \frac{da}{a}, \label{}\\
    \phi\left(a\right)_{higher,3} - \phi\left(a_0\right)_{higher,3} &=& \int_{a_0}^{a}\sqrt{3\Omega_{D}} \times \left(1+\omega_{D,higher,3} \right)  \frac{da}{a}, \label{}
\end{eqnarray}

For the interacting case, we obtain:
\begin{eqnarray}
    \dot{\phi}^2_{higher,I,1} &=& \rho_{D,higher,I,1}\left(1+\omega_{D,higher,I,1} \right)  \label{gcg6mura1model1}\\
    \dot{\phi}^2_{higher,I,2} &=& \rho_{D,higher,I,2}\left(1+\omega_{D,higher,I,2} \right)  \label{gcg6mura1model1-2}\\
    \dot{\phi}^2_{higher,3,I} &=& \rho_{D,higher,I,3}\left(1+\omega_{D,higher,I,3} \right)  \label{gcg6mura1mode1-3}\\
    V\left(  \phi \right)_{higher,I,1} &=& \frac{\rho_{D,higher,I,1}}{2}\left(1-\omega_{D,higher,I,1} \right)   \label{gcg7mura2model1}\\
    V\left(  \phi \right)_{higher,I,2} &=& \frac{\rho_{D,higher,I,2}}{2}\left(1-\omega_{D,higher,I,2} \right)   \label{gcg7mura2model1-2}\\
    V\left(  \phi \right)_{higher,I,3} &=& \frac{\rho_{D,higher,I,3}}{2}\left(1-\omega_{D,higher,I,3} \right)   \label{gcg7mura2model1-3}
\end{eqnarray}
and the evolutionary form of the GCG model are given by:
\begin{eqnarray}
	\phi\left(a\right)_{higher,I,1} - \phi\left(a_0\right)_{higher,I,1} &=& \int_{a_0}^{a}\sqrt{3\Omega_{D}} \times \left(1+\omega_{D,higher,I,1} \right)  \frac{da}{a}, \label{}\\
    \phi\left(a\right)_{higher,I,2} - \phi\left(a_0\right)_{higher,I,2} &=& \int_{a_0}^{a}\sqrt{3\Omega_{D}} \times \left(1+\omega_{D,higher,I,2} \right)  \frac{da}{a}, \label{}\\
    \phi\left(a\right)_{higher,I,3} - \phi\left(a_0\right)_{higher,I,3} &=& \int_{a_0}^{a}\sqrt{3\Omega_{D}} \times \left(1+\omega_{D,higher,I,3} \right)  \frac{da}{a}, \label{}
\end{eqnarray}

In the limiting case for a flat Dark Dominated Universe, we obtain:
\begin{eqnarray}
\dot{\phi}_{Dark,higher}^2\left(t \right) &=&\frac{8\alpha}{\beta}\frac{1}{ \left( t+C_3  \right)^2},\label{dusan1} \\
V_{Dark,higher}\left(t\right)&=&\left(2-\frac{\beta}{6\alpha}   \right) \frac{24\alpha^2}{\left[\beta \left( t+C_3  \right)\right]^2}.\label{murano13gcg}
\end{eqnarray}
Integrating with respect to the time $t$ the expression of $\dot{\phi}_{Dark,higher}^2\left(t \right) $, we obtain:
\begin{eqnarray}
\phi_{Dark,higher}\left(t \right) &=&2\sqrt{\frac{2\alpha}{\beta}}\cdot \ln \left( t+C_3  \right)+constant,\label{dusan1}
\end{eqnarray}
In the limiting case of $C_3=0$, we obtain:
\begin{eqnarray}
\phi_{Dark,higher,lim}\left(t \right) &=&2\sqrt{\frac{2\alpha}{\beta}}\cdot \ln t+constant,\label{dusan1}\\
V_{Dark,higher,lim}\left(t\right)&=&\left(2-\frac{\beta}{6\alpha}   \right) \frac{24\alpha^2}{\left(\beta  t\right)^2}.\label{murano13gcg}
\end{eqnarray}

For the NHDE model, we obtain the following expressions for $\dot{\phi}^2$ and $ V\left(  \phi \right)$:
\begin{eqnarray}
    \dot{\phi}^2_{NHDE} &=& \rho_{D,NHDE}\left(1+\omega_{D,NHDE} \right)  \label{gcg6mura1model1}\\
    V\left(  \phi \right)_{NHDE} &=& \frac{\rho_{D,NHDE,1}}{2}\left(1-\omega_{D,NHDE} \right)   \label{gcg7mura2model1}
\end{eqnarray}
We can obtain the evolutionary form of the GCG model integrating the expressions of $\dot{\phi}^2$ we obtained with respect to the scale factor 
\begin{eqnarray}
	\phi\left(a\right)_{NHDE} - \phi\left(a_0\right)_{NHDE} &=& \int_{a_0}^{a}\sqrt{3\Omega_{D}} \times \left(1+\omega_{D,NHDE} \right)  \frac{da}{a}, \label{}
\end{eqnarray}
For the omteracting NHDE model, we obtain:
\begin{eqnarray}
    \dot{\phi}^2_{NHDE,I} &=& \rho_{D,NHDE,I}\left(1+\omega_{D,NHDE,I} \right)  \label{gcg6mura1model1}\\
    V\left(  \phi \right)_{NHDE,I} &=& \frac{\rho_{D,NHDE,I}}{2}\left(1-\omega_{D,NHDE,I} \right)   \label{gcg7mura2model1}
\end{eqnarray}
and the evolutionary form of the GCG model is given by:
\begin{eqnarray}
	\phi\left(a\right)_{NHDE,I} - \phi\left(a_0\right)_{NHDE,I} &=& \int_{a_0}^{a}\sqrt{3\Omega_{D}} \times \left(1+\omega_{D,NHDE,I} \right)  \frac{da}{a}, \label{}
\end{eqnarray}

In the limiting case for a flat Dark Dominated Universe, we obtain:
\begin{eqnarray}
\dot{\phi}_{Dark,NHDE}^2\left(t \right) &=& \left(\frac{ 2\lambda}{\mu - 1}\right) \frac{1}{(t+C)^2},\label{dusan1} \\
V_{Dark,NHDE}\left(t\right)&=&\frac{ \lambda (3\lambda-\mu+1)}{(\mu - 1)^2 }\cdot \frac{1}{(t + C)^2}  \label{murano13gcg}
\end{eqnarray}
Integrating with respect to the time $t$ the expression of $\dot{\phi}_{Dark,NHDE}^2\left(t \right) $, we obtain:
\begin{eqnarray}
\phi_{Dark,NHDE}\left(t \right) &=&\sqrt{ \frac{ 2\lambda}{\mu - 1}}\cdot \ln (t+C) + const,\label{dusan1}
\end{eqnarray}
In the limiting case corresponding to $\lambda =1$ e $\mu=2$, i.e. the case corresponding to the Ricci scale, we obtain:
\begin{eqnarray}
\phi_{\rm Dark,NHDE,Ricci}(t) &=& \sqrt{2} \, \ln(t + C) + \text{const}, \\
V_{\rm Dark,NHDE,Ricci}(t) &=& \frac{2}{(t + C)^2}.
\end{eqnarray}

In the limiting case of $C=0$, we obtain:
\begin{eqnarray}
\phi_{Dark,NHDE,lim}\left(t \right) &=&\sqrt{ \frac{ 2\lambda}{\mu - 1}}\cdot \ln t + const,\label{dusan1} \\
V_{Dark,NHDE,lim}\left(t\right)&=& \frac{ \lambda (3\lambda-\mu+1)}{(\mu - 1)^2 }\cdot \frac{1}{t^2}  \label{murano13gcg}
\end{eqnarray}

In the limiting case corresponding to $\lambda =1$ e $\mu=2$, i.e. the case corresponding to the Ricci scale, we obtain:
\begin{eqnarray}
\phi_{\rm Dark,NHDE,Ricci,lim}(t) &=& \sqrt{2} \, \ln t  + \text{const}, \\
V_{\rm Dark,NHDE,Ricci,lim}(t) &=& \frac{2}{t ^2}.
\end{eqnarray}

We can easily derive the following relation between $\phi_{Dark,NHDE,Ricci,lim}$ and $V_{Dark,NHDE,Ricci,lim}$:
\begin{eqnarray}
 V(\phi) = 2 \, e^{ -\sqrt{2} \phi } 
\end{eqnarray}

In the limiting case corresponding to the values obtained by Wang \& Xu, the scalar field $\phi_{Dark}$ becomes complex, while the potential reads
\begin{eqnarray}
V_{Dark,NHDE,WX}(t) &\approx& \frac{34.22}{t^2}.
\end{eqnarray}
We find that $\phi_{Dark,NHDE}$ is not a real function.

\subsection{Viscous Generalized Chaplygin Gas (VGCG) Model}

We now consider the fourth model, i.e. the Viscous Generalized Chaplygin Gas (VGCG) model. To be as general as possible with respect to the previous sections, we include a viscous dark energy (DE) component interacting with dark matter (DM). In an isotropic and homogeneous Friedmann–Robertson–Walker (FRW) Universe, dissipative effects arise due to the presence of bulk viscosity in the cosmic fluids \cite{Tawfik:2021rvv,Tawfik:2010bm}. Some dissipative processes in the Universe, including bulk viscosity, shear viscosity, and heat transport, are believed to be present in any realistic theory of cosmic evolution and have been extensively studied in \cite{visc29,visc30,visc31,visc32}.

The role of viscosity has been widely discussed and it is considered a promising candidate to address several cosmological problems such as dark energy. The viscous DE model can provide an explanation for the high photon-to-baryon ratio \cite{visc18} and leads to an inflationary scenario during the early phase of the Universe's evolution \cite{visc19}. The viscosity coefficient is expected to decrease as the Universe expands; moreover, its presence can explain the current accelerated expansion of the Universe \cite{visc21,visc22}. This model is also consistent with astrophysical observations at lower redshifts, and a viscous cosmic fluid favors a standard cold dark matter model with a cosmological constant (\(\Lambda\)CDM) in the late-time cosmic evolution \cite{visc25}. Furthermore, the model allows for the phantom crossing scenario \cite{visc26}.

It is well-known that the general theory of dissipation in a relativistic imperfect fluid was initially formulated by Eckart \cite{visc33} and, in an alternative formulation, by Landau \& Lifshitz \cite{visc34}. This approach corresponds to a first-order deviation from equilibrium but may suffer from causality problems \cite{Tawfik:2021rvv,Tawfik:2010bm}. The full causal theory was later developed by Israel \& Stewart \cite{visc36} and has been applied to the early Universe evolution \cite{visc37}. However, the evolution equations in the full causal theory are highly complicated \cite{Tawfik:2021rvv,Tawfik:2010bm}. Fortunately, since the phenomena considered are quasi-stationary—slowly varying over spatial and temporal scales characterized by the mean free path and mean collision time of fluid particles—the conventional first-order theory remains valid. 

In the context of an isotropic and homogeneous FRW Universe, dissipative processes can be effectively modeled by bulk viscosity within a thermodynamical framework, while shear viscosity can be safely neglected \cite{visc40}. For additional studies on viscous DE models, see \cite{visc41,visc44,visc48,visc49,visc50}.

Dark energy with bulk viscosity exhibits the distinctive property of inducing a phantom-type accelerated expansion during the late stages of cosmic evolution \cite{brevik1,brevik3}. It can also help alleviate cosmological issues such as the coincidence problem, the age problem, and facilitate phantom crossing. Observations suggest that the cosmic medium is not a perfect fluid, and viscosity plays a significant role in the Universe's evolution (see \cite{ren1} and references therein). Within the FRW metric framework, shear viscosity does not contribute to the energy-momentum tensor \( T^{\mu\nu} \), and bulk viscosity manifests as an effective pressure \cite{ren2}.

Bulk viscosity introduces dissipation by modifying the effective pressure as 
\begin{eqnarray}
p_{\text{eff}} = p - 3 \nu H,
\end{eqnarray}
where \(\nu\) is the bulk viscosity coefficient. The condition \(\nu > 0\) guarantees positive entropy production, thereby preserving the second law of thermodynamics \cite{visc51}. A physically motivated case is \(\nu = \tau H\), implying the bulk viscosity is proportional to the fluid velocity vector; this has been considered earlier in astrophysical contexts \cite{visc52}.

The energy conservation equation yields the following expression for the energy density of the VGCG model \(\rho_D\) \cite{visc9}:
\begin{eqnarray}
\rho_D = \left[ \frac{D a^{-3(\theta+1)(1 - \nu \varrho)} - \chi}{1 - \nu \varrho} \right]^{\frac{1}{\theta +1}}, \label{murano72}
\end{eqnarray}
where \(\varrho = m_p^{-1} \sqrt{1 - r_m}\) with \(r_m = \frac{\rho_m}{\rho_D} = \frac{\Omega_m}{\Omega_D}\), and \(D\) is an integration constant.

The energy-momentum tensor corresponding to the bulk viscous fluid is given by
\begin{eqnarray}
T_{\mu \nu} = \left( \rho + \bar{p} \right) u_{\mu} u_{\nu} - \bar{p} g_{\mu \nu}, \label{murano73}
\end{eqnarray}
where
\begin{eqnarray}
\bar{p} = p_D - 3 \varepsilon H, \label{murano74}
\end{eqnarray}
represents the total effective pressure, which includes the proper pressure \( p_D \), the bulk viscosity coefficient \(\varepsilon\), and the Hubble parameter \( H \).

In this scenario, the proper pressure is given by 
\begin{eqnarray}
p_D = \frac{\chi}{\rho_D^{\theta}},
\end{eqnarray}
with \(\chi > 0\). The first term on the right-hand side of Eq. \eqref{murano74} mimics the generalized Chaplygin gas (GCG) model, where the parameter \(\theta\) varies in the range \(0 < \theta < 1\). For \(\theta = 1\), this reduces to the original Chaplygin gas model, whereas if \(\theta < 0\), it corresponds to a polytropic gas.

We choose a bulk viscosity coefficient \(\varepsilon\) depending on the energy density as
\begin{eqnarray}
\varepsilon = \nu \rho_D^{1/2},
\end{eqnarray}
where \(\nu\) is a constant parameter. Using this in Eq. \eqref{murano74}, the effective pressure becomes
\begin{eqnarray}
\bar{p} = \frac{\chi}{\rho_D^{\theta}} - 3 \nu H \rho_D^{1/2}. \label{murano74new}
\end{eqnarray}

The energy density \(\rho_D\) and the proper pressure \(p_D\) of the viscous dark energy model are given by
\begin{eqnarray}
\rho_D &=& \left[ \frac{D a^{-3(\theta+1)(1 - \nu \varrho)} - \chi}{1 - \nu \varrho} \right]^{\frac{1}{\theta + 1}}, \label{schiri1} \\
p_D &=& \chi \left[ \frac{1 - \nu \varrho}{D a^{-3(\theta+1)(1 - \nu \varrho)} - \chi} \right]^{\frac{\theta}{\theta + 1}} - 3 \nu H \left[ \frac{D a^{-3(\theta+1)(1 - \nu \varrho)} - \chi}{1 - \nu \varrho} \right]^{1/2}. \label{schiri2}
\end{eqnarray}

Next, we derive the expressions of the parameters \(\chi\) and \(D\). Dividing Eq. \eqref{murano74new} by \(\rho_D\) and using the general definition of the equation of state (EoS) parameter \(\omega_D = \frac{\bar{p}}{\rho_D}\), we obtain
\begin{eqnarray}
\chi = \rho_D^{\theta + 1} \left( 3 \nu H \rho_D^{-1/2} + \omega_D \right). \label{murano75}
\end{eqnarray}

The parameter \(D\) can be found from Eq. \eqref{schiri1} as
\begin{eqnarray}
D = \left[ \rho_D^{\theta + 1} (1 - \nu \varrho) + \chi \right] a^{3(\theta + 1)(1 - \nu \varrho)}. \label{murano76}
\end{eqnarray}

Substituting \(\chi\) from Eq. \eqref{murano75} into Eq. \eqref{murano76}, we have the final expression
\begin{eqnarray}
D = a^{3(\theta + 1)(1 - \nu \varrho)} \rho_D^{\theta + 1} \left( 1 - \nu \varrho + 3 \nu H \rho_D^{-1/2} + \omega_D \right). \label{murano77}
\end{eqnarray}

For the DE model we are studying in this paper, we obtain the following expressions for $\chi$ and $D$:
\begin{eqnarray}
\chi_{higher,1} &=& \rho_{D,higher,1}^{\theta +1}\left[ 3\nu H \rho_{D,higher,1}^{-1/2}  + \omega_{D,higher,1}  \right] ,\label{}\\
\chi_{higher,2} &=& \rho_{D,higher,2}^{\theta +1}\left[ 3\nu H \rho_{D,higher,2}^{-1/2}  + \omega_{D,higher,2}  \right] ,\label{}\\
\chi_{higher,3} &=& \rho_{D,higher,3}^{\theta +1}\left[ 3\nu H \rho_{D,higher,3}^{-1/2}  + \omega_{D,higher,3}  \right] ,\label{}\\
D_{higher,1}&=& a^{3\left( \theta +1 \right)\left(1-\nu \varrho  \right)}\rho_{D,higher,1}^{\theta +1} \times \nonumber   \\
&&\left[1-\nu \varrho  + 3\nu H \rho_{D,higher,1}^{-1/2} + \omega_{D,higher,1}  \right]\label{}\\
D_{higher,2}&=& a^{3\left( \theta +1 \right)\left(1-\nu \varrho  \right)}\rho_{D,higher,2}^{\theta +1} \times \nonumber   \\
&&\left[1-\nu \varrho  + 3\nu H \rho_{D,higher,2}^{-1/2} + \omega_{D,higher,2}  \right]\label{}\\
D_{higher,3}&=& a^{3\left( \theta +1 \right)\left(1-\nu \varrho  \right)}\rho_{D,higher,3}^{\theta +1} \times \nonumber   \\
&&\left[1-\nu \varrho  + 3\nu H \rho_{D,higher,3}^{-1/2} + \omega_{D,higher,3}  \right]\label{}
\end{eqnarray}
For the interacting case, we obtain:
\begin{eqnarray}
\chi_{higher,I,1} &=& \rho_{D,higher,I,1}^{\theta +1}\left[ 3\nu H \rho_{D,higher,I,1}^{-1/2}  + \omega_{D,higher,I,1}  \right] ,\label{}\\
\chi_{higher,I,2} &=& \rho_{D,higher,I,2}^{\theta +1}\left[ 3\nu H \rho_{D,higher,I,2}^{-1/2}  + \omega_{D,higher,I,2}  \right] ,\label{}\\
\chi_{higher,I,3} &=& \rho_{D,higher,I,3}^{\theta +1}\left[ 3\nu H \rho_{D,higher,I,3}^{-1/2}  + \omega_{D,higher,I,3}  \right] ,\label{}\\
D_{higher,I,1}&=& a^{3\left( \theta +1 \right)\left(1-\nu \varrho  \right)}\rho_{D,higher,I,1}^{\theta +1} \times \nonumber   \\
&&\left[1-\nu \varrho  + 3\nu H \rho_{D,higher,I,1}^{-1/2} + \omega_{D,higher,I,1}  \right]\label{}\\
D_{higher,I,2}&=& a^{3\left( \theta +1 \right)\left(1-\nu \varrho  \right)}\rho_{D,higher,I,2}^{\theta +1} \times \nonumber   \\
&&\left[1-\nu \varrho  + 3\nu H \rho_{D,higher,I,2}^{-1/2} + \omega_{D,higher,I,2}  \right]\label{}\\
D_{higher,I,3}&=& a^{3\left( \theta +1 \right)\left(1-\nu \varrho  \right)}\rho_{D,higher,I,3}^{\theta +1} \times \nonumber   \\
&&\left[1-\nu \varrho  + 3\nu H \rho_{D,higher,I,3}^{-1/2} + \omega_{D,higher,I,3}  \right]\label{}
\end{eqnarray}

In the limiting case of a flat Dark Dominated Universe, we obtain the following expressions for $\chi_{Dark}  $ and $D_{Dark}$
\begin{eqnarray}
\chi_{ Dark,higher} &=&\rho_{DD,higher}^{\theta +1}\left[ 3\nu H_{DD,higher} \rho_{DD.higher}^{-1/2}  + \omega_{DD,higher}  \right],\label{murano82}\\
D_{ Dark,higher}&=& a_{DD,higher}^{3\left( \theta +1 \right)\left(1-\nu \varrho  \right)}\rho_{DD,higher}^{\theta +1} \times \nonumber   \\
&&\left[1-\nu \varrho  + 3\nu H_{DD,higher} \rho_{DD,higher}^{-1/2} + \omega_{DD,higher}  \right]
\end{eqnarray}
In the limiting case of $C_3=0$, we obtain:
\begin{eqnarray}
\chi_{ Dark,higher,lim} &=&\rho_{DD,higher,lim}^{\theta +1}\left[ 3\nu H_{DD,higher,lim} \rho_{DD,higher,lim}^{-1/2}  + \omega_{DD,higher}  \right],\label{murano82}\\
D_{ Dark,higher,lim}&=& a_{DD,higher,lim}^{3\left( \theta +1 \right)\left(1-\nu \varrho  \right)}\rho_{DD,higher,lim}^{\theta +1} \times \nonumber   \\
&&\left[1-\nu \varrho  + 3\nu H_{DD,higher,lim} \rho_{DD,higher,lim}^{-1/2} + \omega_{DD,higher}  \right]
\end{eqnarray}
We must underline that $\omega_{D,DD}$ has the same expression for both case corresponding to the Dark Dominated Universe and the case with Dark Dominated Universe and $C_3=0$.\\

For the non interacting NHDE model, we obtain the following expressions for $\chi$ and $D$:
\begin{eqnarray}
\chi_{NHDE} &=& \rho_{D,NHDE}^{\theta +1}\left[ 3\nu H \rho_{D,NHDE}^{-1/2}  + \omega_{D,NHDE}  \right] ,\label{}\\
D_{NHDE}&=& a^{3\left( \theta +1 \right)\left(1-\nu \varrho  \right)}\rho_{D,NHDE}^{\theta +1} \times \nonumber   \\
&&\left[1-\nu \varrho  + 3\nu H \rho_{D,NHDE}^{-1/2} + \omega_{D,NHDE}  \right]\label{}
\end{eqnarray}
Instead, for the interacting NHDE model, we obtain the following expressions for $\chi$ and $D$:
\begin{eqnarray}
\chi_{NHDE,I} &=& \rho_{D,NHDE,I}^{\theta +1}\left[ 3\nu H \rho_{D,NHDE,I}^{-1/2}  + \omega_{D,NHDE,I}  \right] ,\label{}\\
D_{NHDE,I}&=& a^{3\left( \theta +1 \right)\left(1-\nu \varrho  \right)}\rho_{D,NHDE,I}^{\theta +1} \times \nonumber   \\
&&\left[1-\nu \varrho  + 3\nu H \rho_{D,NHDE,I}^{-1/2} + \omega_{D,NHDE,I}  \right]\label{}
\end{eqnarray}
In the limiting case of a flat Dark Dominated Universe, we obtain the following expressions for $\chi_{Dark}  $ and $D_{Dark}$
\begin{eqnarray}
\chi_{ Dark,NHDE} &=& \left[\frac{3 \lambda^2}{(\mu - 1)^2 (t + C)^2}\right]^{\theta+1}\times \left( \sqrt{3}\nu  -\frac{3\lambda-2\mu+2}{3\lambda}\right),\label{murano82}\\
D_{ Dark,NHDE}&=&  \left[\left( \frac{\mu - 1}{\lambda} \right)(t + C)\right]^{3(\theta + 1)(1 - \nu \varrho)}\left[\frac{3 \lambda^2}{(\mu - 1)^2 (t + C)^2}\right]^{\theta+1}\times\nonumber \\
&&\left[ 1+\nu\left(  \sqrt{3} -\varrho\right)  -\left(\frac{3\lambda-2\mu+2}{3\lambda}\right)  \right]\nonumber \\
&=&3^{\theta+1} 
\left[ 1 + \nu(\sqrt{3} - \varrho) - \left(\frac{3\lambda - 2\mu + 2}{3\lambda} \right)\right]\left[ \left(\frac{\mu - 1}{\lambda}\right) (t + C) \right]^{(\theta+1)(1-3\nu\varrho)} 
\label{murano83}
\end{eqnarray}
In the limiting case corresponding to the Ricci scale, i.e. for $\mu=2$ and $\lambda =1$, we obtain:
\begin{eqnarray}
\chi_{\rm Dark,NHDE,Ricci} &=& 3^{\theta+1} \,\left( \sqrt{3}\,\nu - \frac{1}{3} \right)(t + C)^{-2(\theta+1)} , \\
D_{\rm Dark,NHDE,Ricci} &=& 3^{\theta+1} \left[ \frac{2}{3} + \nu(\sqrt{3} - \varrho) \right] (t + C)^{(\theta+1)(1 - 3\nu\varrho)}.
\end{eqnarray}

In the limiting case of $C=0$, we obtain:
\begin{eqnarray}
\chi_{ Dark,NHDE,lim} &=& \left[\frac{3 \lambda^2}{(\mu - 1)^2 t^2 }\right]^{\theta+1}\times \left( \sqrt{3}\nu  -\frac{3\lambda-2\mu+2}{3\lambda}\right),\label{murano82}\\
D_{ Dark,NHDE,lim}&=&  \left[\left( \frac{\mu - 1}{\lambda} \right)t \right]^{3(\theta + 1)(1 - \nu \varrho)}\left[\frac{3 \lambda^2}{(\mu - 1)^2 t^2 }\right]^{\theta+1}\times\nonumber \\
&&\left[ 1+\nu\left(  \sqrt{3} -\varrho\right)  -\frac{3\lambda-2\mu+2}{3\lambda}  \right]\nonumber \\
&=&3^{\theta+1} 
\left[ 1 + \nu(\sqrt{3} - \varrho) - \left(\frac{3\lambda - 2\mu + 2}{3\lambda} \right)\right]\left[ \left(\frac{\mu - 1}{\lambda}\right) t  \right]^{(\theta+1)(1-3\nu\varrho)} 
\label{murano83}
\end{eqnarray}
In the limiting case corresponding to the Ricci scale, i.e. for $\mu=2$ and $\lambda =1$, and $C=0$, we obtain:
\begin{eqnarray}
\chi_{\rm Dark,NHDE,Ricci,lim} &=& 3^{\theta+1} \,\left( \sqrt{3}\,\nu - \frac{1}{3} \right)t ^{-2(\theta+1)} , \\
D_{\rm Dark,NHDE,Ricci,lim} &=& 3^{\theta+1} \left[ \frac{2}{3} + \nu(\sqrt{3} - \varrho) \right] t ^{(\theta+1)(1 - 3\nu\varrho)}.
\end{eqnarray}
In the limiting case corresponding to the values found by Wang \& Xu, we obtain:
\begin{eqnarray}
\chi_{Dark,NHDE,WX} &=&
\left[\frac{31.020}{(t+C)^2}\right]^{\theta+1}
\left(\sqrt{3}\,\nu - 1.207\right),\\
D_{Dark,NHDE,WX} &=& 3^{\theta+1}\,
\Big[1+\nu(\sqrt{3}-\varrho) - 1.207\Big]\,
\left[-0.311\,(t+C)\right]^{(\theta+1)(1-3\nu\varrho)}.
\end{eqnarray}

Moreover, using the general definition of EoS parameter $\omega_D$,  we can rewrite $ \dot{\phi}^2$ and $V\left(  \phi \right)$ in the following way:
\begin{eqnarray}
    \dot{\phi}^2 &=& \left( 1+ \omega_D \right)\rho_D, \label{gcg6mura1}\\
    V\left(  \phi \right) &=& \frac{1}{2}\left( 1 - \omega_D \right)\rho_D. \label{gcg7mura2}
\end{eqnarray}
For the DE model studied in this paper, we obtain the following expressions for  $\dot{\phi}^2$ and $ V\left(  \phi \right)$:
\begin{eqnarray}
    \dot{\phi}^2_{higher,1} &=& \rho_{D,higher,1}\left(1+\omega_{D,higher,1} \right)  \label{gcg6mura1model1}\\
    \dot{\phi}^2_{higher,2} &=& \rho_{D,higher,2}\left(1+\omega_{D,higher,2} \right)  \label{gcg6mura1model1-2}\\
    \dot{\phi}^2_{higher,3} &=& \rho_{D,higher,3}\left(1+\omega_{D,higher,3} \right)  \label{gcg6mura1mode1-3}\\
    V\left(  \phi \right)_{higher,1} &=& \frac{\rho_{D,higher,1}}{2}\left(1-\omega_{D,higher,1} \right)   \label{gcg7mura2model1}\\
    V\left(  \phi \right)_{higher,2} &=& \frac{\rho_{D,higher,2}}{2}\left(1-\omega_{D,higher,2} \right)   \label{gcg7mura2model1-2}\\
    V\left(  \phi \right)_{higher,3} &=& \frac{\rho_{D,higher,3}}{2}\left(1-\omega_{D,higher,3} \right)   \label{gcg7mura2model1-3}
\end{eqnarray}

We can obtain the evolutionary form of the GCG model integrating the expressions of $\dot{\phi}^2$ we obtained with respect to the scale factor $a\left(t\right)$
\begin{eqnarray}
	\phi\left(a\right)_{higher,1} - \phi\left(a_0\right)_{higher,1} &=& \int_{a_0}^{a}\sqrt{3\Omega_{D}} \times \left(1+\omega_{D,higher,1} \right)  \frac{da}{a}, \label{}\\
    \phi\left(a\right)_{higher,2} - \phi\left(a_0\right)_{higher,2} &=& \int_{a_0}^{a}\sqrt{3\Omega_{D}} \times \left(1+\omega_{D,higher,2} \right)  \frac{da}{a}, \label{}\\
    \phi\left(a\right)_{higher,3} - \phi\left(a_0\right)_{higher,3} &=& \int_{a_0}^{a}\sqrt{3\Omega_{D}} \times \left(1+\omega_{D,higher,3} \right)  \frac{da}{a}, \label{}
\end{eqnarray}

For the interacting case, we obtain:
\begin{eqnarray}
    \dot{\phi}^2_{higher,I,1} &=& \rho_{D,higher,I,1}\left(1+\omega_{D,higher,I,1} \right)  \label{gcg6mura1model1}\\
    \dot{\phi}^2_{higher,I,2} &=& \rho_{D,higher,I,2}\left(1+\omega_{D,higher,I,2} \right)  \label{gcg6mura1model1-2}\\
    \dot{\phi}^2_{higher,3,I} &=& \rho_{D,higher,I,3}\left(1+\omega_{D,higher,I,3} \right)  \label{gcg6mura1mode1-3}\\
    V\left(  \phi \right)_{higher,I,1} &=& \frac{\rho_{D,higher,I,1}}{2}\left(1-\omega_{D,higher,I,1} \right)   \label{gcg7mura2model1}\\
    V\left(  \phi \right)_{higher,I,2} &=& \frac{\rho_{D,higher,I,2}}{2}\left(1-\omega_{D,higher,I,2} \right)   \label{gcg7mura2model1-2}\\
    V\left(  \phi \right)_{higher,I,3} &=& \frac{\rho_{D,higher,I,3}}{2}\left(1-\omega_{D,higher,I,3} \right)   \label{gcg7mura2model1-3}
\end{eqnarray}
and the evolutionary form of the GCG model are given by:
\begin{eqnarray}
	\phi\left(a\right)_{higher,I,1} - \phi\left(a_0\right)_{higher,I,1} &=& \int_{a_0}^{a}\sqrt{3\Omega_{D}} \times \left(1+\omega_{D,higher,I,1} \right)  \frac{da}{a}, \label{}\\
    \phi\left(a\right)_{higher,I,2} - \phi\left(a_0\right)_{higher,I,2} &=& \int_{a_0}^{a}\sqrt{3\Omega_{D}} \times \left(1+\omega_{D,higher,I,2} \right)  \frac{da}{a}, \label{}\\
    \phi\left(a\right)_{higher,I,3} - \phi\left(a_0\right)_{higher,I,3} &=& \int_{a_0}^{a}\sqrt{3\Omega_{D}} \times \left(1+\omega_{D,higher,I,3} \right)  \frac{da}{a}, \label{}
\end{eqnarray}

In the limiting case for a flat Dark Dominated Universe, we obtain:
\begin{eqnarray}
\dot{\phi}_{Dark,higher}^2\left(t \right) &=&\frac{8\alpha}{\beta}\frac{1}{ \left( t+C_3  \right)^2},\label{dusan1} \\
V_{Dark,higher}\left(t\right)&=&\left(2-\frac{\beta}{6\alpha}   \right) \frac{24\alpha^2}{\left[\beta \left( t+C_3  \right)\right]^2}.\label{murano13gcg}
\end{eqnarray}
Integrating with respect to the time $t$ the expression of $\dot{\phi}_{Dark,higher}^2\left(t \right) $, we obtain:
\begin{eqnarray}
\phi_{Dark,higher}\left(t \right) &=&2\sqrt{\frac{2\alpha}{\beta}}\cdot \ln \left( t+C_3  \right)+constant,\label{dusan1}
\end{eqnarray}
In the limiting case of $C_3=0$, we obtain:
\begin{eqnarray}
\phi_{Dark,higher,lim}\left(t \right) &=&2\sqrt{\frac{2\alpha}{\beta}}\cdot \ln t+constant,\label{dusan1}\\
V_{Dark,higher,lim}\left(t\right)&=&\left(2-\frac{\beta}{6\alpha}   \right) \frac{24\alpha^2}{\left(\beta  t\right)^2}.\label{murano13gcg}
\end{eqnarray}

For the NHDE model, we obtain the following expressions for $\dot{\phi}^2$ and $ V\left(  \phi \right)$:
\begin{eqnarray}
    \dot{\phi}^2_{NHDE} &=& \rho_{D,NHDE}\left(1+\omega_{D,NHDE} \right)  \label{gcg6mura1model1}\\
    V\left(  \phi \right)_{NHDE} &=& \frac{\rho_{D,NHDE,1}}{2}\left(1-\omega_{D,NHDE} \right)   \label{gcg7mura2model1}
\end{eqnarray}
We can obtain the evolutionary form of the GCG model integrating the expressions of $\dot{\phi}^2$ we obtained with respect to the scale factor 
\begin{eqnarray}
	\phi\left(a\right)_{NHDE} - \phi\left(a_0\right)_{NHDE} &=& \int_{a_0}^{a}\sqrt{3\Omega_{D}} \times \left(1+\omega_{D,NHDE} \right)  \frac{da}{a}, \label{}
\end{eqnarray}
For the omteracting NHDE model, we obtain:
\begin{eqnarray}
    \dot{\phi}^2_{NHDE,I} &=& \rho_{D,NHDE,I}\left(1+\omega_{D,NHDE,I} \right)  \label{gcg6mura1model1}\\
    V\left(  \phi \right)_{NHDE,I} &=& \frac{\rho_{D,NHDE,I}}{2}\left(1-\omega_{D,NHDE,I} \right)   \label{gcg7mura2model1}
\end{eqnarray}
and the evolutionary form of the GCG model is given by:
\begin{eqnarray}
	\phi\left(a\right)_{NHDE,I} - \phi\left(a_0\right)_{NHDE,I} &=& \int_{a_0}^{a}\sqrt{3\Omega_{D}} \times \left(1+\omega_{D,NHDE,I} \right)  \frac{da}{a}, \label{}
\end{eqnarray}

In the limiting case for a flat Dark Dominated Universe, we obtain:
\begin{eqnarray}
\dot{\phi}_{Dark,NHDE}^2\left(t \right) &=& \left(\frac{ 2\lambda}{\mu - 1}\right) \frac{1}{(t+C)^2},\label{dusan1} \\
V_{Dark,NHDE}\left(t\right)&=&\frac{ \lambda (3\lambda-\mu+1)}{(\mu - 1)^2 }\cdot \frac{1}{(t + C)^2}  \label{murano13gcg}
\end{eqnarray}
Integrating with respect to the time $t$ the expression of $\dot{\phi}_{Dark,NHDE}^2\left(t \right) $, we obtain:
\begin{eqnarray}
\phi_{Dark,NHDE}\left(t \right) &=&\sqrt{ \frac{ 2\lambda}{\mu - 1}}\cdot \ln (t+C) + const,\label{dusan1}
\end{eqnarray}
In the limiting case corresponding to $\lambda =1$ e $\mu=2$, i.e. the case corresponding to the Ricci scale, we obtain:
\begin{eqnarray}
\phi_{\rm Dark,NHDE,Ricci}(t) &=& \sqrt{2} \, \ln(t + C) + \text{const}, \\
V_{\rm Dark,NHDE,Ricci}(t) &=& \frac{2}{(t + C)^2}.
\end{eqnarray}

In the limiting case of $C=0$, we obtain:
\begin{eqnarray}
\phi_{Dark,NHDE,lim}\left(t \right) &=&\sqrt{ \frac{ 2\lambda}{\mu - 1}}\cdot \ln t + const,\label{dusan1} \\
V_{Dark,NHDE,lim}\left(t\right)&=& \frac{ \lambda (3\lambda-\mu+1)}{(\mu - 1)^2 }\cdot \frac{1}{t^2}  \label{murano13gcg}
\end{eqnarray}

In the limiting case corresponding to $\lambda =1$ e $\mu=2$, i.e. the case corresponding to the Ricci scale, we obtain:
\begin{eqnarray}
\phi_{\rm Dark,NHDE,Ricci,lim}(t) &=& \sqrt{2} \, \ln t  + \text{const}, \\
V_{\rm Dark,NHDE,Ricci,lim}(t) &=& \frac{2}{t ^2}.
\end{eqnarray}

We can easily derive the following relation between $\phi_{Dark,NHDE,Ricci,lim}$ and $V_{Dark,NHDE,Ricci,lim}$:
\begin{eqnarray}
 V(\phi) = 2 \, e^{ -\sqrt{2} \phi } 
\end{eqnarray}

In the limiting case corresponding to the values obtained by Wang \& Xu, the scalar field $\phi_{Dark}$ becomes complex, while the potential reads
\begin{eqnarray}
V_{Dark,NHDE,WX}(t) &\approx& \frac{34.22}{t^2}.
\end{eqnarray}
We find that $\phi_{Dark,NHDE}$ is not a real function.

\subsection{Dirac-Born-Infeld (DBI) Model}

We now consider the Dirac-Born-Infeld (DBI) model. Recently, many studies have explored the connection between string theory and inflation. Among these, the concept of branes in string theory has proven especially fruitful. A well-studied scenario is inflation driven by the open string sector through dynamical Dp-branes, known as Dirac-Born-Infeld (DBI) inflation. This model belongs to a special class of K-inflation models.

Considering dark energy (DE) to be driven by a DBI scalar field, the action \( S_{dbi} \) of the field is given by \cite{dbi1,dbi3,dbi5,dbi6,dbi7,dbi8,dbi9,dbi10,dbi11,dbi12}:
\begin{eqnarray}
S_{dbi} = \int d^4x \, a^3(t) \left[ F(\phi) \sqrt{1 - \frac{\dot{\phi}^2}{F(\phi)}} + V(\phi) - F(\phi) \right], \label{murano86}
\end{eqnarray}
where \(F(\phi)\) represents the brane tension, and \(V(\phi)\) is the scalar potential.

From this action, the pressure \( p_{dbi} \) and energy density \( \rho_{dbi} \) of the DBI scalar field are expressed as
\begin{eqnarray}
p_{dbi} &=& \left( \frac{\gamma - 1}{\gamma} \right) F(\phi) - V(\phi), \label{murano87} \\
\rho_{dbi} &=& (\gamma - 1) F(\phi) + V(\phi), \label{murano88}
\end{eqnarray}
where the quantity \(\gamma\) is analogous to the relativistic Lorentz factor and is defined by
\begin{eqnarray}
\gamma = \left[ 1 - \frac{\dot{\phi}^2}{F(\phi)} \right]^{-1/2}. \label{murano89}
\end{eqnarray}
Using the expressions of $p_{dbi}$ and $\rho_{dbi}$ given in Eqs. (\ref{murano87}) and (\ref{murano88}), the EoS parameter $\omega_{dbi}$ can be written as:
\begin{eqnarray}
\omega_{dbi} &=& \frac{\left(\frac{\gamma-1}{\gamma}\right) F\left(\phi\right) -  V\left(\phi\right)}{\left(\gamma-1\right) F\left(\phi\right) +  V\left(\phi\right)} = \frac{\left(\gamma-1\right) F\left(\phi\right) -  \gamma V\left(\phi\right)}{\gamma\left[ \left(\gamma-1\right) F\left(\phi\right) +  V\left(\phi\right)  \right] }.\label{murano90}
\end{eqnarray}

We now ain to derive the relations for the three terms $F$, $\dot{\phi}^2$ and $V$. Adding Eqs. (\ref{murano87}) and (\ref{murano88}) and using the general definition of $\omega_D$, we derive that:
\begin{eqnarray}
F &=& \rho_D\left( \frac{\gamma}{\gamma^2 -1}\right) \left( \omega_D + 1   \right)\label{murano91old}.
\end{eqnarray}
Using the definition of $\gamma$ given in Eq. (\ref{murano89}) and  the result of Eq. (\ref{murano91old}), we obtain the following result:
\begin{eqnarray}
\dot{\phi} &=& \sqrt{\frac{ \rho_D \left( \omega_D + 1   \right) }{\gamma}}.\label{murano92old}
\end{eqnarray}
Subtracting Eqs. (\ref{murano87}) and (\ref{murano88}) and using the general definition of $\omega_D$, we obtain:
\begin{eqnarray}
V &=& - \left(\frac{\rho_D}{\gamma +1}\right) \left( \gamma \omega_D -1   \right).\label{murano93}
\end{eqnarray}
Therefore, we derive the following set of expressions for the three terms $F$, $\dot{\phi}$ and $V$:
\begin{eqnarray}
F &=& \rho_D\left( \frac{\gamma}{\gamma^2 -1}\right) \left(1+ \omega_D \right),\label{murano91}\\
\dot{\phi} &=& \sqrt{\frac{ \rho_D \left( 1+\omega_D    \right) }{\gamma}},\label{murano92}\\
V &=& - \frac{\rho_D}{\gamma +1} \left( \gamma \omega_D  -1   \right) \nonumber \\
&=&  \frac{\rho_D}{\gamma +1} \left(1- \gamma \omega_D     \right).\label{murano93}
\end{eqnarray}

For the DE considered in this work, we obtain the following results for  $F$, $\dot{\phi}$ and $V$:
\begin{eqnarray}
F_{higher,1} &=&\rho_{D,higher,1} \left( \frac{\gamma}{\gamma^2 -1}\right) (1+\omega_{D,higher,1}),\label{}\\
F_{higher,2} &=&\rho_{D,higher,2} \left( \frac{\gamma}{\gamma^2 -1}\right) (1+\omega_{D,higher,2}),\label{}\\
F_{higher,3} &=&\rho_{D,higher,3} \left( \frac{\gamma}{\gamma^2 -1}\right)(1+\omega_{D,higher,3}) ,\label{}\\
\dot{\phi}_{higher,1} &=& \sqrt{\frac{ \rho_{D,higher,1} \left(1+\omega_{D,higher,1}  \right) }{\gamma}},\label{}\\
\dot{\phi}_{higher,2} &=& \sqrt{\frac{ \rho_{D,higher,2} \left(1+\omega_{D,higher,2}  \right) }{\gamma}},\label{}\\
\dot{\phi}_{higher,3} &=& \sqrt{\frac{ \rho_{D,higher,3} \left(1+\omega_{D,higher,3}  \right) }{\gamma}},\label{}\\
V_{ higher,1} &=&  \frac{\rho_{D,higher,1}}{\gamma +1}\left(1-\gamma \omega_{D,higher,1} \right) .\label{}\\
V_{ higher,2 } &=&  \frac{\rho_{D,higher,2}}{\gamma +1} \left(1-\gamma \omega_{D,higher,2} \right).\label{}\\
V_{ higher,3} &=&  \frac{\rho_{D,higher,3}}{\gamma +1} \left(1-\gamma \omega_{D,higher,3} \right).\label{}
\end{eqnarray}

For the interacting case, we obtain:
\begin{eqnarray}
F_{higher,I,1} &=&\rho_{D,higher,I,1} \left( \frac{\gamma}{\gamma^2 -1}\right) (1+\omega_{D,higher,I,1}),\label{}\\
F_{higher,I,2} &=&\rho_{D,higher,I,2} \left( \frac{\gamma}{\gamma^2 -1}\right) (1+\omega_{D,higher,I,2}),\label{}\\
F_{higher,I,3} &=&\rho_{D,higher,I,3} \left( \frac{\gamma}{\gamma^2 -1}\right)(1+\omega_{D,higher,I,3}) ,\label{}\\
\dot{\phi}_{higher,I,1} &=& \sqrt{\frac{ \rho_{D,higher,I,1} \left(1+\omega_{D,higher,I,1}  \right) }{\gamma}},\label{}\\
\dot{\phi}_{higher,I,2} &=& \sqrt{\frac{ \rho_{D,higher,I,2} \left(1+\omega_{D,higher,I,2}  \right) }{\gamma}},\label{}\\
\dot{\phi}_{higher,I,3} &=& \sqrt{\frac{ \rho_{D,higher,I,3} \left(1+\omega_{D,higher,I,3}  \right) }{\gamma}},\label{}\\
V_{ higher,I,1} &=&  \frac{\rho_{D,higher,I,1}}{\gamma +1}\left(1-\gamma \omega_{D,higher,I,1} \right) .\label{}\\
V_{ higher,I,2  } &=&  \frac{\rho_{D,higher,I,2}}{\gamma +1} \left(1-\gamma \omega_{D,higher,I,2} \right).\label{}\\
V_{higher,I,3  } &=&  \frac{\rho_{D,higher,I,3}}{\gamma +1} \left(1-\gamma \omega_{D,higher,I,3} \right).\label{}
\end{eqnarray}

In the limiting case of a flat Dark Dominated Universe, we obtain the following results:
\begin{eqnarray}
F_{Dark,higher} &=&  \left( \frac{\gamma}{\gamma^2 -1}\right)\frac{48\alpha^2}{\left[\beta \left( t+C_3  \right)\right]^2}\left( \frac{\beta}{6\alpha}  \right),\label{murano100}\\
\dot{\phi}_{Dark,higher} &=& \frac{4\alpha}{\beta \left( t+C_3  \right)}\sqrt{\frac{\beta}{2\alpha \gamma}},\label{murano101}\\
V_{Dark,higher} &=& - \frac{48\alpha^2}{(\gamma +1)\left[\beta \left( t+C_3  \right)\right]^2} \left( 1+\gamma -\frac{\beta \gamma}{6\alpha}  \right).\label{murano102}
\end{eqnarray}
In the limiting case of $C_3=0$, we can write:
\begin{eqnarray}
F_{Dark} &=&  \left( \frac{\gamma}{\gamma^2 -1}\right)\frac{8\alpha}{\beta }\left( \frac{1}{t^2}  \right),\label{murano100}\\
\dot{\phi}_{Dark} &=& \frac{4\alpha}{\beta t}\sqrt{\frac{\beta}{2\alpha \gamma}},\label{murano101}\\
V_{Dark} &=& - \frac{48\alpha^2}{(\gamma +1)\left(\beta t\right)^2} \left( 1+\gamma -\frac{\beta \gamma}{6\alpha}  \right).\label{murano102}
\end{eqnarray}

For the non-interating NHDE mode, we obtain:
\begin{eqnarray}
F_{NHDE} &=&\rho_{D,NHDE} \left( \frac{\gamma}{\gamma^2 -1}\right) (1+\omega_{D,NHDE}),\label{}\\
\dot{\phi}_{NHDE} &=& \sqrt{\frac{ \rho_{D,NHDE} \left(1+\omega_{D,NHDE}  \right) }{\gamma}},\label{}\\
V_{NHDE} &=&  \frac{\rho_{D,NHDE}}{\gamma +1}\left(1-\gamma \omega_{D,NHDE} \right) .\label{}
\end{eqnarray}

Instead, forinterating NHDE mode, we obtain:
\begin{eqnarray}
F_{NHDE,I} &=&\rho_{D,NHDE,I} \left( \frac{\gamma}{\gamma^2 -1}\right) (1+\omega_{D,NHDE,I}),\label{}\\
\dot{\phi}_{NHDE,I} &=& \sqrt{\frac{ \rho_{D,NHDE,I} \left(1+\omega_{D,NHDE,I}  \right) }{\gamma}},\label{}\\
V_{NHDE, I} &=&  \frac{\rho_{D,NHDE,I}}{\gamma +1}\left(1-\gamma \omega_{D,NHDE,I} \right) .\label{}
\end{eqnarray}
In the limiting case of a flat Dark Dominated Universe, we obtain the following results:
\begin{eqnarray}
F_{Dark,NHDE} &=&  \left( \frac{2\gamma}{\gamma^2 -1}\right)\frac{ \lambda}{(\mu - 1) (t + C)^2},\label{murano100}\\
\dot{\phi}_{Dark,NHDE} &=&\cdot\sqrt{\frac{ 2\lambda}{\gamma(\mu - 1) }}\cdot \left( \frac{1}{t+C}  \right),\label{murano101}\\
V_{Dark,NHDE} &=&  \frac{3 \lambda^2}{(\gamma+1)(\mu - 1)^2 (t + C)^2}\cdot \left[1+\frac{\gamma\left(3\lambda-2\mu+2\right)}{3\lambda} \right].\label{murano102}
\end{eqnarray}
In the limiting case of Ricci scale, we obtain:
\begin{eqnarray}
F_{\rm Dark,NHDE,Ricci} &=& \left( \frac{2\gamma}{\gamma^2 -1}\right)\frac{1}{(t + C)^2},\\
\dot{\phi}_{\rm Dark,NHDE,Ricci} &=& \sqrt{\frac{ 2}{\gamma }}\left(\frac{1}{t + C}\right),\\
V_{\rm Dark,NHDE,Ricci} &=& \left(\frac{\gamma+3}{\gamma+1}\right) \frac{1}{(t + C)^2}.
\end{eqnarray}

In the limiting case of $C=0$, we obtain:
\begin{eqnarray}
F_{Dark,NHDE,lim} &=&  \left( \frac{2\gamma}{\gamma^2 -1}\right)\frac{ \lambda}{(\mu - 1) t^2  },\label{murano100}\\
\dot{\phi}_{Dark,NHDE,lim} &=&\cdot\sqrt{\frac{ 2\lambda}{\gamma(\mu - 1) ^2 }}\cdot \frac{1}{t},\label{murano101}\\
V_{Dark,NHDE,lim} &=&  \frac{3 \lambda^2}{(\gamma+1)(\mu - 1)^2 t^2}\cdot \left[1+\frac{\gamma\left(3\lambda-2\mu+2\right)}{3\lambda} \right].\label{murano102}
\end{eqnarray}
We now consider the case corresponding to the Ricci DE model and $C=0$. In this case, we have:
\begin{eqnarray}
F_{Dark,NHDE,Ricci,lim} &=&  \left( \frac{2\gamma}{\gamma^2 -1}\right)\frac{ 1}{t^2  },\label{murano100}\\
\dot{\phi}_{Dark,NHDE,Ricci,lim} &=&\cdot\sqrt{\frac{ 2}{\gamma }}\cdot \frac{1}{t},\label{murano101}\\
V_{Dark,NHDE,Ricci,lim} &=&  \frac{\gamma +3}{(\gamma+1)}\frac{1}{t^2}.\label{murano102}
\end{eqnarray}

In the limiting case corresponding to the values found by Wang \& Xu, we obtain:
\begin{eqnarray}
F_{Dark,NHDE,WX} &&\approx -
\left(\frac{2\gamma}{\gamma^2 - 1}\right)\frac{3.215}{(t+C)^2},\\
\dot{\phi}_{Dark,NHDE,WX} &&\approx 
\sqrt{\frac{-6.429}{\gamma}}\cdot \frac{1}{t+C},\\
V_{Dark,NHDE,WX} &&\approx 
\frac{31.020}{(\gamma+1)(t+C)^2}\left(1+1.207\,\gamma\right).
\end{eqnarray}
We then find that $\dot{\phi}_{Dark,NHDE,WX}$ is not a real function.\\
We now consider a particular case of $F\left( \phi  \right)$ given by:
\begin{eqnarray}
F\left( \phi  \right) = F_0 \dot{\phi}^2,\label{murano103}
\end{eqnarray}
with $F_0 >0$ being a constant parameter. For this particular case, we can write $\gamma$ in the following way:
\begin{eqnarray}
\gamma = \sqrt{\frac{F_0}{F_0-1}}.\label{murano104}
\end{eqnarray}
which  implies that $F_0>1$ in order to have a real value of $\gamma$. Negative values of $F_0$ also lead to positive values of $\gamma$ but they cannot be considered since we previously imposed that $F_0 >0$.

Considering the result obtained in Eq. (\ref{murano104}), we can write Eqs. (\ref{murano91}), (\ref{murano92}) and (\ref{murano93}) in the following forms:
\begin{eqnarray}
F &=& \rho_D\sqrt{F_0\left( F_0-1  \right)} \left( 1+\omega_D   \right),\label{murano105} \\
\dot{\phi} &=&\left(\sqrt{\frac{F_0}{F_0-1}}\right)^{-1/4} \sqrt{ \rho_D \left( 1+\omega_D    \right)},\label{murano106} \\
V &=& - \frac{\rho_D }{\sqrt{\frac{F_0}{F_0-1}} + 1} \left[ \left(\sqrt{\frac{F_0}{F_0-1}}\right) \omega_D -1 \right].\label{murano107}
\end{eqnarray}
For the DE model we are considering in this paper, we can write:
\begin{eqnarray}
F_{higher,1} &=& \sqrt{F_0\left( F_0-1  \right)}\cdot\rho_{D,higher,1}\left( 1+\omega_{D,higher,1}\right),\label{} \\
F_{ higher,2} &=& \sqrt{F_0\left( F_0-1  \right)}\cdot\rho_{D,higher,2}\left( 1+\omega_{D,higher,2}\right),\label{} \\
F_{ higher,3} &=& \sqrt{F_0\left( F_0-1  \right)}\cdot\rho_{D,higher,3}\left( 1+\omega_{D,higher,3}\right),\label{} \\
\dot{\phi}_{higher,1} &=&\left(\sqrt{\frac{F_0}{F_0-1}}\right)^{-1/4} \sqrt{ \rho_{D,higher,1}\left( 1+\omega_{D,higher,1}\right)  }\label{} \\
\dot{\phi}_{higher,2} &=&\left(\sqrt{\frac{F_0}{F_0-1}}\right)^{-1/4} \sqrt{ \rho_{D,higher,2}\left( 1+\omega_{D,higher,2}\right)  }\label{} \\
\dot{\phi}_{higher,3} &=&\left(\sqrt{\frac{F_0}{F_0-1}}\right)^{-1/4} \sqrt{ \rho_{D,higher,3}\left( 1+\omega_{D,higher,3}\right)  }\label{} \\
V_{higher,1} &=& -\frac{\rho_{D,higher,1}  }{\sqrt{\frac{F_0}{F_0-1}} + 1} \left[  \left(\sqrt{\frac{F_0}{F_0-1}}\right)\omega_{D,higher,1}-1  \right],\label{}\\
V_{higher,2} &=& -\frac{\rho_{D,higher,2}  }{\sqrt{\frac{F_0}{F_0-1}} + 1} \left[  \left(\sqrt{\frac{F_0}{F_0-1}}\right)\omega_{D,higher,2}-1  \right],\label{}\\
V_{higher,3} &=&- \frac{\rho_{D,higher,3}  }{\sqrt{\frac{F_0}{F_0-1}} + 1} \left[  \left(\sqrt{\frac{F_0}{F_0-1}}\right)\omega_{D,higher,3}-1  \right],\label{}
\end{eqnarray}
For the interacting case, we obtain:
\begin{eqnarray}
F_{ higher,I,1} &=& \sqrt{F_0\left( F_0-1  \right)}\cdot\rho_{D,higher,I,1}\left( 1+\omega_{D,higher,I,1}\right),\label{} \\
F_{ higher,I,2} &=& \sqrt{F_0\left( F_0-1  \right)}\cdot\rho_{D,higher,I,2}\left( 1+\omega_{D,higher,I,2}\right),\label{} \\
F_{ higher,I,3} &=& \sqrt{F_0\left( F_0-1  \right)}\cdot\rho_{D,higher,I,3}\left( 1+\omega_{D,higher,I,3}\right),\label{} \\
\dot{\phi}_{higher,I,1} &=&\left(\sqrt{\frac{F_0}{F_0-1}}\right)^{-1/4} \sqrt{ \rho_{D,higher,I,1}\left( 1+\omega_{D,higher,I,1}\right)  }\label{} \\
\dot{\phi}_{higher,I,2} &=&\left(\sqrt{\frac{F_0}{F_0-1}}\right)^{-1/4} \sqrt{ \rho_{D,higher,I,2}\left( 1+\omega_{D,higher,I,2}\right)  }\label{} \\
\dot{\phi}_{higher,I,3} &=&\left(\sqrt{\frac{F_0}{F_0-1}}\right)^{-1/4} \sqrt{ \rho_{D,higher,I,3}\left( 1+\omega_{D,higher,I,3}\right)  }\label{} \\
V_{higher,I,1} &=& -\frac{\rho_{D,higher,I,1}  }{\sqrt{\frac{F_0}{F_0-1}} + 1} \left[  \left(\sqrt{\frac{F_0}{F_0-1}}\right)\omega_{D,higher,I,1}-1  \right],\label{}\\
V_{higher,I,2} &=& -\frac{\rho_{D,higher,I,2}  }{\sqrt{\frac{F_0}{F_0-1}} + 1} \left[  \left(\sqrt{\frac{F_0}{F_0-1}}\right)\omega_{D,higher,I,2}-1  \right],\label{}\\
V_{higher,I,3} &=&- \frac{\rho_{D,higher,I,3}  }{\sqrt{\frac{F_0}{F_0-1}} + 1} \left[  \left(\sqrt{\frac{F_0}{F_0-1}}\right)\omega_{D,higher,I,3}-1  \right],\label{}
\end{eqnarray}

In the limiting case of a flat Dark Dominated Universe,we obtain the following results:
\begin{eqnarray}
F_{ Dark,higher} &=& \sqrt{F_0\left( F_0-1  \right)}\left(\frac{8\alpha}{\beta}\right)\frac{1}{ \left( t+C_3  \right)^2},\label{murano114} \\
\dot{\phi}_{ Dark,higher} &=& 2 \left(\sqrt{\frac{F_0}{F_0-1}}\right)^{-1/4}\sqrt{\frac{2\alpha}{\beta}}\cdot\frac{1}{ \left( t+C_3  \right)} ,\label{giovanniboh} \\
V_{ Dark,higher} &=& -\left\{ \frac{48\alpha^2}{\left[\beta \left( t+C_3  \right)\right]^2}\right\}\frac{1  }{\sqrt{\frac{F_0}{F_0-1}} + 1}\times \nonumber \\
&&\left[  \left(\sqrt{\frac{F_0}{F_0-1}}\right)\left( \frac{\beta}{6\alpha}-1\right)-1  \right] .\label{murano116}
\end{eqnarray}
Integrating the expression of $\dot{\phi}_{ Dark,higher}$ with respect to the time $t$, we easily derive the following relation for $\phi_{ Dark,higher}$:
\begin{eqnarray}
\phi_{ Dark,higher} =2\left(\sqrt{\frac{F_0}{F_0-1}}\right)^{-1/4}\sqrt{\frac{2\alpha}{\beta}}\cdot\ln (t+C_3) +constant 
\end{eqnarray}
In the limiting case of $C_3=0$, we obtain:
\begin{eqnarray}
F_{ Dark,higher,lim} &=& \sqrt{F_0\left( F_0-1  \right)}\left(\frac{8\alpha}{\beta }\right)\left( \frac{1}{t^2}  \right),\label{murano114} \\
\phi_{ Dark,higher,lim} &=&2\left(\sqrt{\frac{F_0}{F_0-1}}\right)^{-1/4}\sqrt{\frac{2\alpha}{\beta}}\cdot\ln t+constant \\
V_{ Dark,higher,lim} &=& -\left\{ \frac{48\alpha^2}{\left( \beta t\right)^2}\right\}\frac{1  }{\sqrt{\frac{F_0}{F_0-1}} + 1}\times \nonumber \\
&&\left[  \left(\sqrt{\frac{F_0}{F_0-1}}\right)\left( \frac{\beta}{6\alpha}-1\right)-1  \right] .\label{murano116}
\end{eqnarray}

For the non-interacting NHDE model, we obtain the following result:
\begin{eqnarray}
F_{NHDE } &=& \sqrt{F_0\left( F_0-1  \right)}\cdot\rho_{D,NHDE}\left( 1+\omega_{D,NHDE}\right),\label{} \\
\dot{\phi}_{NHDE} &=&\left(\sqrt{\frac{F_0}{F_0-1}}\right)^{-1/4} \sqrt{ \rho_{D,NHDE}\left( 1+\omega_{D,NHDE}\right)  }\label{} \\
V_{NHDE} &=& -\frac{\rho_{D,NHDE}  }{\sqrt{\frac{F_0}{F_0-1}} + 1} \left[  \left(\sqrt{\frac{F_0}{F_0-1}}\right)\omega_{D,NHDE}-1  \right],\label{}
\end{eqnarray}
Instead, for the interacting NHDE model, we obtain the following result:
\begin{eqnarray}
F_{ NHDE,I} &=& \sqrt{F_0\left( F_0-1  \right)}\cdot\rho_{D,NHDE,I,1}\left( 1+\omega_{D,NHDE,I}\right),\label{} \\
\dot{\phi}_{NHDE,I} &=&\left(\sqrt{\frac{F_0}{F_0-1}}\right)^{-1/4} \sqrt{ \rho_{D,NHDE,I}\left( 1+\omega_{D,NHDE,I}\right)  }\label{} \\
V_{NHDE,I} &=& -\frac{\rho_{D,NHDE,I}  }{\sqrt{\frac{F_0}{F_0-1}} + 1} \left[  \left(\sqrt{\frac{F_0}{F_0-1}}\right)\omega_{D,NHDE,I}-1  \right],\label{}
\end{eqnarray}
In the limiting case of a flat Dark Dominated Universe,we obtain the following results:
\begin{eqnarray}
F_{ Dark,NHDE} &=&\frac{  2\lambda\sqrt{F_0\left( F_0-1  \right)}}{(\mu - 1) (t + C)^2},\label{murano114} \\
\dot{\phi}_{ Dark,NHDE} &=&  \left(\sqrt{\frac{F_0}{F_0-1}}\right)^{-1/4}\sqrt{\frac{ 2\lambda}{\mu - 1 }} \cdot \frac{1}{t+C},\label{giovanniboh} \\
V_{ Dark,NHDE} &=& \frac{1  }{\sqrt{\frac{F_0}{F_0-1}} + 1} \cdot \frac{3 \lambda^2}{(\mu - 1)^2 (t + C)^2}\times \nonumber \\
&&  \left[  \left(\sqrt{\frac{F_0}{F_0-1}}\right)\cdot \left(\frac{3\lambda-2\mu+2}{3\lambda}\right)+1  \right]\label{murano116}
\end{eqnarray}
Integrating the expression of $\dot{\phi}_{ Dark,NHDE}$, we easily derive the following relation for $\phi_{ Dark,NHDE}$:
\begin{eqnarray}
\phi_{ Dark,NHDE} =\left(\sqrt{\frac{F_0}{F_0-1}}\right)^{-1/4}\sqrt{\frac{ 2\lambda}{\mu - 1 }}\ln \left(t+C  \right)+const
\end{eqnarray}
In the limiting case of the Ricci scale, i.e. for $\mu=2$ and $\lambda =1$, we obtain:
\begin{align}
F_{\rm Dark,NHDE,Ricci} &= \frac{2 \sqrt{F_0(F_0-1)}}{(t+C)^2}, \\
\phi_{\rm Dark,NHDE,Ricci} &= \left(\sqrt{\frac{F_0}{F_0-1}}\right)^{-1/4} \sqrt{2} \, \ln(t+C) + \text{const}, \\
V_{\rm Dark,NHDE,Ricci} &= \frac{\sqrt{\frac{F_0}{F_0-1}} + 3}{\sqrt{\frac{F_0}{F_0-1}} + 1}\frac{1}{(t+C)^2}
\end{align}

In the limiting case of a $C=0$, we obtain:
\begin{eqnarray}
F_{ Dark,NHDE,lim} &=&\frac{  2\lambda\sqrt{F_0\left( F_0-1  \right)}}{(\mu - 1) t ^2},\label{murano114} \\
\dot{\phi}_{ Dark,NHDE,lim} &=&  \left(\sqrt{\frac{F_0}{F_0-1}}\right)^{-1/4}\sqrt{\frac{ 2\lambda}{(\mu - 1) }} \cdot \frac{1}{t},\label{giovanniboh} \\
V_{ Dark,NHDE,lim} &=& \frac{1  }{\sqrt{\frac{F_0}{F_0-1}} + 1} \cdot \frac{3 \lambda^2}{(\mu - 1)^2 t ^2}\times \nonumber \\
&&  \left[  \left(\sqrt{\frac{F_0}{F_0-1}}\right)\cdot \frac{3\lambda-2\mu+2}{3\lambda}+1  \right]\label{murano116}
\end{eqnarray}
Integrating Eq. (\ref{giovanniboh}), we easily derive the following relation for $\phi$:
\begin{eqnarray}
\phi_{ Dark,lim} =\left(\sqrt{\frac{F_0}{F_0-1}}\right)^{-1/4}\sqrt{\frac{ 2\lambda}{(\mu - 1) }}\ln t + const
\end{eqnarray}
In the limiting case of the Ricci scale, i.e. for $\mu=2$ and $\lambda =1$, and $C=0$, we obtain:
\begin{eqnarray}
F_{\text{Dark,Ricci,NHDE,lim}} &=& \frac{2\sqrt{F_0(F_0 - 1)}}{t^2}, \label{eq:F_Dark_lim} \\
\phi_{\text{Dark,Ricci,NHDE,lim}} &=& \left( \sqrt{ \dfrac{F_0}{F_0 - 1} } \right)^{-1/4} \cdot \sqrt{2} \cdot \ln t + \text{const}. \label{eq:phi_Dark_lim}\\
V_{\text{Dark,Ricci,NHDE,lim}} &=&  \frac{  \sqrt{\frac{F_0}{F_0 - 1}} + 3 }{ \sqrt{\frac{F_0}{F_0 - 1}} + 1\ }  \cdot \frac{1}{t^2} \label{eq:V_Dark_lim}
\end{eqnarray}
In the limiting case corresponding to the values found by Wang \& Xu, we obtain:
\begin{eqnarray}
F_{ Dark,NHDE,WX} &=&-\frac{  6.431\sqrt{F_0\left( F_0-1  \right)}}{ (t + C)^2},\label{murano114} \\
V_{ Dark,NHDE,WX} &=& \frac{1  }{\sqrt{\frac{F_0}{F_0-1}} + 1} \cdot \frac{31.020}{ (t + C)^2}\times \nonumber \\
&&  \left[  1.207\sqrt{\frac{F_0}{F_0-1}}+1  \right]\label{murano116}
\end{eqnarray}
We also obtain that $\dot{\phi}_{ Dark,NHDE}$ is not a real function.

\subsection{Yang-Mills (YM) Model}

We now study the Yang-Mills (YM) model. Recent studies suggest that the Yang-Mills field \cite{ym1,ym9-3,ym9-5,ym9-7} can serve as a viable candidate to describe the nature of dark energy (DE). Two main reasons motivate the consideration of the YM field as a source of DE. First, in contrast to standard scalar field models, the connection of the YM field to fundamental particle physics models is more firmly established. Second, the YM field allows for scenarios where the weak energy condition can be violated, enabling interesting cosmological dynamics.

The YM field we consider exhibits several notable features. It forms an essential cornerstone of particle physics models, where interactions are mediated by gauge bosons, and thus can be naturally incorporated into unified particle physics frameworks. Moreover, the equation of state (EoS) parameter for the effective Yang-Mills condensate (YMC) differs significantly from that of ordinary matter and scalar fields. The YMC can naturally realize states with \(-1 < \omega < 0\) as well as \(\omega < -1\), allowing for phantom-like behavior.

In the effective YMC dark energy model, the effective Yang-Mills field Lagrangian density \( L_{YMC} \) is given by
\begin{eqnarray}
L_{YMC} = \frac{bF}{2} \left( \ln \left| \frac{F}{\kappa^2} \right| - 1 \right), \label{murano117}
\end{eqnarray}
where \(\kappa\) is the renormalization scale with the dimension of squared mass, and \(F\) plays the role of the order parameter of the YMC. The quantity \(F\) is defined as
\begin{eqnarray}
F = -\frac{1}{2} F_{\mu \nu}^\alpha F^{\alpha \mu \nu} = E^2 - B^2, \label{murano118}
\end{eqnarray}
where \(F_{\mu \nu}^\alpha\) is the Yang-Mills field strength tensor, and \(E\) and \(B\) denote the effective electric and magnetic components of the field, respectively.
For the  pure electric case we have, $B = 0$ which implies that $F = E^2$.\\

Moreover, the parameter \(b\) is the Callan-Symanzik coefficient \cite{ym18,ym18-1}, which for the gauge group \(SU(N)\) is given by
\begin{eqnarray}
b = \frac{11N - 2N_f}{24\pi^2}, \label{murano119}
\end{eqnarray}
where \(N_f\) denotes the number of quark flavors.

For the gauge group \(SU(2)\), the coefficient takes the value \(b = 2 \times \frac{11}{24\pi^2}\) when fermionic contributions are neglected, and \(b = 2 \times \frac{5}{24\pi^2}\) when the number of quark flavors is \(N_f = 6\). In the case of \(SU(3)\), the effective Lagrangian given in Eq. \eqref{murano117} provides a phenomenological description of asymptotic freedom for quarks inside hadrons \cite{ym21,ym21-1}.

It is important to emphasize that the \(SU(2)\) Yang-Mills field introduced here as a model for cosmic dark energy should not be directly identified with the QCD gluon fields or the electroweak gauge bosons such as \(Z^0\) and \(W^{\pm}\). The Yang-Mills condensate (YMC) has an energy scale characterized by the parameter \(\kappa^{1/2} \approx 10^{-3} \, \text{eV}\), which is much smaller than the corresponding scales of QCD and the electroweak interactions.

The form of the effective Lagrangian in Eq. \eqref{murano117} can be understood as a 1-loop quantum corrected Lagrangian \cite{ym21,ym21-1}. The classical \(SU(N)\) Yang-Mills Lagrangian is given by
\begin{eqnarray}
L = \frac{1}{2g_0^2} F, \label{murano120}
\end{eqnarray}
where \(g_0\) is the bare coupling constant. Including 1-loop quantum corrections leads to a running coupling constant \(g\) that replaces the bare coupling as
\begin{eqnarray}
g_0^2 \rightarrow g^2 = \frac{4 \times 12 \pi^2}{11N \ln\left(\frac{k}{k_0^2}\right)} = \frac{2}{b \ln\left(\frac{k}{k_0^2}\right)}, \label{murano121}
\end{eqnarray}
where \(k\) is the momentum transfer and \(k_0\) is the corresponding energy scale.

To construct an effective theory, the momentum scale \(k^2\) is replaced by the field strength \(F\) via the substitution
\begin{eqnarray}
\ln\left(\frac{k}{k_0^2}\right) \rightarrow 2 \ln \left| \frac{F}{\kappa^2 e} \right| = 2 \ln \left| \frac{F}{\kappa^2} - 1 \right|. \label{murano122}
\end{eqnarray}
This replacement recovers the effective Lagrangian expression shown in Eq. \eqref{murano117}.

Some of the interesting characteristics of this effective Yang-Mills  (YM) action include asymptotic freedom, gauge invariance, Lorentz invariance and  the correct trace anomaly   \cite{ym16}.

Due to the logarithmic dependence on the field strength, the Lagrangian of the YMC model resembles the Coleman-Weinberg scalar effective potential \cite{ym19} and the Parker-Raval effective gravity Lagrangian \cite{ym20}.

We emphasize that the renormalization scale \(\kappa\) is the only free parameter of this effective YM model. In contrast to scalar field dark energy models, the YM Lagrangian is completely fixed by quantum corrections up to one-loop order, and its functional form cannot be arbitrarily adjusted.

From the effective Lagrangian given in Eq. \eqref{murano117}, the energy density \(\rho_y\) and pressure \(p_y\) of the YMC can be derived as
\begin{eqnarray}
\rho_y &=& \frac{\epsilon E^2}{2} + \frac{b E^2}{2}, \label{murano123}\\
p_y &=& \frac{\epsilon E^2}{6} - \frac{b E^2}{2}, \label{murano124}
\end{eqnarray}
where \(\epsilon\) is the dielectric constant of the YM, defined by
\begin{eqnarray}
\epsilon = 2 \frac{\partial L_{eff}}{\partial F} = b \ln \left| \frac{F}{\kappa^2} \right|. \label{murano125}
\end{eqnarray}

Equations \eqref{murano123} and \eqref{murano124} can be alternatively rewritten as
\begin{eqnarray}
\rho_y &=& \frac{1}{2} b \kappa^2 (y+1) e^{y}, \label{murano126}\\
p_y &=& \frac{1}{6} b \kappa^2 (y-3) e^{y}, \label{murano127}
\end{eqnarray}
or equivalently as
\begin{eqnarray}
\rho_y &=& \frac{1}{2} (y+1) b E^2, \label{murano128} \\
p_y &=& \frac{1}{6} (y-3) b E^2, \label{murano129}
\end{eqnarray}
where the dimensionless parameter \(y\) is defined as
\begin{eqnarray}
y = \frac{\epsilon}{b} = \ln \left| \frac{F}{\kappa^2} \right| = \ln \left| \frac{E^2}{\kappa^2} \right|. \label{defiy}
\end{eqnarray}

Using the expressions for \(\rho_y\) and \(p_y\) in Eqs. \eqref{murano126} and \eqref{murano127} (or equivalently in Eqs. \eqref{murano128} and \eqref{murano129}), the equation of state (EoS) parameter \(\omega_y\) of the YMC model is given by
\begin{eqnarray}
\omega_y = \frac{p_y}{\rho_y} = \frac{y - 3}{3(y + 1)}. \label{omegay}
\end{eqnarray}

At the critical point where \(\epsilon = 0\) (i.e., \(y=0\)), we find \(\omega_y = -1\), corresponding to a de Sitter expansion of the Universe. Near this point, if \(\epsilon < 0\), then \(\omega_y < -1\), while if \(\epsilon > 0\), then \(\omega_y > -1\). Thus, as previously stated, the YMC model naturally realizes the range of EoS values \(0 > \omega_y > -1\) as well as \(\omega_y < -1\).

The expression of \(\omega_y\) given in Eq. \eqref{omegay} can be inverted to yield the following expression for \(y\):
\begin{eqnarray}
y = - \frac{3(\omega_y + 1)}{3\omega_y - 1}. \label{murano130}
\end{eqnarray}

In order to ensure that the energy density \(\rho_y\) is positive in any physically viable model, the parameter \(y\) must satisfy \(y > 1\). This condition implies
\begin{eqnarray}
F > \frac{\kappa^2}{e} \approx 0.368\, \kappa^2.
\end{eqnarray}

Before considering a particular cosmological scenario, it is instructive to study \(\omega_y\) as a function of the condensate strength \(F\). The YMC model exhibits an equation of state characteristic of radiation with
\begin{eqnarray}
p_y = \frac{1}{2} \rho_y
\end{eqnarray}
and the corresponding EoS parameter
\begin{eqnarray}
\omega_y = \frac{1}{2}
\end{eqnarray}
for large values of the dielectric constant, i.e. \(\epsilon \gg b\) (which implies \(F \gg \kappa^2\)).

On the other hand, at the critical point where \(\epsilon = 0\) (i.e. \(F = \kappa^2\)), the YMC behaves like a cosmological constant with
\begin{eqnarray}
\omega_y &=& -1,\\
p_y &=& -\rho_y.
\end{eqnarray}
This critical case corresponds to the YMC energy density taking the value
\begin{eqnarray}
\rho_y = \frac{1}{2} b \kappa^2,
\end{eqnarray}
which acts as the critical energy density \cite{ym1}.

This interesting property of the YMC equation of state — smoothly evolving from \(\omega_y = \frac{1}{3}\) at high energies (\(F \gg \kappa^2\)) to \(\omega_y = -1\) at low energies (\(F = \kappa^2\)) — allows the existence of scaling solutions for the dark energy component in this model \cite{ym10,ym10-1}. Moreover, this transition is smooth, since \(\omega_y\) is a continuous function of \(y\) over the range \(( -1, \infty )\).

We now examine whether the EoS parameter \(\omega_y\) can cross the phantom divide \(-1\). From Eq. \eqref{omegay}, \(\omega_y\) depends solely on the condensate strength \(F\). In principle, \(\omega_y < -1\) can be achieved when
\begin{eqnarray}
F < \kappa^2,
\end{eqnarray}
and this crossing is also smooth with respect to \(F\).

However, when the YMC is incorporated into a cosmological model as a dark energy component, alongside matter and radiation, the value of \(F\) is no longer arbitrary but evolves dynamically as a function of cosmic time \(t\).

Specifically, if the YMC does not decay into matter or radiation, \(\omega_y\) approaches \(-1\) asymptotically but never crosses it. In contrast, if the YMC decays into matter and/or radiation, \(\omega_y\) can cross below \(-1\). Depending on the coupling strength, \(\omega_y\) settles asymptotically to a value around \(-1.17\).

An important advantage of this lower region \(\omega_y < -1\) is that all physical quantities \(\rho_y\), \(p_y\), and \(\omega_y\) remain smooth and free from finite-time singularities, unlike some classes of scalar field models.

Equating the EoS of the YM $\omega_y$ with the EoS parameters of the model we are studying, we can write $y$ as follows.
\begin{eqnarray}
y = - \frac{3\left(\omega_D+1\right)}{3\omega_D-1}.\label{murano131}
\end{eqnarray}
therefore, using the results we obtained, we can write:
\begin{eqnarray}
y_{higher,1} &=& - \frac{3\left(\omega_{D,higher,1}+1\right)}{3\omega_{D,higher,1}-1} \label{}\\
y_{higher,2} &=& - \frac{3\left(\omega_{D,higher,2}+1\right)}{3\omega_{D,higher,2}-1} \label{}\\
y_{higher,3} &=& - \frac{3\left(\omega_{D,higher,3}+1\right)}{3\omega_{D,higher,3}-1} \label{}
\end{eqnarray}
For the interacting case, we obtain:
\begin{eqnarray}
y_{higher,I,1} &=& - \frac{3\left(\omega_{D,higher,I,1}+1\right)}{3\omega_{D,higher,I,1}-1} \label{}\\
y_{higher,I,2} &=& - \frac{3\left(\omega_{D,higher,I,2}+1\right)}{3\omega_{D,higher,I,2}-1} \label{}\\
y_{higher,I,3} &=& - \frac{3\left(\omega_{D,higher,I,3}+1\right)}{3\omega_{D,higher,I,3}-1} \label{}
\end{eqnarray}

In the limiting case of a flat Dark Dominated Universe, we can write:
\begin{eqnarray}
y_{ Dark,higher} =- \frac{3\beta}{3\beta -8\alpha} .\label{murano134}
\end{eqnarray}
In order to have $y>1$, we derive from Eq. (\ref{murano134}) 
\begin{eqnarray}
 \quad \frac{4}{3} \alpha < \beta < \frac{8}{3} \alpha
\end{eqnarray}

For the non-interacting NHDE model, we can write:
\begin{eqnarray}
y_{NHDE} &=& - \frac{3\left(\omega_{D,NHDE}+1\right)}{3\omega_{D,NHDE}-1} \label{}
\end{eqnarray}
Instead, for the interacting NHDE model, we can write:
\begin{eqnarray}
y_{NHDE,I} &=& - \frac{3\left(\omega_{D,NHDE,I}+1\right)}{3\omega_{D,NHDE,I}-1} \label{}
\end{eqnarray}

In the limiting case of a flat Dark Dominated Universe, we can write:
\begin{eqnarray}
y_{ Dark,NHDE} =\frac{\mu-1}{2\lambda-\mu+1}
\end{eqnarray}
We must have that $2\lambda-\mu+1\neq 0$, i.e. $\mu \neq 2\lambda +1$.\\
In the limiting case corresponding to $\lambda =1$ e $\mu=2$, i.e. the case corresponding to the Ricci scale, we obtain:
\begin{eqnarray}
y_{ Dark,NHDE,Ricci} =1
\end{eqnarray}
which is at the limit of the condition $y>1$.\\
In order to have strictly $y>1$ we must have $\lambda + 1 < \mu < 2\lambda + 1$.

In the limiting case corresponding to the values found by Wang \& Xu, we obtain
\begin{eqnarray}
    y_{Dark,NHDE,WX} \approx -0.135
\end{eqnarray}
which does not satisfy the condition $y>1$.

\subsection{Non Linear Electro-Dynamics (NLED) Model}

We now consider the last model we have chosen to study, i.e., the Non Linear Electro-Dynamics (NLED) Model. Recently, a new approach has been considered to avoid the cosmic singularity through a nonlinear extension of Maxwell's electromagnetic theory. Exact solutions of Einstein's field equations coupled with Non Linear Electro-Dynamics (NLED) reveal acceptable nonlinear effects in strong gravitational and magnetic fields. Moreover, General Relativity (GR) coupled with NLED effects can explain primordial inflation.

The Lagrangian density \(L_M\) for free fields in Maxwell electrodynamics can be written as follows \cite{ele1,ele2}:
\begin{eqnarray}
L_M = - \frac{F^{\mu \nu}F_{\mu \nu}}{4\mu}, \label{murano135}
\end{eqnarray}
where \(F^{\mu \nu}\) is the electromagnetic field strength tensor and \(\mu\) is the magnetic permeability.

We now consider the generalization of the Maxwell electromagnetic Lagrangian up to second order terms of the fields as:
\begin{eqnarray}
L = -\frac{F}{4\mu_0} + \omega F^2 + \eta F^{*2}, \label{murano136}
\end{eqnarray}
where \(\omega\) and \(\eta\) are two arbitrary constants, and \(F^*\) is defined as
\begin{eqnarray}
F^* = F_{\mu \nu}^* F^{\mu \nu}, \label{murano137}
\end{eqnarray}
with \(F_{\mu \nu}^*\) being the dual of \(F_{\mu \nu}\).

We consider the particular case when the homogeneous electric field \(E\) in plasma rapidly decays due to electric current of charged particles, so that the squared magnetic field dominates over the electric field, i.e., \(E^2 \approx 0\), hence
\begin{eqnarray}
F = 2B^2.
\end{eqnarray}
Thus, \(F\) depends only on the magnetic field \(B\).

The pressure \(p_{NLED}\) and energy density \(\rho_{NLED}\) of the Nonlinear Electrodynamics Field are given by
\begin{align}
p_{NLED} &= \frac{B^2}{6\mu}\left(1 - 40 \mu \omega B^2 \right), \label{murano138}\\
\rho_{NLED} &= \frac{B^2}{2\mu}\left(1 - 8 \mu \omega B^2 \right). \label{murano139}
\end{align}

The weak energy condition \(\rho_{NLED} > 0\) is satisfied for
\begin{eqnarray}
B < \frac{1}{2\sqrt{2 \mu \omega}},
\end{eqnarray}
and the pressure \(p_{NLED}\) becomes negative for
\begin{eqnarray}
B > \frac{1}{2\sqrt{10 \mu \omega}}.
\end{eqnarray}

The magnetic field can generate dark energy if the strong energy condition is violated, i.e., if \(\rho_B + 3 p_B < 0\), which happens when
\begin{eqnarray}
B > \frac{1}{2\sqrt{6 \mu \omega}}.
\end{eqnarray}

The equation of state (EoS) parameter \(\omega_{NLED}\) for the Nonlinear Electrodynamics Field is
\begin{eqnarray}
\omega_{NLED} = \frac{p_{NLED}}{\rho_{NLED}} = \frac{1 - 40 \mu \omega B^2}{3 \left(1 - 8 \mu \omega B^2\right)}, \label{murano140}
\end{eqnarray}
which leads to the following expression for \(B^2\):
\begin{eqnarray}
B^2 = \frac{1 - 3 \omega_{NLED}}{8 \mu \omega \left(5 - 3 \omega_{NLED}\right)}. \label{murano141}
\end{eqnarray}

By making a correspondence between the EoS parameter of the NLED model and the EoS of the dark energy model under study, we get
\begin{eqnarray}
B^2 = \frac{1 - 3 \omega_D}{8 \mu \omega \left(5 - 3 \omega_D\right)}. \label{murano142}
\end{eqnarray}

For the model we are studying, we obtain the following expressions for $B^2$:
\begin{eqnarray}
B^2_{higher, 1} &=& \frac{1-3\omega_{D,higher,1}}{8\mu \omega\left(5 -3\omega_{D,higher,1}   \right) } \label{}\\
B_{ higher,2} &=& \frac{1-3\omega_{D,higher,2}}{8\mu \omega\left(5 -3\omega_{D,higher,2}   \right) } \label{}\\
B^2_{ higher,3} &=& \frac{1-3\omega_{D,higher,3}}{8\mu \omega\left(5 -3\omega_{D,higher,3}   \right) } \label{}
\end{eqnarray}
For the interacting case, we obtain the following expressions for $B^2$
\begin{eqnarray}
B^2_{higher,I,1} &=& \frac{1-3\omega_{D,higher,I,1}}{8\mu \omega\left(5 -3\omega_{D,higher,I,1}   \right) } \label{}\\
B_{ higher,I,2} &=& \frac{1-3\omega_{D,higher,I,2}}{8\mu \omega\left(5 -3\omega_{D,higher,I,2}   \right) } \label{}\\
B^2_{ higher,I,1} &=& \frac{1-3\omega_{D,higher,I,3}}{8\mu \omega\left(5 -3\omega_{D,higher,I,3}   \right) } \label{}
\end{eqnarray}

Moreover, in the limiting case of a flat Dark Dominated Universe for the higher model, we obtain:
\begin{eqnarray}
B^2_{ Dark,higher}  = \frac{4-\beta/(2\alpha)}{8\mu \omega\left(8- \beta/(2\alpha)  \right)}.\label{murano145}
\end{eqnarray}
We have that $\beta/(2\alpha) \neq 8 \rightarrow \beta \neq 16 \alpha$ in order to avoid singularities in the expression of $B^2_{ Dark}$.

For the non-interacting NHDE model, we obtain obtain the following expressions for $B^2$:
\begin{eqnarray}
B^2_{NHDE} &=& \frac{1-3\omega_{D,NHDE}}{8\mu \omega\left(5 -3\omega_{D,NHDE}   \right) } \label{}
\end{eqnarray}
Instead, for the interacting NHDE model, we obtain obtain the following expressions for $B^2$:
\begin{eqnarray}
B^2_{NHDE, I} &=& \frac{1-3\omega_{D,NHDE,I}}{8\mu \omega\left(5 -3\omega_{D,NHDE,I}   \right) } \label{}
\end{eqnarray}
Moreover, in the limiting case of a flat Dark Dominated Universe, we obtain:
\begin{eqnarray}
B^2_{ Dark,NHDE}  = \frac{\mu_{GO}-\lambda-1}{8\mu \omega(\lambda+\mu_{GO}-1)}.\label{murano145}
\end{eqnarray}
In this case, we indicated with $\mu_{GO}$ the parameter of the Granda-Oliveros model in order to avoid misunderstandings with the other parameter $\mu$ of this scalar field model. 
Therefore we must have that $\mu_{GO} \neq1- \lambda $.\\
In the limiting case corresponding to $\lambda =1$ e $\mu=2$, i.e. the case corresponding to the Ricci scale, we obtain:
\begin{eqnarray}
B^2_{ Dark,NHDE,Ricci}  = 0.\label{murano145}
\end{eqnarray}
which is a value lower than what is expected to produce DE.\\ 
In the limiting case corresponding to the values found by Wang \& Xu, we obtain the following: 
\begin{eqnarray}
B^2_{Dark,NHDE,WX} \approx -\frac{0.238}{\mu \omega}
\end{eqnarray}
which is a value lower than what is expected to produce DE.\\

\section{Conclusions}
In this paper, we have explored the cosmological implications of two different DE models.\\
The first model considered is a Generalized Holographic Dark Energy (GHDE) model proposed by Chen \& Jing \cite{modelhigher}, which is a function of the Hubble parameter $H$ and its first and second derivatives with respect to the cosmic time $t$ as $\rho_{higher} = 3\left( \alpha \ddot{H}H^{-1}+\beta\dot{H}+\gamma H^2   \right)$. The model is characterized by three constant parameters $\alpha$, $\beta$, and $\gamma$.  Moreover, this model can be considered as a generalization of other DE models, i.e. the Ricci DE model and the DE model with Granda-Oliveros cut-off. In fact, in the limiting case of $\alpha=0$, we obtain the DE model with Granda-Oliveros cut-off, while for $\alpha=0$, $\beta =1$ and $\gamma =2$ we obtain the Ricci DE model for a flat Universe.\\
The second model we have analyzed is the New Holographic Dark Energy (NHDE) model, characterized by an energy density depending on both the Hubble parameter squared and its time derivative given by  $\rho_{NHDE}=3(\mu H^2+\lambda\dot{H})$, where $\mu$ and $\lambda$ are two positive constant parameters.   This formulation, originally introduced by Granda \& Oliveros, provides a generalization of the standard holographic approach  and naturally reduces to the Ricci DE case in the limiting case corresponding to $\mu=2$ and $\lambda=1$  in a spatially flat geometry. 

We derived for both models the expressions of the reduced Hubble parameter $h^2$, the equation of state (EoS) parameter of DE $\omega_D$, the pressure of DE $p_D$ and the deceleration parameter $q$ for the case corresponding to non-interacting and interacting DE and DM under various assumptions on the integration constants.  \\
Finally, we established a correspondence between the DE models considered and some scalar fields like the Generalized Chaplygin Gas (GCG), the Modified Chaplygin Gas (MCG), the Modified Variable Chaplygin Gas (MVCG), the Viscous Generalized Chaplygin Gas (VGCG), the Dirac-Born-Infeld (DBI),  the Yang-Mills (YM) and the Non Linear Electro-Dynamics (NLED) ones. These correspondences are important to understand how various candidates of DE are mutually related to each other. Moreover,  we  calculated the quantities we obtained for the limiting case of the flat Dark Dominated Universe, i.e. $\Omega_D =1$ and $\Omega_m= \Omega_k = 0$.


\begin{thebibliography}{999}

\bibitem{cmb1} C. L. Bennett, et al., Astrophys. J. \textbf{148}, 1 (2003).
\bibitem{cmb2} D.N. Spergel, et al., Astrophys. J. Suppl. Ser. \textbf{148}, 175 (2003).
\bibitem{sn2} S. Perlmutter et al., Astrophys. J. \textbf{517}, 565 (1999).
\bibitem{sn4} P. de Bernardis et al, Nature \textbf{404}, 955 (2000).
\bibitem{sds1} M. Tegmark et al., Phys. Rev. D \textbf{69}, 103501 (2004).
\bibitem{sds3} K. Abazajian et al., Astron. J. \textbf{129}, 1755 (2005).
\bibitem{sds4} J.K. Adelman-McCarthy, et al., Astrophys. J. Suppl. Ser. \textbf{175}, 297 (2008).
\bibitem{planck} Planck Collaboration, P.A.R. Ade et al.\ 2013, arXiv:1303.5076
\bibitem{xray} S.W. Allen et al., Mon. Not. Roy. Astron. Soc. \textbf{353}, 457 (2004).
\bibitem{cosm1} E.J. Copeland, M. Sami, S. Tsujikawa, Int. J. Mod. Phys. D \textbf{15}, 1753 (2006).
\bibitem{cosm3} V. Sahni, A. Starobinsky, Int. J. Mod. Phy. D \textbf{9}, 373 (2000).
\bibitem{cosm4} P.J.E. Peebles, B. Ratra, Reviews of Modern Physics, \textbf{75}, 559 (2003).
\bibitem{cosm5} T. Padmanabhan, Phys. Rep. \textbf{380}, 235 (2003).
\bibitem{cosm6} S. D. Odintsov, D. Saez-Chillon Gomez, G. S. Sharov, Eur. Phys. J. C \textbf{77} 862 (2017).
\bibitem{twothirds} H.V. Peiris, et al., Astrophys. J. Suppl. Ser. \textbf{148}, 213 (2003).
\bibitem{quint1} B. Ratra, B., Peebles, P.J.E., Phys. Rev. D, \textbf{37}, 3406 (1988).
\bibitem{quint4} P.J.E. Peebles, B. Ratra, Astrophys. J. Lett., \textbf{325}, L17 (1988).
\bibitem{quint5} C. Wetterich, Nucl. Phys. B \textbf{302}, 668 (1988).
\bibitem{quint6} R.R. Caldwell, R. Dave, P.J. Steinhardt, Phys. Rev. Lett. \textbf{80}, 1582 (1998).
\bibitem{quint7} A. Pasqua, Astrophys. Space Sci. \textbf{346}, 531 (2013). DOI: https://doi.org/10.1007/s10509-013-1464-8
\bibitem{kess2} C. Armendariz-Picon, V. Mukhanov,  P.J. Steinhardt, Phys. Rev. Lett. \textbf{85}, 4438 (2000).
\bibitem{kess3} T. Chiba, T. Okabe, M. Yamaguchi, Phys. Rev. D \textbf{62}, 023511 (2000).
\bibitem{kess4} C.A. Picon, T. Damour, V. Mukhanov, Phys. Lett. B \textbf{458}, 209 (1999).
\bibitem{kess5} A. Pasqua, A. Khodam-Mohammadi, M. Jamil, R. Myrzakulov, Astrophys Space Sci \textbf{340}, 199-208 (2012). DOI: https://doi.org/10.1007/s10509-012-1031-8
\bibitem{tac1} A. Sen, Journal of High Energy Physics \textbf{4}, 48 (2002).
\bibitem{tac2} T. Padmanabhan, Phys. Rev. D \textbf{66}, 021301 (2002).
\bibitem{tac4} A. Pasqua, M. Jamil, R. Myrzakulov B.  Majeed,  Physica Scripta \textbf{86}, 045004 (2012). DOI: 10.1088/0031-8949/86/04/045004
\bibitem{pha2} S. Nojiri, S.D. Odintsov, Phys. Lett. B \textbf{565}, 1 (2003).
\bibitem{pha5} L.P. Chimento, R. Lazkoz, Phys. Rev. Lett. \textbf{91}, 211301 (2003).
\bibitem{pha6} B. Boisseau, G. Esposito-Farese, D. Polarski, Alexei A. Starobinsky,Phys. Rev. Lett. \textbf{85}, 2236 (2000).
\bibitem{dil1} N. Arkani-Hamed, P. Creminelli, S. Mukohyama, M. Zaldarriaga, J. Cosmol. Astropart. Phys. \textbf{4}, 1 (2004).
\bibitem{dil2} M. Gasperini, F. Piazza, G. Veneziano, Phys. Rev. D \textbf{65}, 023508 (2002).
\bibitem{dil3} F. Piazza, S. Tsujikawa, J. Cosmol. Astropart. Phys. \textbf{7}, 4 (2004).
\bibitem{qui2} E. Elizalde, S. Nojiri, S.D. Odintsov, Phys. Rev. D \textbf{70}, 043539 (2004).
\bibitem{qui3} S. Nojiri, S.D. Odintsov, S. Tsujikawa, Phys. Rev. D \textbf{71}, 063004 (2005).
\bibitem{qui8} W. Zhao, Y. Zhang, Phys. Rev. D \textbf{73}, 123509 (2006).
\bibitem{qui12} M.R. Setare, E.N. Saridakis, J. Cosmol. Astropart. Phys. \textbf{0809}, 026 (2008).
\bibitem{cgas2} M.C. Bento, O. Bertolami, A.A. Sen, Phys. Rev. D \textbf{66}, 043507 (2002).
\bibitem{cgas3} M.R. Setare, European Physical Journal C \textbf{52}, 689 (2007).
\bibitem{ade1} H. Wei, R. G. Cai, Physics Letters B \textbf{660}, 113 (2008).
\bibitem{ade2} R.G. Cai, Phys. Lett. B \textbf{657}, 228 (2007).
\bibitem{holo1} W. Fischler, L. Susskind, (arXiv:hep-th/9806039).
\bibitem{holo2} L. Susskind, J. Math. Phys. \textbf{36}, 6377 (1995).
\bibitem{holo5} Q.G. Huang, M. Li, J. Cosmol. Astropart. Phys. \textbf{8}, 13 (2004).
 \bibitem{li} M. Li, Phys. Lett. B \textbf{603}, 1 (2004).
\bibitem{nood2} M. Li, X.D. Li, S. Wang, Y. Wang, X. Zhang, J. Cosmol. Astropart. Phys. \textbf{12}, 014 (2009).
\bibitem{nood1} Y.S. Myung, arXiv:1005.2240.  
\bibitem{nood4} M. Khurshudyan, Astrophys. Space Sci. \textbf{361}, 7 (2016).   
\bibitem{nood6} S.I. Nojiri, S.D. Odintsov, V.K. Oikonomou, T. Paul, Phys. Rev. D \textbf{102}, 023540 (2020).
\bibitem{nood7} S. Nojiri, S.D. Odintsov, T. Paul, Symmetry \textbf{13}, 928 (2021). 

\bibitem{coh1} A. Cohen, D. Kaplan, A. Nelson, Phys. Rev. Lett. \textbf{82}, 4971 (1999).
\bibitem{gaoprimo} C. Gao, F. Wu, X. Chen, Y. G. Shen, Phys. Rev. D \textbf{79}, 043511 (2009).
\bibitem{go1} L.N. Granda,  International Journal of Modern Physics D \textbf{18}, 1749 (2009).
\bibitem{go4} L.N. Granda, A. Oliveros, Physics Letters B \textbf{671}, 199 (2009).

\bibitem{cons1} K. Enqvist, S. Hannestad, M. S. Sloth, J. Cosmol. Astropart. Phys. \textbf{2}, 4 (2005).
\bibitem{cons2} X. Zhang, F.Q. Wu, Phys. Rev. D \textbf{76}, 023502 (2007).
\bibitem{cons3} X. Zhang, F.Q. Wu, Phys. Rev. D \textbf{72}, 043524 (2005).
\bibitem{cons4} X. Zhang, Phys. Rev. D \textbf{79}, 103509 (2009).
\bibitem{cons6} S.M.R. Micheletti, J. Cosmol. Astropart. Phys. \textbf{4}, 9 (2010).
\bibitem{cons7} H.C. Kao, W.L. Lee, F.L. Lin, Phys. Rev. D \textbf{71}, 123518 (2005).
\bibitem{cons8} B. Feng, X. Wang,  X. Zhang, Physics Letters B \textbf{607}, 35 (2005).
\bibitem{cons9} J. Shen, B. Wang, E. Abdalla, R. K. Su, Physics Letters B \textbf{609}, 200 (2005).


\bibitem{hde1} S. Nojiri, S. D. Odintsov, E. N. Saridakis, R. Myrzakulov, Nuclear Physics B \textbf{950} 114850 (2020).
\bibitem{hde2} M.R. Setare, Physics Letters B \textbf{642}, 421 (2006).
\bibitem{hde7} M.R. Setare, European Physical Journal C \textbf{50}, 991 (2007).
\bibitem{hde9} S. Nojiri, S. D. Odintsov, E. N. Saridakis, Nuclear Physics B  \textbf{949}, 114790 (2019).
\bibitem{hde10} S. Nojiri, S. D. Odintsov, Eur. Phys. J. C \textbf{77} 528 (2017).
\bibitem{hde13} Y. Gong, Phys. Rev. D \textbf{70}, 064029 (2004).
\bibitem{hde17} M. Jamil, E.N. Saridakis, M.R. Setare, Physics Letters B \textbf{679}, 172 (2009).
\bibitem{hde18} M. Jamil, M.U. Farooq, M. A. Rashid, European Physical Journal C \textbf{61}, 471 (2009).
\bibitem{hde22} H.M. Sadjadi, M. Jamil, General Relativity and Gravitation \textbf{43}, 1759 (2011).
\bibitem{hde23} M.R. Setare, M. Jamil, J. Cosmol. Astropart. Phys. \textbf{2}, 10 (2010).
\bibitem{hde24} M.R. Setare, M. Jamil, General Relativity and Gravitation \textbf{43}, 293 (2011).
\bibitem{hde26} A. Sheykhi, Physics Letters B \textbf{681}, 205 (2009).
\bibitem{hde28} M.R. Setare, S. Shafei, J. Cosmol. Astropart. Phys. \textbf{9}, 11 (2006).
\bibitem{hde30} M.R. Setare, M. Jamil, Physics Letters B \textbf{690},1 (2010).
\bibitem{mioviscuous} A. Pasqua, arXiv:2508.21110v2
\bibitem{hde33} E. Elizalde, S. Nojiri, S.D. Odintsov, P. Wang, Phys. Rev. D \textbf{71}, 103504 (2005).
\bibitem{hde34} M.U. Farooq, M. Jamil, M. A. Rashid, Int. J. Theor. Phys. \textbf{49}, 2334 (2010).
\bibitem{hde35} X. Zhang, Phys. Rev. D \textbf{74}, 103505 (2006).
\bibitem{saridakis11}  M.R. Setare, E.N. Saridakis, Phys. Lett. B \textbf{671}, 331  (2009).
\bibitem{saridakis22} J. Lu, E.N. Saridakis, M.R. Setare, L. Xu,  J. Cosmol. Astropart. Phys. \textbf{03}, 031 (2010).
\bibitem{33} B. Wang, C.Y. Lin, D. Pav{\'o}n, E. Abdalla, Physics Letters B,\textbf{662}, 1 (2008).
\bibitem{15} K. Karami, J. Fehri, Physics Letters B \textbf{684}, 61 (2010).
\bibitem{30} A. Sheykhi, Classical and Quantum Gravity \textbf{27}, 025007 (2010).
\bibitem{42} H.M. Sadjadi,  J. Cosmol. Astropart. Phys. \textbf{02},  026    (2007).
\bibitem{41} K. Karami, J. Fehri, Int. J. Theor. Phys. \textbf{49},  1118  (2010).
\bibitem{miogup1} A. Pasqua, S. Chattopadhyay, \& Khomenko, I.  Int J Theor Phys \textbf{52}, 2496–2507 (2013). https://doi.org/10.1007/s10773-013-1537-z
\bibitem{saridakis2} E.N. Saridakis, J. Cosmol. Astropart. Phys. \textbf{0804}, 020 (2008).
\bibitem{modelhigher} S. Chen, J. Jing, Physics Letters B \textbf{679}, 144 (2009).

\bibitem{gohnde} L.N. Granda, A. Oliveros, Phys. Lett. B \textbf{669}, 275 (2008)

\bibitem{hnde1} M. Sharif, A. Jawad, J. C \textbf{72}, 2097 (2012)

\bibitem{hnde2} S. Wang, Y. Wang, M. Li, Phys. Rep. \textbf{696}, 1 (2017)

\bibitem{hnde3} Y.L. Bolotin, A. Kostenko, O.A. Lemets, D.A. Yerokhin, Int. J. Mod. Phys. D \textbf{24}, 1530007 (2015)


\bibitem{mio1} A. Pasqua, S. Chattopadhyay, R. Myrzakulov, Eur. Phys. J. Plus \textbf{131}, 408 (2016). DOI: https://doi.org/10.1140/epjp/i2016-16408-8

\bibitem{mio2} A. Pasqua, S. Chattopadhyay, Astrophys Space Sci \textbf{348}, 541-551 (2013). DOI: https://doi.org/10.1007/s10509-013-1570-7

\bibitem{mio3} A. Khodam-Mohammadi, A. Pasqua, M. Malekjani, I. Khomenko, M. Monshizadeh, Astrophys Space Sci \textbf{345}, 415-420 (2013). DOI: https://doi.org/10.1007/s10509-013-1400-y
\bibitem{valerossi} S. Nojiri,   S.D. Odintsov,  Gen. Rel. Grav. \textbf{38}, 1285 (2006).

\bibitem{wangalfa} Y. Wang, L. Xu,  Phys. Rev. D, \textbf{81}, 083523 (2010).
\bibitem{hsunuovo} S. D. H. Hsu,  Phys. Lett. B \textbf{669}, 275 (2008).
\bibitem{ref145d} A. Sheykhi, M. Jamil, Phys. Lett. B \textbf{694}, 284 (2011).
\bibitem{ref144d} L. Amendola, D. Tocchini-Valentini, Phys. Rev. D \textbf{64}, 043509 (2001).
\bibitem{ref146d} W. Zimdahl, D. Pavon, Gen. Rel. Grav. \textbf{35}, 413 (2003).
\bibitem{ref147d} C. Feng, et al., Phys. Lett. B \textbf{665}, 111 (2008).
\bibitem{ref148d} K. Ichiki, et al., J. Cosmol. Astropart. Phys. \textbf{06}, 005 (2008).





\bibitem{altri1} D. Jain, A. Dev, Phys. Lett. B \textbf{633}, 436 (2006).
\bibitem{altri2} S. Nojiri, S.D. Odintsov, Phys. Rev. D \textbf{72}, 023003 (2005).
\bibitem{altri3} S. Capozziello, V. Cardone, E. Elizalde, S. Nojiri, S.D. Odintsov, Phys. Rev. D \textbf{73}, 043512 (2006).
\bibitem{gcg1} A. Kamenshchik, U. Moschella, V.  Pasquier, Phys. Lett. \textbf{511}  265 (2001).
\bibitem{gcg6} Li, M., Li, X., \& Zhang, X., Science in China G: Physics and Astronomy, \textbf{53}, 1631 (2010)

 
\bibitem{gcg9} M.C. Bento, O. Bertolami, A.A. Sen, Phys. Rev. D \textbf{70}, 083519 (2004).  
\bibitem{gcg16} V. Gorini, A. Kamenshchik, U. Moschella, Phys. Rev. D \textbf{67}, 063509 (2003).  
\bibitem{gcg17} U. Alam, V. Sahni, T.D. Saini, A.A. Starobinsky, Mon. Not. R. Astron. Soc. \textbf{344}, 1057 (2003).  
\bibitem{gcg4} N. Bilic, G.B. Tupper, R.D. Viollier, Phys. Lett. B \textbf{535}, 17 (2002).  
\bibitem{gcg5} J.C. Fabris, S.B.V. Goncalves, P.E. de Souza, Gen. Rel. Grav. \textbf{34}, 53 (2002).  


\bibitem{gcg15} V. Gorini, A.Y. Kamenshchik, U. Moschella, O.F. Piattella, A.A. Starobinsky, J. Cosmol. Astropart. Phys. \textbf{02}, 016 (2008).  
\bibitem{gcg7} X. Zhang, J. Zhang, J. Cui, L. Zhang, Mod. Phys. Lett. A \textbf{24}, 1763 (2009).  

\bibitem{mcg1} M. Jamil, Y. Myrzakulov, O. Razina, R. Myrzakulov, Astrophys. Space Sci. \textbf{336}, 315 (2011).  
\bibitem{mcg2} M. Jamil, U. Debnath, Astrophys. Space Sci. \textbf{333}, 3 (2011).  
\bibitem{mcg3} U. Debnath, M. Jamil, Astrophys. Space Sci. \textbf{335}, 545 (2011).  
\bibitem{mcg4} M.U. Farooq, M. Jamil, U. Debnath, Astrophys. Space Sci. \textbf{334}, 243 (2011).  
\bibitem{mcg5} U. Debnath, A. Banerjee, S. Chakraborty, Class. Quantum Grav. \textbf{21}, 5609 (2004).  

\bibitem{mvcg1} Z.K. Guo, Y.Z. Zhang, arXiv:astro-ph/0506091.  
\bibitem{mvcg2} M.C. Bento, O. Bertolami, A.A. Sen, Phys. Lett. B \textbf{575}, 172 (2003).  
\bibitem{mvcg3} G. Sethi, S.K. Singh, P. Kumar, D. Jain, A. Dev, arXiv:astro-ph/0508491 (2005).  
\bibitem{mvcg4} Z.K. Guo, Y.Z. Zhang, arXiv:astro-ph/0509790 (2005).  
\bibitem{mvcg5} U. Debnath, Astrophys. Space Sci. \textbf{312}, 295 (2007).  
\bibitem{mvcg8} O. Bertolami et al., Mon. Not. R. Astron. Soc. \textbf{353}, 329 (2004).  
\bibitem{mvcg9} A.G. Tekola, arXiv:0706.0804 [gr-qc].  
\bibitem{mvcg6} Z.K. Guo, N. Ohta, Y.Z. Zhang, arXiv:astro-ph/0505253.  












\bibitem{Tawfik:2021rvv} A. Tawfik, C. Greiner, Entropy \textbf{23} 295 (2021).
\bibitem{Tawfik:2010bm} A. Tawfik, T. Harko,  H. Mansour, M. Wahba,  Annalen Phys. \textbf{523} 194 (2011).
%
\bibitem{visc29} B. Li and J. D. Barrow, Phys. Rev. D \textbf{79}, 103521 (2009)
\bibitem{visc30} J. D. Barrow, Phys. Lett. B \textbf{180}, 335 (1987)
\bibitem{visc31} I. Brevik, S. D. Odintsov, Phys. Rev. D \textbf{65}, 067302 (2002)
\bibitem{visc32} D. J. Liu and X. Z. Li, Phys. Lett. B \textbf{611}, 8 (2005)
\bibitem{visc18} J.D. Nightingale, Astroph. J. \textbf{185} (1973) 105.
\bibitem{visc19} I. Waga et al, Phys. Rev. D \textbf{33} (1986) 1839.
\bibitem{visc21} C.P. Singh et al, Class. Quantum Gravit. \textbf{24} (2007) 455.
\bibitem{visc22} R. Colistete et al, Phys. Rev. D \textbf{76} (2007) 103516.
\bibitem{visc25} M-Guang Hu and X-He Meng, Phys. Lett. B \textbf{635} (2006) 186.
\bibitem{visc26} N. Cruz et al, Phys. Lett. B \textbf{646} (2007) 177.
\bibitem{visc33}  C. Eckart, Phys. Rev. \textbf{58}, 919 (1940).
\bibitem{visc34}  L. D. Landau and E. M. Lifshitz, Fluid Mechanics (Butterworth Heinemann,1987).
\bibitem{visc36}  W. Israel and J. M. Stewart, Phys. Lett. A \textbf{58}, 213 (1976).
\bibitem{visc37} T. Harko and M. K. Mak, Class. Quantum Grav. \textbf{20}, 407 (2003) .
\bibitem{visc40}  I. H. Brevik and O. Gorbunova, Gen. Rel. Grav. \textbf{37}, 2039 (2005) .
\bibitem{visc41}  X. H. Zhai, Y. D. Xu and X. Z. Li, Int. J. Mod. Phys. D \textbf{15}, 1151 (2006)..
\bibitem{visc44} X. H. Meng, J. Ren and M. G. Hu, Commun. Theor. Phys. \textbf{47}, 379 (2007).
\bibitem{visc48} D. F. Mota, J. R. Kristiansen, T. Koivisto and N. E. Groeneboom, Mon. Not. Roy. Astron. Soc. \textbf{382}, 793 (2007).
\bibitem{visc49} T. Koivisto and D. F. Mota, Phys. Rev. D \textbf{73}, 083502 (2006) .
\bibitem{visc50} T.Padmanabhan and S.M. Chitre, Phys. Letts. A , \textbf{120}, 433 (1987).







\bibitem{brevik1} I. Brevik, \& O. Gorbunova, European Physical Journal C, \textbf{56}, 425 (2008).
\bibitem{brevik3} I. Brevik, O. Gorbunova, Y. A. \& Shaido,   International Journal of Modern Physics D, \textbf{14}, 1899 (2005)
\bibitem{ren1} J. Ren, \& X. H. Meng,  Physics Letters B, \textbf{633}, 1 (2006)
\bibitem{ren2} J. Ren, \& X. H. Meng, Physics Letters B, \textbf{636}, 5 (2006)
\bibitem{visc51}  W. Zimdahl and D. Pavon, Phys. Rev. D \textbf{61}, 108301 (2000).
\bibitem{visc52}  O. Gron, Astrophys. Space Sci. \textbf{173}, 191 (1990).
\bibitem{visc9} M. Jamil, M. A. Rashid, Eur. Phys. J. C \textbf{56} (2008) 429.







\bibitem{dbi1} Y. F. Cai, J. B.  Dent, D. A. Easson, Phys. Rev. D. \textbf{83}, 101301 (2011).
\bibitem{dbi3} S. Chattopadhyay, U. Debnath,   Int. J. Theor. Phys. \textbf{49}, 1465 (2010).
\bibitem{dbi5} J. E. Lidsey and I. Huston, JCAP \textbf{0707} 002 (2007).
\bibitem{dbi6} W. H. Kinney and K. Tzirakis, Phys. Rev. D \textbf{77} 103517 (2008).
\bibitem{dbi7} J. Martin and M. Yamaguchi, Phys. Rev. D \textbf{77} 123508 (2008).
\bibitem{dbi8} M. Spalinski, JCAP 017 05 (2007)  Phys. Lett. B \textbf{650} 313 (2007).
\bibitem{dbi9} X. Chen, M. X. Huang, S. Kachru and G. Shiu, JCAP \textbf{0701} 002 (2007).
\bibitem{dbi10} S. Kecskemeti, J. Maiden, G. Shiu and B. Underwood, JHEP \textbf{0609} 076 (2006).
\bibitem{dbi11} M. X. Huang, G. Shiu and B. Underwood, Phys. Rev. D \textbf{77}023511 (2008).
\bibitem{dbi12} D. Seery and J. E. Lidsey, Phys. Rev. D \textbf{75} 043505 (2007).
\bibitem{ym1} Y.Zhang, Phys.Lett.B \textbf{340} (1994) 18


\bibitem{ym9-3} M. Tong, Y. Zhang and T. Xia, Int. J. Mod. Phys. D \textbf{18} 797 (2009).
\bibitem{ym9-5} W. Zhao, Astron. Astrophys. \textbf{9} 874 (2009)
\bibitem{ym9-7} Y. Zhang, T. Y. Xia , and W. Zhao, Class. Quant. Grav.\textbf{24} 3309 (2007).
\bibitem{ym18} H.Politzer, Phys.Rev.Lett. \textbf{30} (1973) 1346
\bibitem{ym18-1}D.J.Gross and F.Wilzcek, Phys.Rev.Lett. \textbf{30} (1973) 1343
\bibitem{ym21} S. L. Adler, Phys.Rev.D \textbf{23}, 2905 (1981).
\bibitem{ym21-1} S. L. Adler, Nucl.Phys.B \textbf{217}, 3881 (1983).
\bibitem{ym16} H.Pagels \& E.Tomboulis, Nucl.Phys.B \textbf{143}  485 (1978).
\bibitem{ym19} S.Coleman and E.Weinberg, Phys.Rev.D \textbf{7}  1888 (1973). 
\bibitem{ym20} L.Parker and A.Raval, Phys.Rev.D \textbf{60} 063512 (1999) .
\bibitem{ym10} E. J. Copeland, A. R. Liddle and D. Wands, Phys. Rev. D \textbf{57}, 4686 (1989).
\bibitem{ym10-1} T. Barreiro, E.J. Copeland, and N.J. Nunes, Phys.Rev.D \textbf{61}, 127301 (2002) .
\bibitem{ele1} N. Breton, R. Garcia-Salcedo, arXiv:hep-th/0702008 . DOI:
https://doi.org/10.48550/arXiv.hep-th/0702008
\bibitem{ele2} V. A. De Lorenci et al, Phys. Rev. D \textbf{65} 063501 (2002).




\end{thebibliography}
\end{document}